# Information Theoretic Resources in Quantum Theory

Sebastian Meznaric
Lincoln College, Oxford

A thesis submitted to the Mathematical and Physical Sciences Division for the degree of Doctor of Philosophy in the University of Oxford

*"Contemplation is a luxury." -Jean Paul Sartre*

# Abstract

Resource identification and quantification is an essential element of both classical and quantum information theory. Entanglement is one of these resources, arising when quantum communication and nonlocal operations are expensive to perform. In the first part of this thesis we quantify the effective entanglement when operations are additionally restricted to account for both fundamental restrictions on operations, such as those arising from superselection rules, as well as experimental errors arising from the imperfections in the apparatus. For an important class of errors we find a linear relationship between the usual and effective higher dimensional generalization of concurrence, a measure of entanglement.

Following the treatment of effective entanglement, we focus on a related concept of nonlocality in the presence of superselection rules (SSR). Here we propose a scheme that may be used to activate nongenuinely multipartite nonlocality, in that a single copy of a state is not multipartite nonlocal, while two or more copies exhibit nongenuinely multipartite nonlocality. The states used exhibit the more powerful genuinely multipartite nonlocality when SSR are not enforced, but not when they are, raising the question of what is needed for genuinely multipartite nonlocality. We show that whenever the number of particles is insufficient, the degrading of genuinely multipartite to nongenuinely multipartite nonlocality is necessary.

While in the first few chapters we focus our attention on understanding the resources present in quantum states, in the final part we turn the picture around and instead treat operations themselves as a resource. We provide our observers with free access to classical operations - ie. those that cannot detect or generate quantum coherence. We show that the operation of interest can then be used to either generate or detect quantum coherence if and only if it violates a particular commutation relation. Using the relative entropy, the commutation relation provides us with a measure of nonclassicality of operations. We show that the measure is a sum of two contributions, the generating power and the distinguishing power, each of which is separately an essential ingredient in quantum communication and information processing. The measure also sheds light on the operational meaning of quantum discord - we show it can be interpreted as the difference in superdense coding capacity between a quantum state and a classical state.

# Acknowledgements

First of all, it was a privilege to have had the opportunity to work with Prof Dieter Jaksch. His guidance and numerous comments on my research have been invaluable. Without the benefit of his expertise the task would have been considerably more difficult.

I would like to thank Stephen Clarke, whose excellent ideas and insight have led to numerous interesting discussions as well as continuing fruitful collaborations. His great ability to motivate, deep intuition and rigurous approach to physics have been instrumental. Stephen also introduced me to Animesh Datta, whose attention to detail and a seemingly inexhaustible source of useful ideas have greatly enhanced the quality of our work. I have also collaborated with Libby Heaney, whom I am grateful to for her guidance and for introducing me to the topics of superselection rules as well as their impact on mode entanglement and nonlocality. I am grateful to Jacob Biamonte for the most excellent course he has delivered in the physics department in Oxford on the topic of Penrose string diagrams, which led to fruitful research on their application to evolution of entanglement. I am also grateful to all of above the mentioned people for reading and reviewing my thesis and providing valuable comments.

For financial support I would like to thank EPSRC research council, who have provided me with financial support during the first three years and to Lincoln College and Department of Physics for providing funds enabling me to attend conferences.

Finally, I would like to thank my parents and my wife, Xiang Zeng, whose support and encouragement have made my task considerably easier.

# Contents

CHAPTER 1

# Introduction

The history of classical information theory began when Claude Shannon published his seminal paper in 1948 [1] in which he defined the terminology and proved the basic results that were to form the foundation of the field. Since its inception, the goal of information theory has been to quantify the amount of information transmitted during a communication in an *encoding-independent* manner. This means that it does not matter whether the information is written onto a piece of paper, carved into stone, transmitted audibly using the Morse code or with the help of electromagnetic radiation, the method of quantification must stay the same.

However, the independence of encoding also presents an interesting conundrum for physicists in that it suggests that information theory should then also be independent of physics itself. As long as there exist mutable physical objects whose various forms can be reliably distinguished from one another, the independence of encoding makes the underlying physics irrelevant. This is reflected by the fact that until relatively recently, information theory has been mainly a mathematical endeavour. However, with the advent of quantum theory and our increasing understanding of its functioning, there came a realization that quantum objects are so fundamentally different from the classical objects, that they actually violate some of the assumptions made by the classical information theory. This meant classical information was no longer sufficient to describe all information transfers we can observe in nature, placing physics firmly into the centre ground of the study of information.

Classically, information is usually quantified by counting the *minimum* number of mutable physical objects, each with a fixed number of distinguishable forms, that must be prepared in order to record some information. Because the number of forms is fixed, we may distinguish exponentially many of them with the increasing number of objects. This immediately suggests that quantity of information is related to the logarithm of the number of possible states. Encoding-independence enters through the fact that we take the minimum possible *number* of such objects, regardless of how they are actually realized.

Quantum mechanically, however, objects may be found not only in a fixed number of states, but also in quantum superpositions of those possibilities. Thus, we are now faced with infinitely many possible forms some of which are not even perfectly distinguishable from one another. Noncommutativity of quantum observables ensures that our classical intuition about information is immediately thrown out the window!

Particularly curious consequences of the ability to form superpositions were initially examined by Einstein, Podolsky and Rosen in an effort to demonstrate that quantum mechanics is not a complete theory [2]. They required of any complete theory to have an *element of reality* corresponding to each physical quantity. This meant that if a particular observable is measured, then another observable must still have a well defined value even if it was not actually measured. However, noncommuting observables, such as momentum and position, do not satisfy this criterion, a fact that they made apparent using entangled states.

Simultaneous reality for different observables was such a fundamental and crucial concept to them that they took its violation as proof that quantum theory was not a complete theory, beginning the quest for finding a theory of which quantum mechanics is a special case and which may be called complete. Almost 30 years later, John Bell devised a now famous thought experiment that for the first time showed that finding such a theory is in fact impossible [3–5]. A consequence of Bell's work is that all complete theories with simultaneous realities for all observables satisfy inequalities now known as the Bell inequalities. Quantum theory, however, predicts that these inequalities are violated.

The states that violate Bell inequalities are not just interesting for fundamental reasons, they can also be very useful. Namely, some of them come with the capacity to outstrip the communication performance of classical states and generally do not conform to the notions of classical information theory. For instance, to fully specify a quantum state one needs to write down several complex valued parameters. Quantum mechanically it is possible to transfer this information to another party by only transmitting a small integer and without ever knowing what the state actually is! Classically, this is mind boggling. Quantum mechanically, it is a protocol known as the quantum teleportation developed in early 90s [6]. As an application of quantum indistinguishability, quantum cryptography was developped allowing the communicating parties to detect any attempt at eavesdropping on their effort to generate a joint cryptographic key while located far apart [7, 8]. Classical systems have determined realities in line with Einstein, Podolsky, Rosen reasoning and can be measured without altering that reality, making quantum cryptography unattainable classically.

The discovery of better performance in these and other tasks provided motivation to determine what is the magical ingredient in quantum states allowing them to display this nonclassical behaviour. This was and still is guided by several goals, including these identified by Nielsen and Chuang [9]:

(i) *Identify the static information resources provided by quantum mechanics.* Fundamental examples of these resources include a quantum state in two-dimensional Hilbert space, named a *qubit*, as well as the classical information resources such as the *bit*, with the von Neumann entropy providing a method of quantification. Other novel resources identified so far include entanglement [10–12] and more recently the more general quantum discord [13–17].

(ii) *Identify the dynamical processes of quantum mechanics that can aid in the quantum information tasks.* These processes include foundational quantum gates, measurements, channels used in transmission of qubits and so on.

(iii) *Quantify resource trade-offs when performing dynamical processes using static resources.* Examples include using local operations and classical communication on entangled states, discovering how other resources change while a class of operations is applied, etc.

In quantifying the static information resources, the first of the above goals, we usually proceed by fixing a certain set of dynamical processes that are allowed during this procedure. The allowed processes are usually cheap, while those that are not allowed are expensive. Without this distinction, it is impossible to consider some states as resources. To see why this is the case, consider the reverse scenario - all dynamical processes are free or at least cheap. Given any two quantum states $\rho$ and $\sigma$, there exists some dynamical process that maps $\rho$ to $\sigma$. Therefore it does not matter which state we are given, as we can easily transform it to any other state, meaning that no state is particularly more useful than any other.

However, some states do become more important when a restriction is made concerning the allowed dynamical processes. As an example, suppose that a state $\rho$ is shared among two observers who are located very far away from one another and where transferring of one observer's part of the state to the other observer is very expensive, because quantum states are very fragile and difficult to transfer. Then the operations that are now cheap are precisely those that can be performed locally by the observers, allowing only classical communication between them. This defines the now ubiquitously used class of local operations and classical communication (LOCC). It has been shown that some states, such as separable states, cannot be transformed into all other states using only LOCC. This establishes a natural hieararchy of states, where those states that can be transformed into other states are higher up on the hierarchy and vice versa (see figure 1.1). The ability of the superior states to transfer into a greater number of states makes them a valuable resource, while the value of the states further down the chain is diminished. We therefore see that in order to understand the value of a given state, it is essential to understand which dynamical processes are readily available.

The states that are found closer to the top of the hierarchy are also better at performing those quantum information tasks that use only the operations from the allowed class. This is because the ability to transform the state $\rho$ into $\sigma$ implies that the state $\rho$ can be used to perform the task at least as well as the state $\sigma$ by first transforming it into $\sigma$ and then completing the task. This suggests another way to attain a hierarchy of states - compare their performance using a specific set of tasks one would like to use them for and place the states that perform better further up in the hierarchy structure.

However, in an experimental setup it is rather rare that one can peform all the operations in the often used classes of operations (such as LOCC). Almost always the apparatus will posses some kind of imperfections, such as for example making errors during measurement [18]. In practice this imposes a further restriction on which dynamical processes one is able to perform. In our example, those processes that never make an experimental error cannot be performed. Since the allowed set of dynamical processes is therefore now smaller, we must correspondingly update our state hierarchy and potentially change the way resources are counted. Particularly, one would like to compare the resource with restricted operations to that attainable without the additional restrictions. One potential method, and one we will adopt in this thesis, is to compare the performance of states in certain tasks with and without restrictions and then use the results to place the state at an appropriate position in the hierarchy.

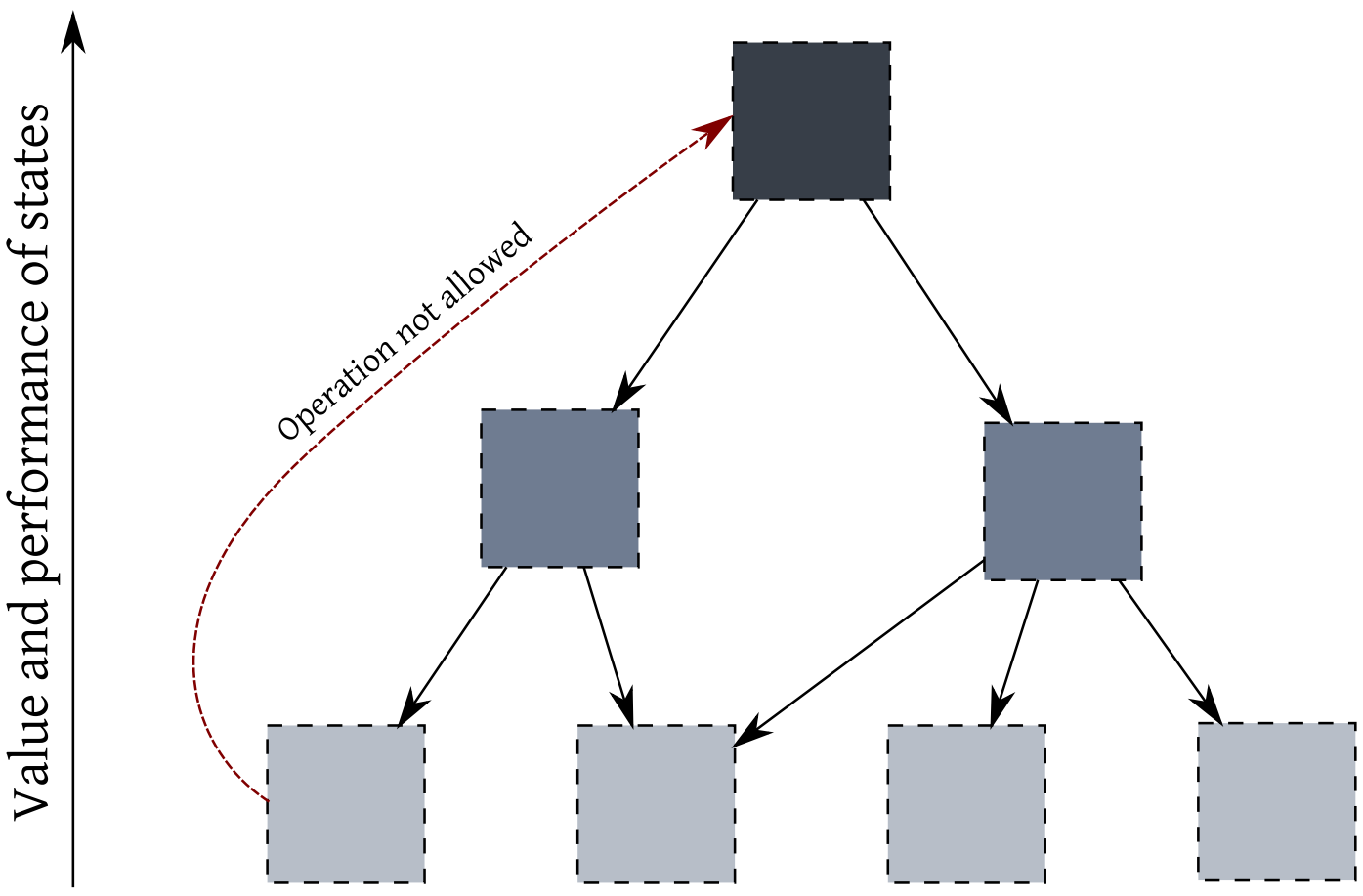

Figure 1.1: A potential state hierarchy established by a set of dynamical processes, where the states further down the hierarchy can be obtained from some states further up (denoted by solid black line), but the inferior states cannot be transformed into the superior ones (denoted by red dotted line). This ordering makes the superior states a valuable resource. In the paradigm of local operations and classical communication the less entangled states cannot be transformed into the more entangled states.

A particular example of this is LOCC class becoming more limited by experimental errors. We expect that with operations becoming more restricted, the power of the entangled states should diminish and with it the effective entanglement should also decrease. This process is partially understood for the special case of operations being required to satisfy certain symmetry requirements. The symmetry gives rise to operational restrictions known as the superselection rules [19] which, unless they can somehow be overcome, reduce the effective entanglement [20–22]. In fact this occurs to the extent that some states that are otherwise maximally entangled become effectively separable.

This suggests that restricting the allowed operations has a powerful effect on our capability to extract quantum advantage from a state, even if the given state is otherwise placed at the top of the hierarchy. This consideration leads to the question of whether we can turn things around by taking certain *states* as being cheap and others as expensive and asking what properties should our set of *operations* posses in order to extract quantum advantage from the expensive states. Firstly, given that expensive states are likely to be expensive because of their relative rarity and/or fragility in the environment, we would like our operations to be able to generate the expensive states out of cheap ones. Secondly, our operations must be able to distinguish the expensive states from cheap states. This is particularly important - if the operations do not differentiate between the cheap and expensive states, then possessing the expensive states is not useful and one might as well use the cheap states. If our set of available operations can neither generate nor distinguish expensive states from cheap ones, it prevents us from using any potential performance boosting capability of the expensive states. We will make use of this idea later in the thesis to determine what properties operations must posses in order to take advantage of quantum coherence. We now proceed to give the general layout of the thesis.

## 1.1 Thesis overview

In chapter 2, we provide background information that will be needed later in the thesis. We provide an overview of a useful diagrammatic language followed by a brief review of the formalism of generalized quantum measurement. We complete the chapter by reviewing the quantum operations behind the evolution of quantum states, encompassing pure unitary evolution as well as noisy open-system evolution.

We provide a basic review of quantum information theory in chapter 3. Here we begin by discussing the fundamental differences between quantum and classical world in terms of probability distributions and give an overview of how information is quantified. Rather than take the standard route of beginning with the Shannon entropy and then generalizing to other quantities, we place the relative entropy at the centre of our discussion and derive other information theoretic quantities from it, similarly to how many thermodynamic quantities follow from the canonical partition function.

We continue the chapter by examining the generalization of classical information theory to the quantum context. Here we encounter quantities that have recently been at the centre of attention for the quantum information - nonclassical correlations. These can be found in section 3.2 and include a brief review of entanglement and quantum discord. Finally, section 3.3 provides an introduction to quantum nonlocality and Bell inequalities, exposing the fundamental physical and mathematical reasons of why quantum information theory cannot be reduced to the classical information theory.

In chapter 4 we show how entanglement, a ubiquitous quantum resource, can be quantified when operations are further limited either by possessing an imperfect experimental apparatus or through fundamental considerations. We are greatly aided in this task by the remarkable progress recently [23, 24] in obtaining a complete characterisation of a quantum optical experiments by performing tomography not only of the input state and dynamics of an apparatus, but also of the photon detector itself. Experimentalists are now able to understand all three components of a quantum communication protocol, input, process and measurement [25–27] allowing them to make more effective use of their device [28, 29].

We take the approach of using nonlocal games [30], introduced at the beginning of the chapter, as a gauge of the amount of effective entanglement. It was shown in [30] that entangled quantum states are always strictly better at playing these games than separable states and moreover, those states that perform better in these games must also be more entangled. Since we limit the parties to be able to perform only the measurements from a certain limited set, we can generate a restriction-dependent entanglement measure by comparing the thus attained performance to that attainable without the restriction.

In chapter 5 we study the behaviour of multipartite nonlocality when operations are not allowed to form or detect coherent superpositions of states with different particle numbers, known as the particle number superselection rule (SSR). Since multipartite nonlocality can be either the more powerful genuinely multipartite or form or the less powerful nongenuinely multipartite, we examine whether enforcing the SSR degrades the nature of nonlocality. After studying several states, we provide a gen-eral no-go theorem, where we provide a necessary condition that must be satisfied in order for the nonlocality to not be degraded.

In chapter 6 we turn our attention away from the states and examine the quantumness of operations themselves. We take the cheap, classical states to be those that easily survive in the classical environment while expensive, quantum states contain fragile quantum superpositions, while a classical observer is one that is not fast enough to beat the effects of the environment. We then suppose that the observer is given some quantum operation as a black box and study its usefulness. We show that the observer can use the operation to generate quantum states or distinguish them from classical states if and only if the operation satisfies an evironment dependent commutation relation. This relation allows us to quantify quantumness of an operation by measuring how far away an operation is from satisfying the relation.

Finally, we conclude the thesis in chapter 7.

# CHAPTER 2

# Quantum Theory

As one of the pillars of modern physics, quantum theory still plays a major role in our understanding of physics. Here we review some of the basic mathematical components of quantum theory. We begin by introducing the Penrose's graphical representation of tensors and operations thereon. These will play an important role in elucidating the nature of quantum operations as well as later on in the thesis in illustrating important concepts. Quantum mechanically a pure state is a state vector $|\psi\rangle$ in some Hilbert space $\mathcal{H}$, while a mixed state may be written as $\rho = \sum_i p_i |\psi_i\rangle \langle\psi_i|$, where the state $|\psi_i\rangle$ occurs with probability $p_i$. As bounded linear operators, quantum states are particularly amenable to a representation via a tensor structure. A pure state, being isomorphic to a vector, may be represented as a single index tensor $\psi_\mu$ while mixed states can be represented as two-index tensors $\rho_{\mu\nu}$ [9, 31]. A particular feature of tensor spaces is that the they are equipped with an inner product, defined using a metric tensor $g_{\mu\nu}$. Namely for any two vectors $\psi_\mu, \phi_\mu$, inner product is evaluated by computing $g^{\mu\nu}\psi_\mu\phi_\nu$, where $g^{\mu\nu}$ is the inverse matrix of $g_{\mu\nu}$. We can write this shortly as $\psi^\mu\phi_\mu$, where $\psi^\mu = g^{\mu\nu}\psi_\nu$. In sharp contrast with this picture, inner product in complex Hilbert spaces, written in the Dirac notation as $\langle\phi\,|\,\psi\rangle$, is the Euclidean inner product between one vector $|\psi\rangle$ and the complex conjugate transpose of the vector $|\phi\rangle$. The vector $\langle\phi|$, complex conjugate of vector $|\phi\rangle$, is also known as the dual vector of $|\phi\rangle$. Denote $\boldsymbol{C}(|\phi\rangle) = \langle\phi|$ as the operator mapping the Hilbert space $\mathcal{H}$ to its dual. Importantly, notice that $\boldsymbol{C}$ is not a linear operator as $\boldsymbol{C}(\alpha\,|\phi\rangle + \beta\,|\psi\rangle) = \alpha^*\,\langle\phi| + \beta^*\,\langle\psi| \neq \alpha\,\langle\phi| + \beta\,\langle\psi|$ unless $\alpha$ and $\beta$ happen to be real. This must immediately imply that the inner product in Hilbert spaces does not admit a description using the metric tensor, highlighting an important difference between tensors and Hilbert spaces.

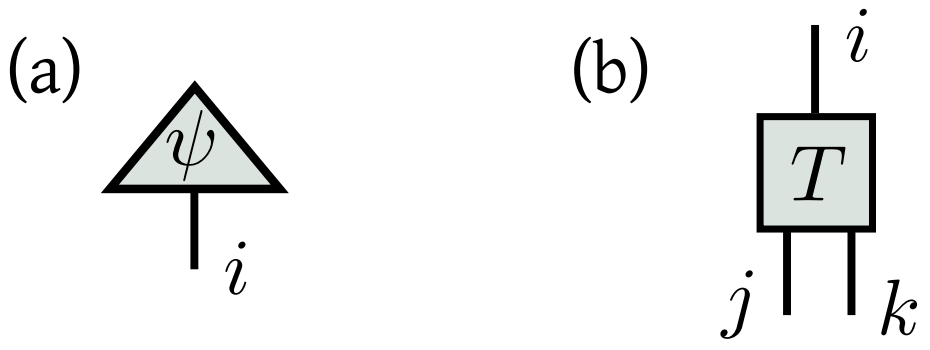

Figure 2.1: In this example we show tensors (a) $\psi_i$, representing a quantum state $|\psi\rangle$ and (b) $T^{i}_{\ jk}$, representing a more general operator.

In this section we therefore take a slightly more general and relaxed attitude towards tensors, where raising and lowering of indices is done by taking the complex conjugate and where care must be taken to not assume linearity of the raising and lowering operation. With this in mind, we shall present here a particularly simple and appealing graphical language, developed originally by Penrose [32] and known as Penrose tensor networks or string diagrams, which will be particularly useful for depicting results about quantum operations. These diagrams can aid in intuition and also represent mathematical equations [33], with much work done recently on the theory and expresiveness of the tensor network diagrams [34–37].

A multi-index tensor in Penrose notation is presented as a box or a triangle with wires, or legs, extending outwards from it. Each of the legs represents an index, with those extending towards the right or downwards being the lowered indices and those extending towards the left or upwards being the raised indices (for example see figure 2.1).

Graphically a two party quantum state $|\psi\rangle$ would have one wire for each index. We can write $|\psi\rangle$ in abstract index notation as $\psi_{ij}$ or in terms of the Dirac notation convention (which we mainly adopt here) as $|\psi\rangle = \sum_{ij}\psi_{ij}\,|i\rangle \otimes |j\rangle$. Appropriately joining a flipped (transpose) and conjugated (star or overbar) copy of the quantum state $|\psi\rangle$ allows one to represent the density operator $\rho = \sum \psi_{ij}c^{kl}\,|ij\rangle\,\langle kl|$ as follows [32]. Here a box with two inputs $(i, j)$ and two outputs $(k, l)$ would be used to represent

$$\rho = |\psi\rangle\langle\psi| =$$

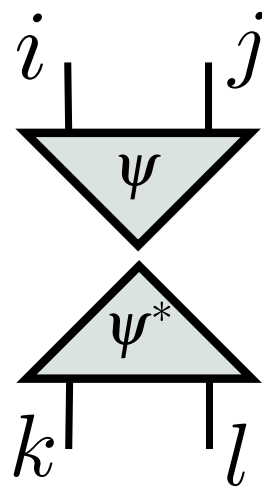

Figure 2.2: Penrose's graphical representation of a pure density matrix.

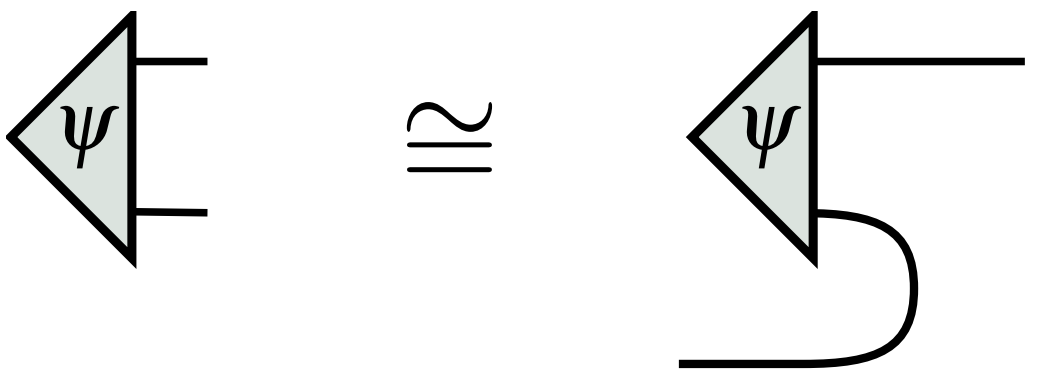

Figure 2.3: Bending one of the legs of the state tensor creates a matrix, dual of the state.

a general density operator $\rho = \sum \rho_{ij}^{\,kl} |ij\rangle \langle kl|$ with no known additional structure (see figure 2.2). The Penrose graphical calculus is a generalization of the language of quantum circuits [34].

We are interested in the following operations: (i) tensor index contraction by connecting legs of different tensors, (ii) raising and lowering indices by bending a leg upwards or downwards, respectively and thus taking the appropriate conjugate transpose and (iii) a duality between maps, states and linear maps in general, called *Penrose wire bending duality*. This duality will play a major role in our application of the language to entanglement evolution. The duality is obtained by noticing that states are tensors with all open wires pointing in the same direction. Bending some of the wires in the opposite direction makes various linear maps out of a given state. Given the computational basis, this amounts to turning all kets belonging to one of the Hilbert spaces into bras and vice versa. A bipartite state in the fixed standard basis $|\psi\rangle = \sum_{i,j} \psi_{i,j} |i\rangle \otimes |j\rangle$ for example, is dual to the operator $\sum_{i,j} \psi_{i,j} |i\rangle \langle j|$. Notice we did not take a complex conjugate here. This is because we can treat $\psi_{i,j}$ as belonging to either $|i\rangle$, or $|j\rangle$. This suggests the duality is not unique and care must be taken to maintain consistency. See also figure 2.3.

Particular example of interest are Penrose's cups, caps and identity wires. As in

[32], these three tensors are given diagrammatically as

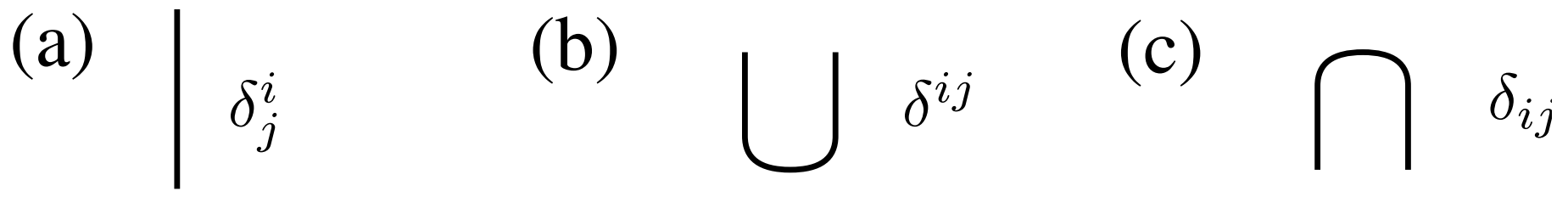

By thinking of these tensors now in terms of components, e.g. $\delta_{ij}$ is 1 whenever $i = j$ and 0 otherwise, we note that

$$\mathbf{1} = \sum_{ij} \delta^i_j \left|i\right\rangle \left\langle j\right| = \sum_k \left|k\right\rangle \left\langle k\right| \tag{2.1}$$

$$\left\langle 00\right| + \left\langle 11\right| + \cdots + \left\langle nn\right| = \sum_{ij} \delta^{ij} \left\langle ij\right| = \sum_k \left\langle kk\right| \tag{2.2}$$

$$\left|00\right\rangle + \left|11\right\rangle + \cdots + \left|nn\right\rangle = \sum_{ij} \delta_{ij} \left|ij\right\rangle = \sum_k \left|kk\right\rangle \tag{2.3}$$

where the identity map (a) corresponds to Equation (2.1), the cup (b) to (2.2) and the cap (c) to (2.3). The three equations are all different representations of the same tensor. The relation between these three equations is again given by Penrose's wire bending duality: in a basis, bending a wire corresponds to changing a bra to a ket, and vise versa, allowing one to translate between Equations (2.1), (2.2) and (2.3) at will.

As an example, consider quantum teleportation (figure 2.4). We start with a Bell state as in Eq. (2.3) and some arbitrary state $\left|\psi\right\rangle$. Then we take a joint inner product of this state with another Bell state, but this time represented as the one in Eq. (2.2), up to a unitary operation which depends on the Bell measurement outcome. Unbending all the wires leads to a single straight line, giving us the original state $\left|\psi\right\rangle$, up to a unitary operation.

The contraction of two tensor indices diagrammatically amounts to appropriately joining open wires. Given tensors $T^i{}_{jk}$, $A^l{}_n$ and $B^m_q$ we form a contraction by multiplying by $\delta^j_l \delta^q_k$ resulting in the tensor

$$T^i{}_{jk} A^j{}_l B^k{}_m := \Gamma^i{}_{lm} \tag{2.4}$$

where we use the Einstein summation convention (repeated indices are summed over) and the tensor $\Gamma^i{}_{lm}$ is introduced per definition to simplify notation. In quantum

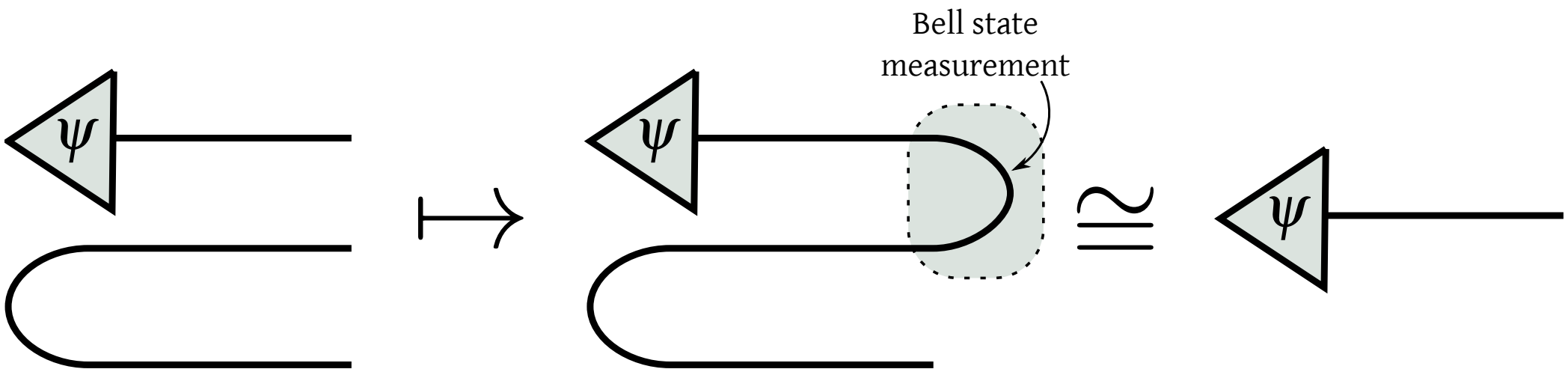

Figure 2.4: Graphical Penrose representation of quantum teleportation. Starting with a Bell state and some other state $|\psi\rangle$ and unbending the wires gives us, up to a unitary operation, the original state.

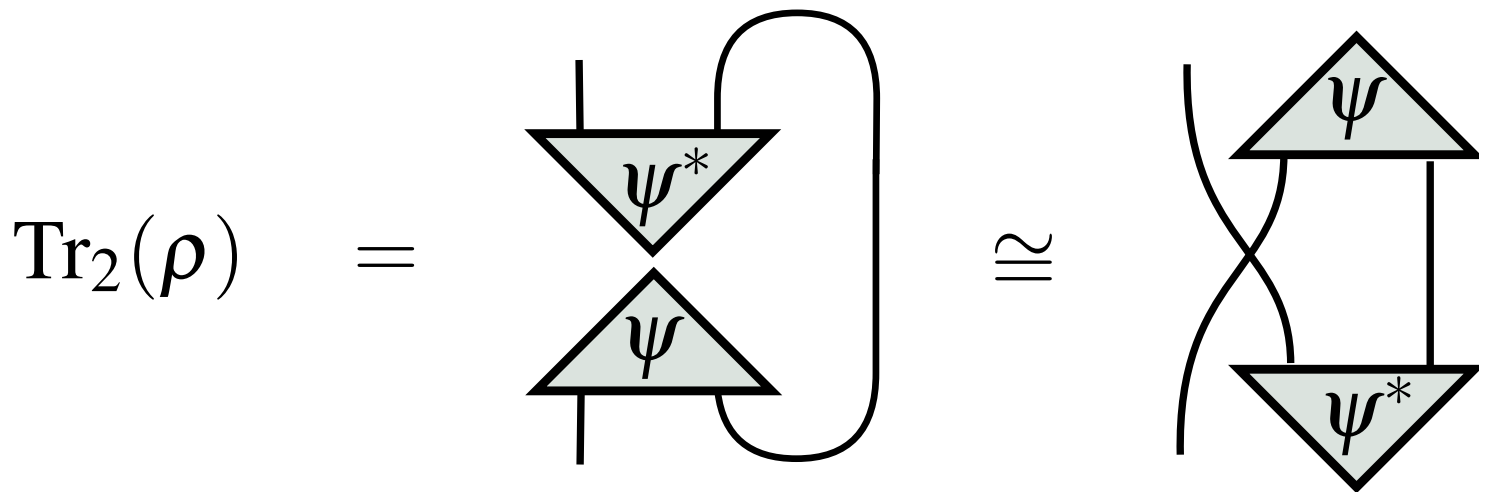

Figure 2.5: Graphical representation of a partial trace of a pure state. For a mixed state the diagram would look very similar, but with an additional leg connecting the states $\psi$ and $\psi^*$ to denote the internal structure of the four legged tensor $\rho$. The network on the far right is due to Penrose. (Note that Penrose used reflection across the page to represent adjoint, yet we have placed a star on $\psi$ to represent conjugate for illustrative purposes.)

physics notation, this is typically expressed in equational form as

$$\Gamma = \sum_{ilm} \Gamma^{i}{}_{lm} |lm\rangle \langle i| = \sum_{ijklm} T^{i}{}_{jk} A^{j}{}_{l} B^{k}{}_{m} |lm\rangle \langle i| \,. \tag{2.5}$$

**Remark 2.1** (Graphical trace - Penrose)**.** *Graphically the trace is performed by appropriately joining wires. The following depiction in Figure 2.5 (b) is Penrose's representation of a reduced density operator, where the stars on the $\Psi$'s represents complex conjugation.*

We have now presented the key tensor network building blocks used here. In practice, tensor networks contain an increasing number of tensors, making it difficult to form expressions using (inherently one-dimensional) equations. The two-dimensional diagrammatic depiction of tensor networks can simplify such expressions and often reduce calculations. A key component of this unified view relies on the *natural equivalence* induced by the so called *snake equations*, which we will review next.

## 2.1 Measurement in quantum mechanics

The conditions under which an experimenter may obtain information about a quantum system are typically presented as one of the postulates underpinning quantum mechanics.

**Postulate 2.1.** *Measurement is described by a set of measurement operators $\{\boldsymbol{M}_k\}$ such that $\sum_k \boldsymbol{M}_k^\dagger \boldsymbol{M}_k = \mathbf{1}$ [9]. The measured system, initially and just before the measurement prepared in the state $|\psi\rangle$, is found immediately after the measurement in the state*

$$\frac{\boldsymbol{M}_k |\psi\rangle}{\langle\psi| \boldsymbol{M}_k^\dagger \boldsymbol{M}_k |\psi\rangle^{1/2}}, \tag{2.6}$$

*provided measurement result $k$ is obtained and the denominator is here purely for normalization. The probability of obtaining the measurement result $k$ is given by $\langle\psi| \boldsymbol{M}_k^\dagger \boldsymbol{M}_k |\psi\rangle$. For mixed states the post measurement state when obtaining measurement outcome $k$ is*

$$\frac{\boldsymbol{M}_k \rho \boldsymbol{M}_k^\dagger}{\mathrm{Tr}[\boldsymbol{M}_k^\dagger \boldsymbol{M}_k\, \rho]} \tag{2.7}$$

*and the corresponding measurement probability is given by* $\mathrm{Tr}[\boldsymbol{M}_k^\dagger \boldsymbol{M}_k\, \rho]$.

The operators $\boldsymbol{M}_k$ may be any operators in $\mathcal{H}$, provided they satisfy the condition in the postulate. We can therefore see that there may be many operators corresponding yielding the same probabilities for any state. For example, $\boldsymbol{M}_k$ and $\boldsymbol{U}\boldsymbol{M}_k$, where $\boldsymbol{U}$ is a unitary operation, give the same measurement probabilities for all $k$ since $\boldsymbol{M}_k^\dagger \boldsymbol{U}^\dagger \boldsymbol{U} \boldsymbol{M}_k = \boldsymbol{M}_k^\dagger \boldsymbol{M}_k$. The role of the operator $\boldsymbol{U}$ here is to alter the post-measurement state but not the measurement statistics.

There may be occassions when we are not interested in the post-measurement states. For such occassions, a poweful formalism known as the positive operator valued measure (POVM) has been developed. A POVM is a set of positive operators $\{\boldsymbol{P}_k\}$ such that $\sum_k \boldsymbol{P}_k = \mathbf{1}$. Measurement probabilities are then given by $\langle\psi| \boldsymbol{P}_k |\psi\rangle$ for pure states, or $\mathrm{Tr}[P_k \rho]$ in general. Since a positive operator is by definition one that

can be written as $\boldsymbol{P}_k = \boldsymbol{A}_k^\dagger \boldsymbol{A}_k$, we see that the POVM formalism is fully compatible with the measurement postulate 2.1.

It appears as though the state changes rapidly and in a non-unitary fashion during the measurement process, apparently in contradiction with the Schrodinger equation. However, provided that one treats the measurement apparatus and the state of its pointer as itself being quantum, the contradiction disappears. Let the states of the apparatus corresponding to measurement outcome $k$ be denoted as $|k\rangle$. Then, the measurement process can be described as the state transformation $|\psi\rangle \otimes |0\rangle \mapsto \sum_k \boldsymbol{M}_k |\psi\rangle \otimes |k\rangle$. This transformation preserves the inner product as

$$\sum_{k,l} \langle\phi| \boldsymbol{M}_l^\dagger \boldsymbol{M}_k |\psi\rangle \langle l \,|\, k\rangle = \langle\phi \,|\, \psi\rangle \tag{2.8}$$

and can therefore be described as a joint unitary operation $\boldsymbol{U} |\psi\rangle \otimes |0\rangle$ (for more details on this unitary extension see [9], p.95). However, one might counter that an observer never actually sees this superposition, instead they see a single collapsed state. To see how that occurs, one must treat the observer as a quantum object, whose memory of the past events may be found in a number of states, each corresponding to a particular measurement outcome. Let us label the state where the observer remembers measurement result $k$ with the state $|\mu_k\rangle$, and having no memory at all about the measurement is the state $|0\rangle$. Then observing the apparatus leads to an interaction of the apparatus and the observer so that the joint state of the apparatus and the observer $|k\rangle \otimes |0\rangle$ transforms to $|k\rangle \otimes |\mu_k\rangle$.

The net result of measurement of the system by the apparatus, followed by the observation of the apparatus by the observer implies that the system-apparatus-observer transform jointly as $|\psi\rangle \otimes |0\rangle \otimes |0\rangle \mapsto \sum_k \boldsymbol{M}_k |\psi\rangle \otimes |k\rangle \otimes |0\rangle \mapsto \sum_k |\psi\rangle \otimes |k\rangle \otimes |\mu_k\rangle$. We see that the observer actually finds him/herself in a quantum superposition. The observation of the apparent collapse can be seen as merely a consequence of finding out which of the states in the superposition one is experiencing [38]. While the existence of the superposition implies all are actually experienced, similarly to how an electron or a photon traverses each of the trajectories in the double slit experiment, the very distinct memory states in each of the terms in the superposition imply that the observer can be aware of only one of the terms "at a time". We can thus conclude that a measurement process is a natural ingredient of quantum mechanics, arising from the nature of quantum evolution in general and in particular requires us to treat the observer him/herself as a quantum object in order to obtain a coherent and consistent picture. However, the emergence of the Born rule from this picture is still lacking a physical foundation and must be taken as a postulate [39]. The emergence of the classical world is then simply a consequence of joint unitary evolution of the observer, environment, and the measured system, more on which in the next section.

## 2.2 Evolution of quantum states

We have looked at quantum measurement and briefly at the unitary evolution of quantum states, a ubiquitous quantum operation taking a state $\rho$ to $\boldsymbol{U}\rho\boldsymbol{U}^{\dagger}$. We will now look at physical state transformations in general.

More generally, we admit a quantum operation $\$ : \mathcal{B}(\mathcal{H}) \to \mathcal{B}(\mathcal{H})$, where $\$$ is a linear *super-operator* mapping density operators to density operators. Conventionally the following properties are required of such an operation [9], and when they are both satisfied the operation is termed a *completely positive map* (CPM):

(i) *Linearity*, $\$(p_1\rho_1 + p_2\rho_2) = p_1\$(\rho_1) + p_2\$(\rho_2)$. This is to ensure that a probabilistic mixture of states is transformed to the probabilistic mixture of corresponding transformed states.

(ii) *Complete positivity*, meaning that introducing an arbitrary dimensional auxiliary Hilbert space, $\mathbf{1} \otimes \$$ maps positive operators on the joint Hilbert space to positive operators. This ensures that if a map acts on only a subsystem of a larger quantum state, its output remains physical.

It is often convenient to further assume that $\$$ is trace preserving, so that $\text{Tr}[\$(\rho)] = \text{Tr}[\rho]$, ensuring that density operators are mapped to density operators. Trace preserving maps are the only operations that can be executed deterministically. Complete positivity is normally taken as synonymous with physicality, although under certain conditions this requirement can be relaxed [40]. However, for our purposes we will assume complete positivity and trace preservation throughout this thesis.

Mathematically, there are many ways quantum operations can be represented. We will briefly look at the most important of these below, and show how they are related.

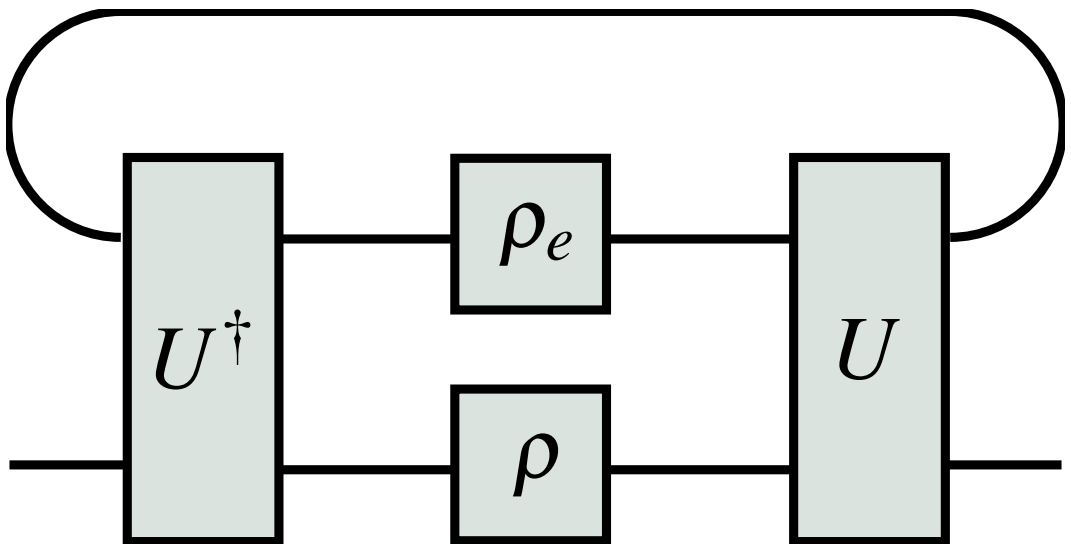


Figure 2.6: Tracing out the environment after joint unitary evolution leaves the evolution of the state $\rho$ alone in a non-unitary, but linear, form.

### 2.2.1 Evolution with environment

A typical way for effectively non-unitary evolution to arise in quantum mechanics is for a joint state of our system of interest and the environment to evolve unitarily in a joint manner through interaction. The evolution of the system alone is then not in general unitary. We assume that the initial state of the joint system is $\rho \otimes \rho_e$, where $\rho_e$ is the state of the environment. An operation \$ may be written as

$$\$(\rho) = \mathrm{Tr}_e \left[ \boldsymbol{U} \rho \otimes \rho_e \boldsymbol{U}^\dagger \right] . \tag{2.9}$$

If the initial state in the system is a product state of the form $\rho \otimes \rho_e$, the resulting operation \$ is completely positive. See figure 2.6. It has, however, recently been shown that a necessary and sufficient condition for complete positivity is for the initial joint state to be of zero discord, with measurements conducted on the state of the system (not the environment) [41]. We will come back to quantum discord in section 3.2.3. However, as shown in [42], it is generally true that a completely positive and trace preserving \$ can always be represented in the form of Eq. (2.9), a fact first noticed by Stinespring [43].

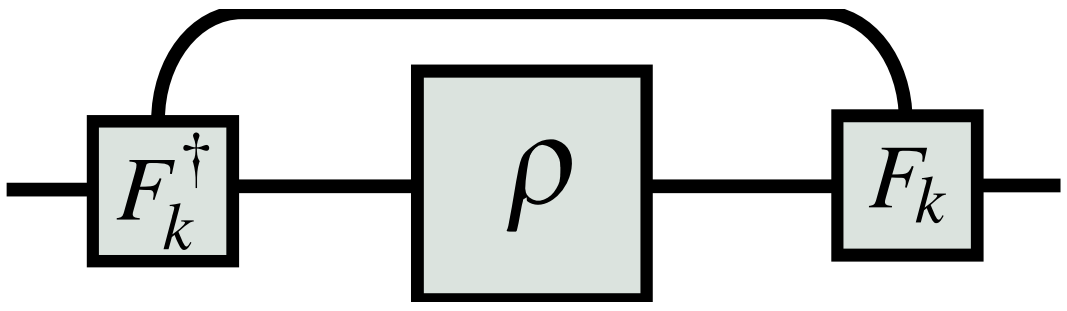


Figure 2.7: Diagrammatic representation of Kraus operators. Comparing with the figure 2.6, the formal equivalence of the two approaches to quantum operations can be read out.

### 2.2.2 Kraus operator representation

Given a *general*, not necessarily completely positive, operation \$ mapping bounded operators $\mathcal{B}(\mathcal{H}_1) \mapsto \mathcal{B}(\mathcal{H}_2)$, \$ is linear if and only if it can be represented as

$$\$(\boldsymbol{\chi}) = \sum_\alpha \boldsymbol{F}_\alpha \boldsymbol{\chi} \boldsymbol{G}_\alpha^\dagger, \tag{2.10}$$

where $\boldsymbol{\chi} \in \mathcal{B}(\mathcal{H}_1)$ and $\boldsymbol{F}_\alpha, \boldsymbol{G}_\alpha$ are the left and right operation elements, respectively. The map \$ is trace preserving if $\sum_\alpha \boldsymbol{F}_\alpha^\dagger \boldsymbol{G}_\alpha = \mathbf{1}$, where $\mathbf{1}$ is the identity operator living in $\mathcal{B}(\mathcal{H}_2)$, which can be seen by taking trace of both sides.

The operation is *Hermitian* if it maps Hermitian operators in $\mathcal{B}(\mathcal{H}_1)$ to Hermitian operators in $\mathcal{B}(\mathcal{H}_2)$, which is the case if and only if \$ can be represented as

$$\$(\boldsymbol{\chi}) = \sum_\alpha c_\alpha \boldsymbol{F}_\alpha \boldsymbol{\chi} \boldsymbol{F}_\alpha^\dagger, \tag{2.11}$$

where $c_\alpha \in \mathbb{R}$ (see figure 2.7). For complete positivity we must only require that $c_\alpha \geq 0$ for all $\alpha$. A proof can be found in [44]. This representation when the operation \$ is completely positive is known as the Kraus representation and the operators $\boldsymbol{F}_\alpha$ are known as the Kraus operators [45]. However, the result was already known before by Choi [46], building on ideas by Stinespring [43].

The Kraus operator formulation is equivalent to the formulation using the joint evolution with the environment. While this is always true, the correspondence is particularly simple when the initial system-environment joint state is a product state

$\rho \otimes \rho_e$. Then

$$\$(\rho) = \text{Tr}_e \left[ \boldsymbol{U} \rho \otimes \rho_e \boldsymbol{U}^\dagger \right] = \text{Tr}_e \left[ \sum_l \mu_l \boldsymbol{U} \rho \otimes |e_l\rangle \langle e_l| \boldsymbol{U}^\dagger \right] \tag{2.12}$$

$$= \sum_{k,l} \mu_l \langle e_k| \boldsymbol{U} \rho \otimes |e_l\rangle \langle e_l| \boldsymbol{U}^\dagger |e_k\rangle = \sum_k \boldsymbol{E}_k \rho \boldsymbol{E}_k^\dagger, \tag{2.13}$$

where $\boldsymbol{E}_k = \sum_l \mu_l \langle e_k| \boldsymbol{U} |e_l\rangle$ and where $|e_l\rangle$ are the eigenstates of the density operator $\rho_e$.

### 2.2.3 Choi matrix

The final way to faithfully represent a quantum operation rests on the map-state duality we considered before (figure 2.3). The Choi matrix of a map \$ is defined as [46]

$$\rho_{\$} = \frac{1}{d} \sum_{i,j} \$(|i\rangle \langle j|) \otimes |i\rangle \langle j| = \frac{1}{d} \$ \otimes \mathbf{1} \big( \sum_{i,j} |ii\rangle \langle jj| \big). \tag{2.14}$$

Notice that $\sum_{i,j} |ii\rangle \langle jj|$ is the maximally entangled state, and so Eq. (2.14) can be understood as the map \$ acting on one side of a maximally entangled state. The output is a density matrix. What makes the Choi matrix important is the fact that the action of \$ on a part of a general state $\rho$ can be fully understood in terms of the matrix $\rho_{\$} \cong \$$, seen in figure 2.8, where $\cong$ denotes duality. Specifically, we can show that

$$\mathbf{1} \otimes \$_\rho(\rho_{\$}) = \$ \otimes \mathbf{1}(\rho), \tag{2.15}$$

where $\$_\rho \cong \rho$. Thus, in the dual space the roles of the density matrices and operations on density matrices are reversed.

To see that this is the case, see figure 2.9, where we apply the map dual $\$_\rho \cong \rho$ to the state $\rho_{\$}$. Formally, we can write any density matrix as $\rho = \sum_{klmn} \rho_{klmn} |k\rangle \langle l| \otimes |m\rangle \langle n|$. Then the dual operation can be written as $\$_\rho(\sigma) = \sum_{klmn} \rho_{klmn} |m\rangle \langle k| \sigma |l\rangle \langle n|$, where $\sigma$, a density matrix, denotes the argument of the operation $\$_\rho$. We then obtain

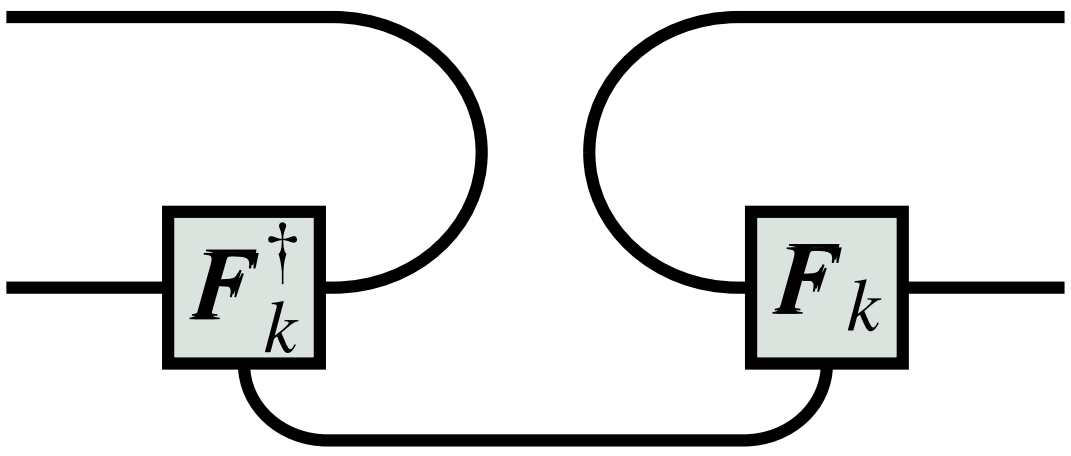

Figure 2.8: The state $\rho_\$$, Eq. 2.14, corresponds to the Choi matrix of the map $\$$.

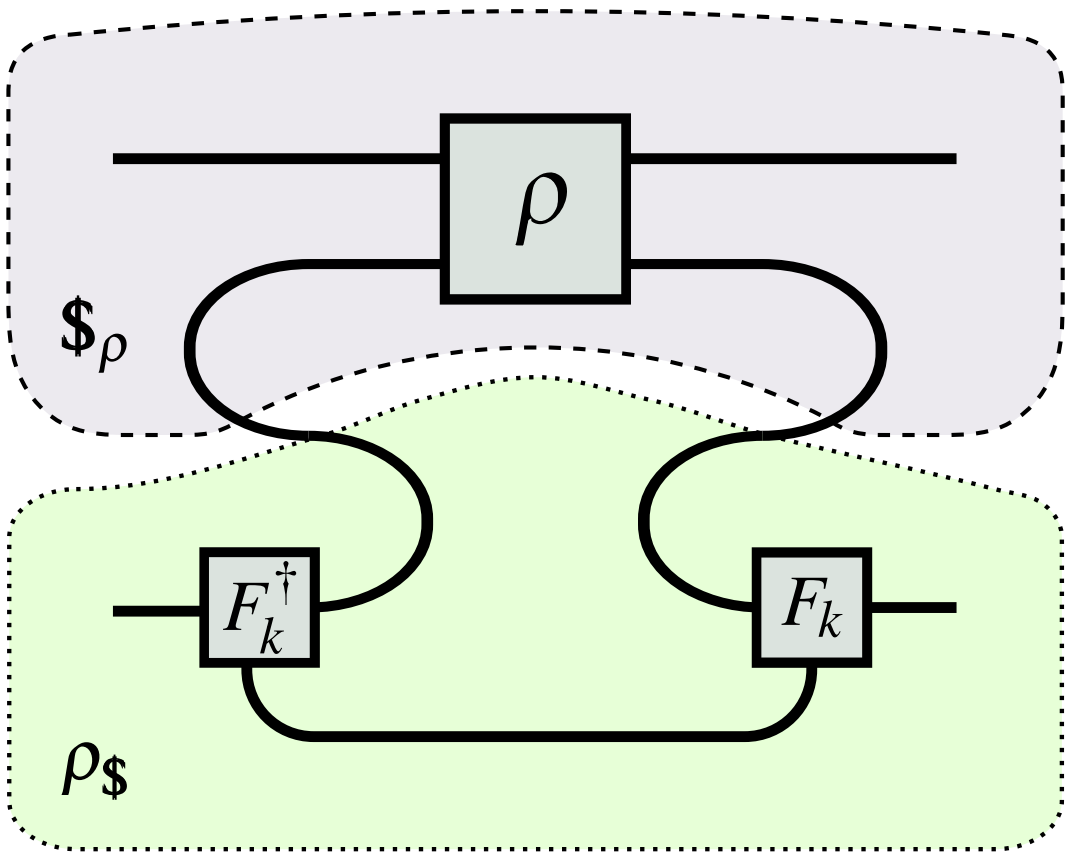

Figure 2.9: The equivalence of the Kraus operator and Choi matrix representation of an operation can be seen in the above diagram. The action of $\$$ on a state $\rho$ is equivalent to the action of $\$_\rho$ on state $\rho_\$$, as formally shown in Eq. (2.16). The top blue shaded area shows the operation $\$_\rho$, while the bottom green shaded area represents the state $\rho_\$$.

equation (2.15) from

$$\begin{aligned}\mathbf{1} \otimes \$_\rho(\rho_\$) &= \sum_{ij} \$(|i\rangle\langle j|) \otimes \sum_{klmn} \rho_{klmn} |m\rangle\langle k|i\rangle\langle j|l\rangle\langle n| \\ &= \sum_{ijmn} \rho_{ijmn} \$(|i\rangle\langle j|) \otimes |m\rangle\langle n| \,. \end{aligned} \tag{2.16}$$

The above Eq. (2.16) tells us that doing calculations with either $\rho$ and $\$$ or their duals $\$_\rho$ and $\boldsymbol{\rho}_\$$, respectively, is equivalent. This justifies our choice of notation, where we write $\rho \cong \$_\rho$ and $\$ \cong \boldsymbol{\rho}_\$$. The formal equivalence between states and operations is known as the Choi-Jamiołkowski isomorphism, which will prove to be a very important computational aid in subequent chapters.

# CHAPTER 3

## Quantum Information Theory

Quantum systems are fundamentally different from their classical counterparts. Classically, in order to facilitate communication between two parties, correlated probability distributions are used. Suppose we have two observers, Alice and Bob. In order to send some information to Bob, Alice would prepare a physical system in one of the states $\{1, \ldots, n\}$ and send it to Bob. Bob is able to distinguish between any of the states Alice sends him and has thus received the information Alice intended to send. Notice that from Bob's point of view this is completely equivalent to Alice and Bob being in possession of a joint probability distribution $p(k_A, k_B) = \delta_{k_A,k_B}/n$, where $\delta_{k_A,k_B}$ is the Kronecker delta, saying that Alice prepares one of the states with probability $1/n$, while Bob's subsequent measurement of the state is guaranteed to be completely correlated with Alice's preparation[1].

If we understand classical communication resources as probability distributions, we should consider how quantum states differ from that. A simple probability distribution such as $p(k_A, k_B)$ is not sufficient to fully describe a quantum system. As we saw in the previous chapter, probability distributions arise from a quantum state through measurement. We might write $p_{a,b}(k_A, k_B) = \text{Tr}[\boldsymbol{P}_{a,k_A} \otimes \boldsymbol{P}_{a,k_B}\rho]$, where $(a, b)$ are some indices allowing Alice and Bob to choose which respective POVMs $\boldsymbol{P}_{a,k_A}, \boldsymbol{P}_{b,k_B}$ they would like to measure. Thus, a single state allows us to select among infinitely many probability distributions through our choice of measurement.

[1]For completeness we note that in a more general scenario errors may occur during the transmission which may reduce the correlation between Bob's and Alice's random variable.

Once they have made their choice of measurement and executed it, they have fixed the single distribution they will get. Importantly however, they are able to select the distribution *without* communicating their respective choices to the other party. Could they simulate the ability to select such a distribution classically? They might attempt to do so by preparing infinitely many classical physical systems simulating possible choices of measurement $(a, b)$ with corresponding classical systems exhibiting probability distributions $p_{a,b}(k_A, k_B)$. But there are therefore many, in general infinitely many, joint classical systems corresponding to a particular choice of $b$ but with differing values of $a$ so for Bob to know which one of them he must examine, he must somehow be informed of Alice's choice of index $a$. Thus, while this classical scheme can simulate the selection of probability distributions, it is unable to do so without communication between the parties. This illustrates that quantum states are different from classical probability distributions on a very fundamental level and that the differences may be understood from the perspective of theory of information and communication. In this chapter we will present the mathematical framework allowing us to understand precisely how we may exploit the unique features of quantum mechanics for the purposes of communication.

## 3.1 Quantification of information

The purpose of quantum information theory is to quantify information, mainly for the purposes of communication but also on some occasions to better understand information processing. During a communication protocol we usually imagine at least two spatially separated parties attempting to reproduce a message selected by one party at the location of the other party. Note that it is not necessary for every 0 to be turned into a 0 and every 1 to be turned into a 1, as long as the output and input are correlated. Every 0 might, for example, be consistently turned into a 1 and vice versa, and the message itself could still be reproduced perfectly.

The amount of information the other party receives when the message is repro-duced is inherently tied to the degree to which the second party is surprised by the message they receive. If the party has already expected the message, the amount of new information they have received is very small. We will see that the notion of surprise is further tied with the minimum length of a message sufficient to impart information to another party.

In this section we will give a short introduction to classical and quantum information theory and the relevant concepts that will be important in the later chapters. In particular, we will briefly review the relative entropy and the quantities following from it such as the Shannon and von Neumann entropies, conditional entropy and the mutual information.

### 3.1.1 Classical surprise, relative and absolute

We will treat the message the party, Bob, receives as a random variable $X$ with outcomes $x_1, \dots x_n$ for which Bob believes the associated probabilities to be $q_1, \dots q_n$. The amount by which Bob is surprised when he receives the message $x_k$ must be an increasing function of $1/q_k$, which we will for now denote as $f(1/q_k)$. We will see that adding another, independent message determines this function uniquely by requiring surprise to be additive under composition with independent messages.

Thus, take $Y$, with outcomes $y_1, \dots y_m$ to be another possible set of messages with Bob's probability estimates being $s_1, \dots s_m$. These are independent of the distribution $q_k$. Then, if he receives messages $x_k, y_l$ his surprise will be $f\big(1/(q_k s_l)\big)$. Since we want surprise to be additive under composition of independent messages, we require $f\big(1/(q_k s_l)\big) = f(1/q_k) + f(1/s_l)$. There is a unique differentiable mathematical function satisfying this requirment, it is $f(x) \propto \ln(x)$ [47]. In accordance with convention, we will use $f(x) = \log_2(x)$.

Now we said that the probabilities $q_k$ are only estimates of the actual probabilities, denoted as $p_k$, for the message $x_k$ to be received. They describe Bob's prior knowledge about the message Alice is to send. Then the average surprise for Bob upon receiving Alice's message is given by $-\sum_k p_k \log_2(q_k)$, where we take $0\log(0) = 0$. The difference between this surprise and the surprise if he had had correct probabilities is termed the relative entropy [48],

$$H\left(p_k \| q_k\right) = \sum_k p_k(\log_2(p_k) - \log_2(q_k)). \tag{3.1}$$

The importance of this quantity to information theory cannot be overstated. It plays a similar role in information theory that the canonical partition function plays in thermodynamics in that many of the important quantities in information theory can be expressed in terms of the relative entropy.

It is important to note that the quantity $-\log_2(p_k)$ represents the minimum number of bits one should use to represent event $k$ if such event occurs with probability $p_k$. This is optimal in the sense that the average number of bits used, $\sum_k p_k \log_2(p_k)$, is minimal [1, 49, 50]. Relative entropy therefore tells us how much shorter our messages become when we learn the true distribution $p_k$, where before we had assumed $q_k$, or equally the amount of information we gain when we learn the true distribution $p_k$. In this sense information takes on a physical role in terms of the least number of physical systems we must use to encode a message.

In going from an inaccurate probability distribution to an accurate distribution we expect that our messages should shorten and that therefore $H\left(p_k \| q_k\right) \geq 0$. The result known as the Gibbs' inequality tells us that this is indeed the case, with equality if and only if $p_k = q_k$ for all $k$. As shown in [9], this can be proved by using the inequality $\ln(x) \leq x - 1$, with equality if and only if $x = 1$. Notice that if our prior distribution takes an event that is actually possible, $p_k > 0$ to be almost impossible, $q_k = 0$, then relative entropy becomes infinite.

Suppose now that Bob had previously held no information about the actual probability distribution $p_k$ and had therefore simply assumed it to be the uniform distribution $q_k = 1/n$, where we assume the number of events $n$ is finite. The amount of information he then learns on finding out the true distribution is given by

$$H\left(p_k \| 1/n\right) = \log_2(n) + \sum_k p_k \log_2(p_k) = \log_2(n) - H(p_k), \tag{3.2}$$

where $H(p_k) = -\sum_k p_k \log_2(p_k)$, or $H(X)$ for short, is termed the entropy of the distribution $p_k$.

As an example, suppose Alice wants to send Bob several pages of English text. The symbols she uses can be considered to be the set of letters of the alphabet, assumed to appear according to the probability distribution in the English language, or it may be the set of two-letter combinations, also appearing according to the probability distribution in the English language. An example of random letters generated according to the distribution of letters in English, due to Shannon [1], might be

> OCRO HLI RGWR NMIELWIS EU LL NBNESEBYA TH EEI ALHENHTTPA OOBTTVA NAH BRL,

while random words generated according to the distribution in English might be

> REPRESENTING AND SPEEDILY IS AN GOOD APT OR COME CAN DIFFERENT NATURAL HERE HE THE A IN CAME THE TO OF TO EXPERT GRAY COME TO FURNISHES THE LINE MESSAGE HAD BE THESE.

Clearly, the second is closer to the actual distribution in English and can, as a result, be compressed to shorter messages. How much shorter, precisely, can be evaluated using the difference in the relative entropies. Redundancies in the English language were also computed experimentally by Shannon in [51].

Another scenario of interest may be to suppose that Bob had already received a message from Alice, imparting upon him partial information about the complete message. In other words, she sent him a message corresponding to a random variable $X$, which is correlated with another message $Y$ he is yet to receive. How much additional information, on average, will he receive when finally obtaining $Y$? The joint probability distribution is $p(x, y)$ and so the information Bob has available before receiving $Y$ corresponds to the probability distribution $p(x) \cdot 1/n_Y$, where $n_Y$ is the total number of possible outcomes of $Y$ and $p(x) = \sum_y p(x, y)$ is the marginal distribution. Information imparted on him when receiving $Y$ is thus given by

$$H\left(p(x, y) \| p(x) \cdot 1/n_Y\right) = \log_2(n_Y) - H(X, Y) + H(X). \tag{3.3}$$

The quantity $H(Y, X) - H(X) = H(Y|X)$ is known as the conditional entropy.

The third, and final, scenario we will consider for the classical part here will be to introduce mutual information. Suppose Bob receives two correlated messages $X$ and $Y$. How much information in them is redundant due to the correlation, or asked another way how much information do they have in common? As above, the joint probability distribution is $p(x, y)$. If we pretend they have nothing in common, we would assign $p(x)p(y)$ as their joint distribution. The additional information we gain on finding out the true joint distribution is given by the relative entropy

$$p(x, y) = H\left(p(x, y) \| p(x)p(y)\right) = H(X) + H(Y) - H(X, Y) = I(X : Y), \quad (3.4)$$

where $I(X : Y)$ is the mutual information. We have thus derived the major quantities of the information theory from a single quantity, the relative entropy. Next we will consider how these results generalize when the underlying medium used to transmit messages is composed of quantum states.

### 3.1.2 Quantum surprise, relative and absolute

In the previous section we saw how many of the important results in classical information theory follow from the concept of relative entropy. Relative entropy generalizes naturally to the quantum setting, leading to the analogous quantum results.

Suppose Bob is given a mixed quantum state $\rho$ but believes it to be $\sigma$ instead. The quantum case exhibits certain differences from the classical case. Firstly, a state $\rho$ does not represent a single probability distribution, it represents infinitely many probability distributions, one for each choice of measurements. Therefore, each choice of measurements will lead to a different value of surprise. Instead of a simple number, we should therefore treat surprise as an observable and therefore an operator. In analogy with the classical case, we take the operator to be $\log_2(\sigma)$, where we have taken the matrix logarithm. This is a natural way to define surprise, since if Bob expects the state $\sigma$ to have the spectral decomposition $\sigma = \sum_j \sigma_j |\phi_j\rangle \langle\phi_j|$, then obtaining the state $|\phi_j\rangle$ through measurement should yield $\log_2(\sigma_j)$ of classical surprise.

Given that the actual state is $\rho$, the expectation of the surprise operator is therefore given by

$$-\operatorname{Tr}[\rho \log_2(\sigma)]. \quad (3.5)$$

The difference between this surprise and that obtained if he had expected the correct state $\rho$ is the relative entropy

$$S(\rho\|\sigma) = \text{Tr}[\rho(\log_2(\rho) - \log_2(\sigma)]. \tag{3.6}$$

Just like the classical version, quantum relative entropy satisfies $S(\rho\|\sigma) \geq 0$ with equality if and only if $\rho = \sigma$. Proof can be found in [9]. An operational interpretation of quantum relative entropy can be obtained by considering what is the probability of mistakenly believing the state is $\sigma$, where in fact it is $\rho$, after conducting $N$ measurements for large $N$. This can be found to be $2^{-NS(\rho\|\sigma)}$ [42, 52].

The conventional von Neumann entropy [53] can be obtained from the above by computing $S(\rho\|\mathbf{1}/d) = \log_2(d) + \text{Tr}[\rho\log_2(\rho)] = \log_2(d) - S(\rho)$, as in the classical case, where $d$ is the dimension of the Hilbert space. Similarly for the conditional entropy

$$S(A|B) = \log_2(d_A) - S\left(\rho^{AB}\|\rho^A \otimes \mathbf{1}/d_B\right) = S(\rho^{AB}) - S(\rho^B) \tag{3.7}$$

and the mutual information $I(A:B) = S\left(\rho^{AB}\|\rho^A \otimes \rho^B\right)$. The $AB$ in $\rho^{AB}$ is used to explicitly denote which parts of the state $\rho$ we are referring to and will be suppressed when there is no room for confusion. We will see later that quantum relative entropy never increases under the evolution of quantum states. We can thus interpret von Neumann entropy of state $\rho$ as the maximum uncertainty about the outcome of rank-1 measurements conducted on state $\rho$, conditional entropy as the maximum remaining uncertainty about the state $\rho^{AB}$, when a part of the state, $\rho^B$ is known and the mutual information as the minimum amount of uncertainty remaining about the total state $\rho^{AB}$ when the statistics about the states $\rho^A$ and $\rho^B$ are known separately. Unlike the classical conditional entropy, the quantum conditional entropy can be negative. We will discuss this feature in section 3.2.

An important operational interpretation of the von Neumann entropy was given by Schumacher [9, 54]. Namely, given a large number $n$ of states $\rho$, they may be reliably represented by $R \cdot n$ qubits if $R > S(\rho)$. Conversely, if $R < S(\rho)$, then any scheme attempting to represent the states using $R \cdot n$ qubits must be unreliable in the sense that the output states can no longer be decompressed back into $n$ copies of the state $\rho$. The proof of the result relies on writing the state $\rho^{\otimes n}$ in its eigenbasis and treating the resulting state similarly to a classical probability distribution.

We have looked at the definitions and intuitive underpinnings of entropic quantities. Next we shall review the inequalities satisfied by these quantities.

### 3.1.3 Entropic inequalities

The following theorem, due to Klein [9, 55], implies that the relative entropy is positive since $\mathrm{Tr}[\rho] = 1$ for all density matrices.

**Theorem 3.1** (Klein's Inequality). *For any positive operators $\boldsymbol{A}, \boldsymbol{B}$, we have that*

$$\mathrm{Tr}\big[\boldsymbol{A}\big(\log_2(\boldsymbol{A}) - \log_2(\boldsymbol{B})\big)\big] \geq \mathrm{Tr}\big[\boldsymbol{A} - \boldsymbol{B}\big], \tag{3.8}$$

*with equality if and only if $\boldsymbol{A} = \boldsymbol{B}$.*

The relative entropy is monotone under discarding a subsystem, ie. taking partial trace. This is intuitively plausible as discaring information should not increase our ability to distinguish quantum states.

**Theorem 3.2** (Monotonicity under partial trace [52, 55]). *Let $\rho_{12}, \sigma_{12} \in \mathcal{B}(\mathcal{H}_1)\otimes\mathcal{B}(\mathcal{H}_2)$ and let $\rho_1 = \mathrm{Tr}_2[\rho_{12}]$, $\sigma_1 = \mathrm{Tr}_2[\sigma_{12}]$. Then*

$$S\left(\rho_{12}\|\sigma_{12}\right) \geq S\left(\rho_1\|\sigma_1\right), \tag{3.9}$$

*with equality if and only if the operator equation $\log_2(\rho_{12}) - \log_2(\sigma_{12}) = \log_2(\rho_1) - \log_2(\sigma_1)$ is satisfied. Here the implicit tensor products are suppressed so that $\log_2(\sigma_{12})$ means $\log_2(\sigma_{12}) \otimes \mathbf{1}_3$.*

A simple consequence of the equality condition of theorem 3.2 is that for any three density matrices $\pi$, $\rho$ and $\sigma$ we have that $S\left(\rho\otimes\pi\|\sigma\otimes\pi\right) = S\left(\rho\|\sigma\right)$. It may be obtained through a direct calculation and by using the result that $\log_2(\boldsymbol{A}\otimes\boldsymbol{B}) = \log_2(\boldsymbol{A})\otimes\mathbf{1} + \mathbf{1}\otimes\log_2(\boldsymbol{B})$.

The theorem 3.2, together with Eq. (3.7), immediately implies another entropic inequality, $S\left(\rho_{123}\|\rho_{12}\otimes\mathbf{1}/d_3\right) \geq S\left(\rho_{23}\|\rho_2\otimes\mathbf{1}/d_3\right)$, where $d_3$ is the dimension of the third subspace. This gives us the following important inequality.

**Theorem 3.3** (Strong subadditivity [52, 55]). *Given $\rho_{123} \in \mathcal{B}(\mathcal{H}_1)\otimes\mathcal{B}(\mathcal{H}_2)\otimes\mathcal{B}(\mathcal{H}_3)$,*

$$S(\rho_{123}) - S(\rho_{12}) \leq S(\rho_{23}) - S(\rho_2), \tag{3.10}$$

*with equality if and only if* $\log_2(\rho_{123}) - \log_2(\rho_{12}) = \log_2(\rho_{23}) - \log_2(\rho_2)$.

Partial trace monotonicity together with the representation of quantum evolution as a joint unitary evolution with the environment, Eq. (2.9), leads to the following result. Because of the high importance of this result and its recurrent use in chapter on quantumness of operations, we will provide a proof of it.

**Theorem 3.4** (Monotonicity of relative entropy under quantum operations [52, 55]). *Given a quantum operation* \$ *and states* $\rho, \sigma$, *relative entropy satisfies*

$$S\left(\$\rho \| \$\sigma\right) \leq S\left(\rho \| \sigma\right). \tag{3.11}$$

*Proof.* We know that trace, being the sum of the eigenvalues of a matrix, is invariant under unitary operations $\text{Tr}[\boldsymbol{U}\rho\boldsymbol{U}^\dagger] = \text{Tr}[\rho]$. We also know that for any analytic function $f$, such as $\log_2$, $f(\boldsymbol{U}\rho\boldsymbol{U}^\dagger) = Uf(\rho)U^\dagger$. Using these properties together with the monotonicity of relative entropy under partial trace and the representation of quantum operations using joint unitary evolution with the environment, Eq. (2.9), we arrive at the following

$$\begin{aligned} S\left(\$\rho \| \$\sigma\right) &= S\left(\text{Tr}_e[\boldsymbol{U}\rho \otimes \pi_e \boldsymbol{U}^\dagger] \| \text{Tr}_e[\boldsymbol{U}\sigma \otimes \pi_e \boldsymbol{U}^\dagger]\right) \\ &\leq S\left(\rho \otimes \pi_e \| \sigma \otimes \pi_e\right) = S\left(\rho \| \sigma\right), \end{aligned} \tag{3.12}$$

where $\text{Tr}_e$ is the trace over the environment and $\pi$ represents the state of the environment. □

Finally, we state without proof that the relative entropy is jointly convex in both of its arguments.

**Theorem 3.5** (Joint convexity [9, 55]). *Given density matrices* $\rho_1, \rho_2 \in \mathcal{B}(\mathcal{H})$ *and* $\sigma_1, \sigma_2 \in \mathcal{B}(\mathcal{H})$, *and a real number* $0 \leq p \leq 1$, *we have that*

$$S\left(p\rho_1 + (1-p)\rho_2 \| p\sigma_1 + (1-p)\sigma_2\right) \leq pS\left(\rho_1 \| \sigma_1\right) + (1-p)S\left(\rho_2 \| \sigma_2\right), \tag{3.13}$$

*with equality if and only if* $\log_2(\rho_1) - \log_2(\sigma_1) = \log_2(\rho_2) - \log_2(\sigma_2)$.

In particular, joint convexity implies convexity of each of the arguments separately. This can be seen by setting either $\sigma_1 = \sigma_2$ for the first argument or $\rho_1 = \rho_2$ for the second.

We stated in the introduction to this chapter that joint quantum states and the corresponding joint probability distributions are a particularly interesting facet of quantum theory. We will now look at how the derivatives of relative entropy can be used to measure and understand the non-classical features offered by quantum states.

## 3.2 Nonclassical correlations

We consider in particular the scenario where we have two parties, each possessing their part of the joint state $\rho$. They are allowed to communicate classically and perform local quantum operations composed of evolution and measurement. Classically, a typical problem in such scenarios is as follows. Suppose Alice and Bob live in two separate towns with their own weather histories. Each day they record whether the weather is sunny or cloudy (for simplicity assume there are no other weather patterns, although the following discussion generalizes easily) and each town has the same, but independent, probability of sunny or cloudy on any given day. Now Alice wants to communicate to Bob the weather patterns at her location for the past year and use the least possible length of the message by exploiting the correlations between the weather patterns at two locations. Slepian and Wolf theorem then states that the minimum possible length of the message she can attain is given by the conditional entropy $S(A|B)$ [49, 56, 57].

Quantum mechanically, an analogue problem might be that Alice and Bob share a quantum state $\rho$ and Alice would like Bob to posses full knowledge of the state $\rho$, including the part local to her. What is the minimum number of qubits of quantum information that Alice needs to transfmit? The apparently obvious answer, the quantum conditional entropy $S(A|B)$, has a slight problem - it can be negative. Particularly, consider the joint state to be $|\psi\rangle = \sum_l \sqrt{\omega_l}\, |l\rangle \otimes |l\rangle$. Pure states can always be written in this form by using the Schmidt decomposition [9]. Then $S(A|B) = -S(A) = \sum_l \omega_l \log_2(\omega_l) \leq 0$, with equality if and only if one of $\omega_l = 1$ for some $l = k$ and $\omega_l = 0$ for all other $l \neq k$. The negativity of conditional entropy had long troubled quantum information theorists until it was realized that negative conditional entropy means that Bob can gain full knowledge of the state $\rho$ with only classical communication, leaving $-S(A|B)$ as the potential for future quantum communication [58, 59].

In the case of the state $|\psi\rangle$ as above, Bob can reconstruct the state $|\psi\rangle$ locally with Alice transmitting classical information only, while leaving the state $|\psi\rangle$ to be used as a resource for future quantum communication through protocols such as quantum teleportation [6, 9]. This alleviates the problem of negative conditional entropy by giving it an operational interpretation.

We shall now consider another interpretation of quantum conditional entropy through the coding capacity of $\rho$. Alice can encode information by applying unitary operations on her part of the state $\rho$ and sending her part of the state to Bob. How much information does Bob gain? Bob's knowledge of the state gives him the prior state $\mathbf{1}/d_A \otimes \rho_B$. The maximum he may learn by receiving Alice's part of the state as well is therefore given precisely by the relative entropy $S\left(\rho \| \mathbf{1}/d_A \otimes \rho_B\right)$, as argued in section 3.1.2. The coding capacity of $\rho$ is thus

$$S\left(\rho \| \mathbf{1}/d_A \otimes \rho_B\right) = \log_2(d_A) - S(\rho^{A|B}). \tag{3.14}$$

A detailed argument for this fact from first principlescan be found in [60]. Here we use $S(\rho^{A|B})$ to denote the conditional entropy $S(A|B)$ of state $\rho$ and do not suggest the existence of a conditional state $\rho^{A|B}$. The negative conditional entropy thus entails additional capacity to encode information in a quantum state beyond what is possible classically, where the maximum achievable rate is $\log_2(d_A)$. The effect was first noticed by Bennett and named superdense coding [61]. The quantity $-S(\rho^{A|B})$ is sometimes given the name *coherent information*. This is just one of the cases where infinitely many probability distributions encoded in a single state offer a tangible advantage in communication.

The coherent information has played a fundamental role in quantum information theory. The quantity $\max\{0, -S(\rho^{A|B})\}$, defined only on pure states, is as we saw above positive only for states which are not of the separable states of the form $\left|\psi^A\right\rangle \otimes \left|\psi^B\right\rangle$. As such, it is a measure of entanglement for pure states. We will review entanglement measures next, followed by more recent measures of quantum correlations such as quantum discord.

### 3.2.1 Measures of entanglement

Entangled states are those that cannot be understood as a random choice of product states, meaning that they cannot be written in the form

$$\rho = \sum_k p_k \rho_k^A \otimes \rho_k^B, \tag{3.15}$$

for some $p_k \geq 0$ such that $\sum_k p_k = 1$. Such states were first noticed in the seminal paper by Einstein, Podolsky and Rosen (EPR) [2], where they famously called it *spooky action at a distance*, ironically in their attempt to find a complete and classical description of quantum mechanics by assigning definite quantities to each variable prior to measurement. Entanglement plays a role in many key discoveries, including quantum cryptography based on Bell's theorem [8] and quantum teleportation [6] as well as, as we saw before, superdense coding [61].

Entanglement measures are functionals $E : \mathcal{B}(\mathcal{H}) \to \mathbb{R}$ measuring the extent to which a state $\rho$ is not of the form of Eq. (3.15). We will here review the properties such functionals need to satisfy and give the explicit formulations of the most prominent measures. However, we will not provide proofs for most of the claims, as they are now generally considered to be standard repertoirs of quantum information theory. Extensive and detailed reviews are available [10, 11].

Typically one requires the following properties of a measure of entanglement [10–12]:

(i) *Vanishing for separable states.* $E(\rho) = 0$ if and only if $\rho$ is of the form of Eq. (3.15).

(ii) *Unitary invariance.* The measure is invariant under local unitary operations, $E(U_A \otimes U_B\, \rho\, U_A^\dagger \otimes U_B^\dagger) = E(\rho)$.

(iii) *LOCC monotonicity.*It is non-increasing under local operations and classical communication.

It is often further required that the measure $E$ reduces to $-S(\rho^{A|B}) = S(\rho^A) = S(\rho^B)$ on pure states. This quantity is known as the entropy of entanglement. However, we will only require the above three properties here.

Two particularly important measures of entanglement are entanglement of formation and entanglement of distillation [62]. Entanglement of formation tells us the rate at which maximally entangled states, such as $d^{-1/2}\sum_{k=0}^{d-1}|k\rangle\otimes|k\rangle$, can be transformed into states $\rho$ in the limit of infinitely many copies of the states. More precisely, given $m$ maximally entangled states, suppose they can be converted into $n(m)$ copies of $\rho$. Then $E_F(\rho) = \lim_{m\to\infty} m/n(m)$. We will describe the transformation in more detail in the next paragraphs. The entanglement of distillation $E_D$, on the other hand, is the opposite process where one measures the rate at which $n$ states $\rho$ may be converted into $m(n)$ approximately maximally entangled states, i.e. $E_D(\rho) = \lim_{n\to\infty} m(n)/n$.

For pure states they are both equal to $S(\rho^A)$. This can be seen by the following protocol: Alice prepares the state $|\psi\rangle^{\otimes n}$ locally and compresses the partial density matrix $(\rho^A)^{\otimes n}$ to $nS(\rho^A)$ qubits using the Schumacher protocol. She then teleports it to Bob, who decompresses it. They end up sharing $n$ approximate states $|\psi\rangle$ and have used $nS(\rho^A)$ maximally entangled pairs to accomplish the task. For mixed states, however, the picture is more complex. It is known that $E_F(\rho) \geq E_D(\rho)$ [10, 11, 42]. It is possible to express $E_F$ in terms of the entropy of entanglement, however, by the equation

$$E_F(\rho) = \inf_{\{p_k,|\psi_k\rangle\}} \sum_k p_k S\big(\mathrm{Tr}_A[|\psi_k\rangle\langle\psi_k|]\big), \tag{3.16}$$

where the infimum is over all convex decompositions of $\rho$, i.e. all $\{p_k, |\psi_k\rangle\}$ such that $\sum_k p_k |\psi_k\rangle\langle\psi_k| = \rho$. Here $|\psi_k\rangle$ are not necessarily orthogonal. This formula is obtained in a similar way as the pure state version. Given a particular decomposition $\epsilon = \{p_k, |\psi_k\rangle\}$, Alice can locally generate a pure state $\Phi_k^\epsilon = \sum_k \sqrt{p_k}\,|\psi_k\rangle \otimes |k\rangle$, where $|k\rangle$ represent the memory containing the which-state information. Tracing out the memory gives us the state $\rho$ independently of $\epsilon$. Alice can now compress the states $|\psi_k\rangle$ as in the pure case and sending them to Bob along with the memory, which consists of only the values $k$ and can therefore be communicated classically.

The Eq. (3.16) is known as the convex roof extension and can be used to extend an arbitrary measure that is defined for pure states to a definition for mixed states. If the measure is a monotone of the entropy of entanglement on pure states, then it will satisfy the properties required of the entanglement measure [10, 11]. One should think of $E_F$ and $E_D$ as the wholesale exchange rate between given states $\rho$ and maximally entangled states. Going through the entire cycle of distillation and formation leaves us with less entanglement than we started with, similar to the lossy nature of thermo-dynamic cycles. Although outside the scope of this thesis, further parallels between the theory of entanglement and thermodynamics can be found in [63, 64].

For the case of qubits an exact expression exists for entanglement of formation in terms of concurrence, an easily computable entanglement measure for a pair of qubits [65]. Concurrence is given by

$$C(\rho) = \max\{0, \lambda_1 - \lambda_2 - \lambda_3 - \lambda_4\}, \tag{3.17}$$

where $\lambda_k$ are the eigenvalues of the matrix $\sqrt{\sqrt{\rho}\widetilde{\rho}\sqrt{\rho}}$ listed in the decreasing order and where $\widetilde{\rho} = (\sigma_y \otimes \sigma_y)\rho^*(\sigma_y \otimes \sigma_y)$, with $\sigma_y$ being the Pauli $Y$ matrix. Entanglement of formation can be computed in terms of concurrence as

$$E_F(\rho) = h\left(\frac{1 + \sqrt{1 - C(\rho)^2}}{2}\right), \tag{3.18}$$

where $h(x) = -x\log_2(x) - (1-x)\log_2(1-x)$ is the binary entropy function. Given a pure state $|\psi\rangle = \sum_{ij} A_{ij} |i\rangle \otimes |j\rangle$ concurrence is given by

$$C(|\psi\rangle\langle\psi|) = 2\left[\det(AA^\dagger)\right]^{1/2}, \tag{3.19}$$

which is the geometric mean of the Schmidt numbers of the state $|\psi\rangle$. For mixed states it can be expressed as the convex roof extension of the concurrence defined on pure states, as proved in [65].

Concurrence has a natural extension to qudits by using the geometric mean of the Schmidt numbers as its definition for pure states and employing the convex roof construction. The measure obtained in this way is known as G-concurrence, given by

$$G_d(|\psi\rangle\langle\psi|) = d\left[\det(AA^\dagger)\right]^{1/d}, \tag{3.20}$$

where $d \times d$ is the dimension of the joint Hilbert space [66]. G-concurrence has a number of desirable properties

(i) *Multiplicativity:* Given a state $|\psi_1\rangle \otimes |\psi_2\rangle$ in the Hilbert space of dimensions $(d_1 \times d_1) \otimes (d_2 \times d_2)$, then $G_{d_1 \cdot d_2}(|\psi_1\rangle \otimes |\psi_2\rangle = G_{d_1}(|\psi_1\rangle)G_{d_2}(|\psi_2\rangle)$.

(ii) *Homogeneity:* The measure is homogenous of degree 1, namely for any $s \in \mathbb{C}$,

$G_d(s\rho) = |s|G_d(\rho)$.

(iii) *SL-invariance:* For two operators $\boldsymbol{A}$ and $\boldsymbol{B}$ acting on $\mathcal{H}_A$ and $\mathcal{H}_B$, respectively, we have that $G_d\big((\boldsymbol{A}\otimes\boldsymbol{B})\rho(\boldsymbol{A}^\dagger\otimes\boldsymbol{B}^\dagger)\big) = \det(A)^{2/d}\det(B)^{2/d}G_d(\rho)$. The property is called SL-invariance because it implies that whenever $\boldsymbol{A},\boldsymbol{B} \in SL(d)$, G-concurrence is preserved.

The property (iii) in particular will come useful when we look at the evolution of quantum entanglement under quantum operations.

Finally, a natural measure of entanglement arising out of the definition of relative entropy is the relative entropy of entanglement. It is given by

$$E_{RE}(\rho) = \min_{\sigma\in\mathcal{D}} S\left(\rho\|\sigma\right), \tag{3.21}$$

where $\mathcal{D}$ is the set of separable states of the form (3.15). It is known that $E_D(\rho) \leq E_{RE}(\rho) \leq E_F(\rho)$ [42, 67]. A closed form of the relative entropy of entanglement for all dimensions and for multipartite, as well as bipartite, states is known for a special set of states [68]. The relative entropy has been shown to be able to interpolate between $E_D$ and $E_F$. The entanglement of formation can be expressed in terms of the relative entropy of entanglement, with details available in [42, 67].

### 3.2.2 Evolution of entanglement

We consider bipartite states $\rho \in \mathcal{H}_A \otimes \mathcal{H}_B$. Since such states may be entangled, one is often particularly concerned with how entanglement evolves or changes under completely positive maps [69]. Thus, given an entanglement measure $C$ and a *local* completely positive map $\$ = \$_A \otimes \$_B$ with the Kraus decomposition $\$\left[\rho\right] = \sum_k (\boldsymbol{F}_k \otimes \boldsymbol{G}_k)\rho(\boldsymbol{F}_k^\dagger \otimes \boldsymbol{G}_k^\dagger)$. We will want to compute $C[(\$_A \otimes \$_B)\rho]$, for some entanglement measure $C$. An elementary, but still critically important, example of such maps are one-sided maps of the form $\$ = \$_A \otimes \mathbf{1}_B$.

In [70] Konrad et al. showed that for pure bipartite qubit states

$$C\big[(\$ \otimes \mathbf{1})\left|\psi\right\rangle\left\langle\psi\right|\big] = C\left[(\$ \otimes \mathbf{1})|\phi^+\rangle\langle\phi^+|\right] \cdot C\big[\left|\psi\right\rangle\left\langle\psi\right|\big], \tag{3.22}$$

where $|\phi^+\rangle = \left(|00\rangle + |11\rangle\right)/\sqrt{2}$ is one of the Bell states and $C$ is the concurrence.

The function $C[\$, |\psi\rangle]$ of two variable arguments has thus been factored into a product of functions, one depending only on the starting state and the other only on the completely positive map characterising the one-sided evolution. Thus, calculating the evolution of entanglement for a single initial state, $|\Phi^+\rangle$, allows us to compute it for all other pure initial states. As will be seen later, this provides an upper bound on the quantity of evolved entanglement for all initial mixed states. For mixed states they found the upper bound to be given by the product [70]

$$C\big[(\$ \otimes \mathbf{1})\rho\big] \leq C\left[(\$ \otimes \mathbf{1})|\Phi^+\rangle\langle\Phi^+|\right] \cdot C[\rho]. \tag{3.23}$$

In [71] the analogous result was obtained for higher-dimensional systems, but where $C$ was replaced by the $G$-concurrence (see the previous section for definition of $G$-concurrence). These results were further extended to the multipartite case in [72].

To obtain the Eqs. (3.22),(3.23) for $G$-concurrence, we can make use of the Penrose notation and the state-map duality Eq. (2.15). First write the initial pure state in its Schmidt decomposed form $|\psi\rangle = \sum_l \sqrt{\omega_l}\,|l\rangle \otimes |l\rangle$. The operation dual to $|\psi\rangle$ is then $\$_\psi(\sigma) = \boldsymbol{M}\sigma\boldsymbol{M}^\dagger$, where $\boldsymbol{M} = \sum_l \sqrt{\omega_l}\,|l\rangle\langle l|$ (see figure 3.1). Conversely, the operation $\$$ itself has a dual state $\rho_\$$, it's Choi matrix, Eq. (2.14). The state-map duality, Eq. (2.15), guarantees that $\$ \otimes \mathbf{1}(|\psi\rangle\langle\psi|) = \mathbf{1} \otimes \$_\psi(\rho_\$) = (\mathbf{1} \otimes \boldsymbol{M})\rho_\$(\mathbf{1} \otimes \boldsymbol{M}^\dagger)$. Now since the $G$-concurrence is SL-invariant, we deduce that $G_d\big((\mathbf{1} \otimes \boldsymbol{M})\rho_\$(\mathbf{1} \otimes \boldsymbol{M}^\dagger)\big) = \det(\boldsymbol{M})^{2/d} G_d(\rho_\$)$. By the definition of $G$-concurrence in Eq. (3.20) we know that $\det(\boldsymbol{M})^{2/d} = G_d(|\psi\rangle\langle\psi|)$, whereas $G_d\big(\rho_\$\big) = G_d\big(\mathbf{1} \otimes \$(|\phi_d\rangle\langle\phi_d|)\big)$, where $|\phi_d\rangle = \sum_k |k\rangle \otimes |k\rangle / \sqrt{d}$ by the definition of Choi matrix. Thus,

$$G_d\big(\$ \otimes \mathbf{1}(|\psi\rangle\langle\psi|)\big) = G_d\big[(\$ \otimes \mathbf{1})|\phi_d\rangle\langle\phi_d|\big] \cdot G_d\big[|\psi\rangle\langle\psi|\big]. \tag{3.24}$$

The entire sequence of reasoning can also be seen in the graphical Penrose notation in figure 3.2.

Having established the evolution equation (3.24), we will now consider an upper bound. Since the measure $G_d$ is convex, we can find an upper bound on the entanglement evolution for mixed states. This upper bound is given by

$$G_d\big[(\$ \otimes \mathbf{1})\rho\big] \leq G_d\big[(\$ \otimes \mathbf{1})|\phi_d\rangle\langle\phi_d|\big] \cdot G_d[\rho]. \tag{3.25}$$

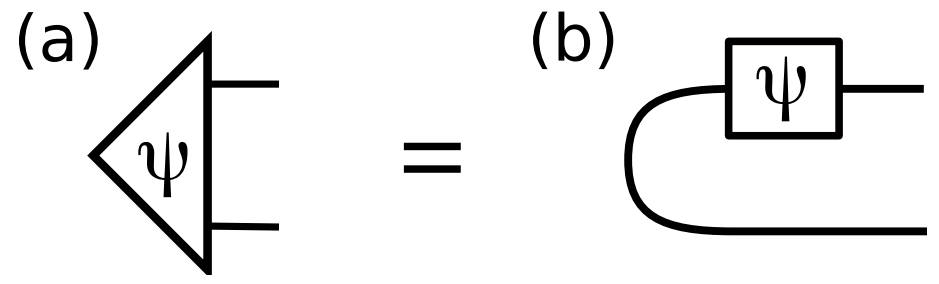

Figure 3.1: The diagram shows the Penrose notation for finding the dual operation $\$_\psi \cong |\psi\rangle\langle\psi|$.

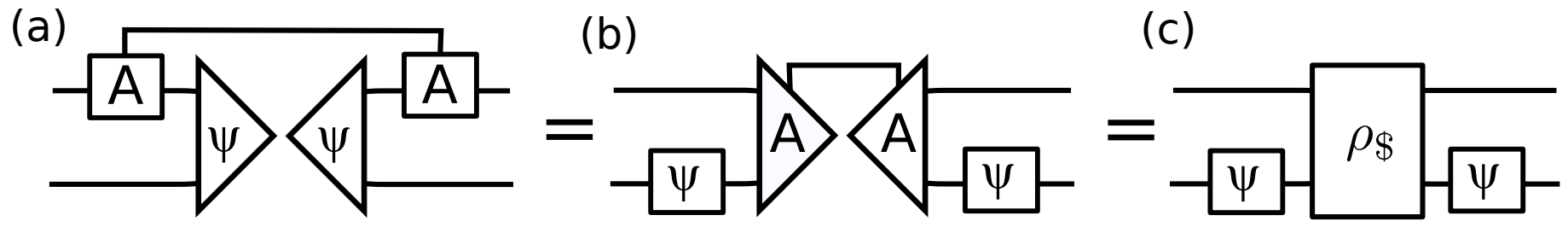

Figure 3.2: Tensor network summary of the crucial duality behind the theory of entanglement evolution. (a) State $|\psi\rangle$ acted on by a single sided superoperator $\$$. (b) Duality equation can be used to transform (a) into the view where $|\psi\rangle$ becomes a single sided superoperator $\$_\psi$ which acts on the bipartite mixed state $\rho_\$$. (c) An equivalent expression for the valence four tensor $\rho_\$$.

Equally, an upper bound can be given for two-sided operations $\$_A \otimes \$_B$ by considering it as a sequence of one-sided operations $\$_A \otimes \$_B = (\$_A \otimes \mathbf{1}) \circ (\mathbf{1} \otimes \$_B)$, where $\circ$ stands for functional composition. Then

$$G_d\big[(\$_A \otimes \$_B)\rho\big] \leq G_d\big[(\$_A \otimes \mathbf{1})|\phi_d\rangle\langle\phi_d|\big] \cdot G_d\big[(\mathbf{1} \otimes \$_B)|\phi_d\rangle\langle\phi_d|\big] \cdot G_d[\rho]. \tag{3.26}$$

In the future equations, we shall denote $F[\$] = G_d\big[(\$ \otimes \mathbf{1})|\phi_d\rangle\langle\phi_d|\big]$, and $F$ shall be called the quality factor. Since the Choi matrix $\rho_\$$ uniquely determines the quantum operation $\$$ through the duality relationship, it is not unexpected that entanglement evolution can be expressed through the use of the isomorphism. The above equations help us understand how entanglement transforms under completely positive operations. Importantly, notice that an operation $\$ \otimes \mathbf{1}$ either breaks *all* entangled pure states into separable states - this is when $F[\$] = 0$, or it breaks *none* of them and $F[\$] > 0$. Eq. (3.24) thus completely characterizes local entanglement breaking operations on pure states.

### 3.2.3 Quantum discord

While entanglement is an important feature of quantum states, it was realized already in the late 90s that it is not sufficient to capture the entirity of nonclassical behaviour specific to distributed quantum states. More recently, quantum enhancements have been exhibited in certain types of quantum computation with limited amounts of entanglement or even none at all when the involved quantum state is mixed [73–77], and universal quantum computation with *pure* states is possible with only little amounts of entanglement [78].

As an example, it is possible to find an orthogonal collection of product states $\{|\psi_k^A\rangle \otimes |\psi_k^B\rangle\}$ such that the observers Alice and Bob cannot determine which of the states they posses by using only local operations and classical communication [79]. This occurs because while the states $\{|\psi_k^A\rangle \otimes |\psi_k^B\rangle\}$ are globally orthogonal, they are locally nonorthogonal. Therefore the phenomenon is essentially quantum, but since the states used are product states, we know that entanglement cannot be responsible. This means that the state $\rho_{nl} = \sum_k \mu_k |\psi_k^A\rangle\langle\psi_k^A| \otimes |\psi_k^B\rangle\langle\psi_k^B|$ must in some sense be nonclassical.

Soon after the discovery of such states, Zurek noticed that two classically equivalent expressions for conditional entropy, or equivalently mutual information, give different results for some quantum states [15]. One way to compute conditional entropy is to use the expression $S(\rho^{A|B}) = S(\rho^{AB}) - S(\rho^B)$. On the other hand, suppose Bob measures his part of the state with rank-1 projective measurement operators $\{\mathbf{\Pi}_k\}$. We can then compute the average entropy of the conditional density matrix left to Alice. Upon obtaining outcome $k_0$, Alice is left with a conditional density matrix $\rho_{k_0}^A$, given by

$$\rho_{k_0}^A = \frac{\mathrm{Tr}_B[\mathbf{\Pi}_{k_0}\rho^{AB}\mathbf{\Pi}_{k_0}]}{\mathrm{Tr}[\mathbf{\Pi}_{k_0}\rho^{AB}\mathbf{\Pi}_{k_0}]}. \tag{3.27}$$

Denoting $p_k = \mathrm{Tr}[\mathbf{\Pi}_{k_0}\rho^{AB}\mathbf{\Pi}_{k_0}]$, the average entropy on Alice's side after measurement is therefore given by

$$S_c^{\{\mathbf{\Pi}_k\}}(\rho^{A|B}) = \sum_k p_k S\Big(\frac{1}{p_k}\mathrm{Tr}_B[\mathbf{\Pi}_k\rho^{AB}\mathbf{\Pi}_k]\Big), \tag{3.28}$$

telling us the remaining Alice's ignorance about her part of the state after Bob has communicated the result of his measurement to her. For classical probability distributions only one set of projectors $\{\mathbf{\Pi}_k\}$ is meaningful and the two expressions are equal. However, for quantum states in general $S_c^{\{\mathbf{\Pi}_k\}}(\rho^{A|B}) \geq S(\rho^{A|B})$. Zurek's original definition of quantum discord is therefore given by

$$Q_z(\rho^{A|B}) = S_c^{\{\mathbf{\Pi}_k\}}(\rho^{A|B}) - S(\rho^{A|B}) \tag{3.29}$$

and it implicitly depends on Bob's choice of measurement. For this reason quantum discord was later defined by Ollivier-Zurek and Henderson-Vedral to be [13, 14, 80–82]

$$Q(\rho^{A|B}) = \inf_{\{\mathbf{\Pi}_k\}} \Big( S_c^{\{\mathbf{\Pi}_k\}}(\rho^{A|B}) - S(\rho^{A|B}) \Big), \tag{3.30}$$

where the infimum is over all choices of rank-1 projectors.

The difference between the two quantities can perhaps most intuitively be understood in terms of the relative entropy. We already saw how this is done for the conditional entropy term. For this purpose define the measurement operation $\mathbf{\Gamma}_B(\rho) = \sum_k (\mathbf{1}_A \otimes \mathbf{\Pi}_k)\rho(\mathbf{1}_A \otimes \mathbf{\Pi}_k)$, which gives us the average post-measurement state $\sum_k \rho_k^A \otimes \mathbf{\Pi}_k$. When an operation of this form is the result of an interaction with the environment, it is also referred to as the *einselection*, short for environment induced superselection [80]. For the classical conditional entropy term we have that

$$S_c^{\{\mathbf{\Pi}_k\}}(\rho^{A|B}) = \log_2(d_A) - S\left(\mathbf{\Gamma}_B(\rho^{AB}) \| \mathbf{\Gamma}_B(\mathbf{1}/d_A \otimes \rho^B)\right). \tag{3.31}$$

To show this, we start with the relative entropy

$$S\left(\rho^{AB} \| \mathbf{\Gamma}_B(\mathbf{1}/d_A \otimes \rho^B)\right) = -S\big(\mathbf{\Gamma}_B(\rho)\big) - \mathrm{Tr}\big[\mathbf{\Gamma}_B(\rho)\log_2\big(\mathbf{\Gamma}_B(\mathbf{1}/d_A \otimes \rho^B)\big)\big] \tag{3.32}$$

$$= -S\Big(\sum_k p_k \rho_k^A \otimes \mathbf{\Pi}_k\Big) + \log_2(d_A) - \mathrm{Tr}\Big[\sum_k p_k \mathbf{\Pi}_k \log_2\big(\sum_l p_l \mathbf{\Pi}_l\big)\Big] \tag{3.33}$$

$$= \log_2(d_A) - \sum_k p_k S(\rho_k^A) = \log_2(d_A) - S_c^{\{\mathbf{\Pi}_k\}}(\rho^{A|B}), \tag{3.34}$$

where the second line follows by using $\log_2(\rho \otimes \sigma) = \log_2(\rho) \otimes \mathbf{1} + \mathbf{1} \otimes \log_2(\sigma)$ and for the last line we used that $S\left(\sum_k p_k \rho_k^A \otimes \mathbf{\Pi}_k\right) = \sum_k p_k S(\rho_k) + H(p_k)$, with $H$ being

the classical entropy of the distribution $p_k$. We can therefore write quantum discord as

$$Q(\rho^{A|B}) = \inf_{\mathbf{\Gamma}_B}\Big(S\left(\rho^{AB}\|\mathbf{1}/d_A\otimes\rho^B\right) - S\left(\mathbf{\Gamma}_B(\rho^{AB})\|\mathbf{\Gamma}_B(\mathbf{1}/d_A\otimes\rho^B)\right)\Big). \tag{3.35}$$

This equation has a natural interpretation. Suppose Alice and Bob share the state $\rho^{AB}$. If Alice sends to Bob her part of the state, he can gain $S\left(\rho^{AB}\|\mathbf{1}/d_A\otimes\rho^B\right)$ amount of information through superdense coding. However, if he was impatient and had already measured his part of the state to try to find out something about Alice's part, the amount he can learn is decreased to only $S\left(\mathbf{\Gamma}_B(\rho^{AB})\|\mathbf{\Gamma}_B(\mathbf{1}/d_A\otimes\rho^B)\right)$. Classically this is an unintuitive result, which occurs quantum mechanically due to the disturbance quantum measurement imparts on even a joint state. A similar argument, but without relative entropy and in terms of quantum state merging, was made in [83]. Other cases of quantum communication exhibiting advantages over classical communication in the absence of entanglement can be found in [17, 84–87]. We will next look at a subject very much related to quantum correlations, quantum nonlocality.

## 3.3 Tests of quantum nonlocality

Entangled states such as those above, where a joint probability distribution dependent on two parameters, one chosen by each party, can be selected out of an infinite set without any communication between the parties would seem to run against our classical intuition based on the special theory of relativity. Einstein-Podolsky-Rosen (EPR) in their seminal paper [2] required that a fundamental principle be satisfied by any *complete* physical theory. This principle was named *local realism* and as the name suggests, it is composed of two requirements

(i) *Locality.* This is the requirement that any physical object is influenced only by its immediate surroundings. More precisely, suppose we have two objects, one at point $A$ and one at point $B$. If an object travelling with the speed of light could not move from point $A$ to point $B$ within the time $t_0$ to $t_1$, then the state of the object at point $A$ at time $t_0$ cannot influence the state of the object at point $B$ at time $t_1$.

(ii) *Realism* or *counterfactual definiteness.* This is the requirement that outcomes of measurements that have not been performed nonetheless posses definite values. In a counterfactually definite theory, measuring momentum and obtaining the result $\vec{p}$ means that the measured particle's momentum was $\vec{p}$ already before the measurement, even though we may have been ignorant of the fact.

A major breakthrough in our understanding of local realism came in the 1960s, when Bell devised an experiment capable of testing whether or not a theory obeys local realism [3, 5]. The test, which we will describe below, can be applied to nature itself in a real experiment.

The experiment is as follows. Firstly, suppose Alice and Bob are located at a considerable distance from one another. They each have a number of settings their local apparatus can be configured to measure, which we shall label as $\boldsymbol{X}_1 \in \{\boldsymbol{A}_1, \boldsymbol{B}_1, \ldots\}$ for Alice and $\boldsymbol{X}_2 \in \{\boldsymbol{A}_2, \boldsymbol{B}_2, \ldots\}$ for Bob. Both $\boldsymbol{X}_1, \boldsymbol{X}_2$ may be considered to be random variables, with their values representing measurement outcomes. The objects they perform their measurements are prepared by a source, where the source chose the state of the objects to send, labelled using parameter $\lambda$, according to some probability distribution $\rho(\lambda)$. The parameter $\lambda$ is known as the hidden variable. Given the local realism assumption we ask the question: *What are the allowed probability distributions $p(x_1, x_2)$ of the measurement results Alice and Bob can obtain?* Here $x_1, x_2$ are particular outcomes of their measurements.

Locality requires that when Alice and Bob perform their measurements, the results depend only on the local state of the object measured. Counterfactual definiteness requires further that the joint state of the two objects was local already before the measurement. Applying Bayes' rule, we have that $p(x_1, x_2, \lambda) = p(x_1, x_2|\lambda)\rho(\lambda) = p_1(x_1|\lambda)p_2(x_2|\lambda)\rho(\lambda)$. Counterfactual definiteness would in the most strict sense also require that $p_1(x_1|\lambda)$ takes only the values $0$ or $1$. However, if we allow for the existence of further hidden variables not under the direct control of the source, then the requirement may be relaxed. Therefore, the answer to the question asked in the previous paragraph is that the joint outcome distribution is limited to take the form

$$p(x_1, x_2) = \int d\lambda \rho(\lambda) p_1(x_1|\lambda) p_2(x_2|\lambda). \tag{3.36}$$

Notice how similar this equation is to Eq. (3.15), where we defined the separable states. While from the form of the two equations it follows that separable states do satisfy the above equation, we cannot from this automatically conclude that they are local realist. This is because while violation of the Eq. (3.36) implies that local realism is not satisfied, nonviolation does not imply that it is satisfied. Most importantly, however, Bell showed that quantum mechanics predicts correlations that violate the above equation, and thus local realism.

### 3.3.1 CHSH inequality

Another necessary condition for local realism comes in the form of an inequality, CHSH or *Clauser-Horne-Shimony-Holt* inequality [4]. Before we introduce the CHSH inequality, we shall look more deeply at the meaning of counterfactual definiteness.

Firstly, notice that joint probability distributions do not always exist for particular marginals. Namely, given probability distributions $p(a_1, a_2), p(a_1, b_2)$ etc, a joint probability distribution $p(a_1, b_1, a_2, b_2)$ is *not* guaranteed to exist. As an example consider three random variables describing coin tosses $\boldsymbol{C}_1, \boldsymbol{C}_2, \boldsymbol{C}_3$, with values either heads $H$ or tails $T$. The coins can only be flipped pairwise and not all at a time - as soon as two of them are flipped the third one disappears. When flipping the coins $\boldsymbol{C}_1$ and $\boldsymbol{C}_2$ together you find that they always give the same outcome, with heads and tails occuring with equal probability $1/2$. The same behaviour occurs when flipping $\boldsymbol{C}_2$ and $\boldsymbol{C}_3$ together. However, flipping $\boldsymbol{C}_1$ and $\boldsymbol{C}_3$ gives the opposite behaviour so that the outcomes are always different. So

$$p(\boldsymbol{C}_1 = H, \boldsymbol{C}_2 = H) = p(\boldsymbol{C}_1 = T, \boldsymbol{C}_2 = T) = \frac{1}{2}, \tag{3.37}$$

$$p(\boldsymbol{C}_2 = H, \boldsymbol{C}_3 = H) = p(\boldsymbol{C}_2 = T, \boldsymbol{C}_3 = T) = \frac{1}{2}, \tag{3.38}$$

$$p(\boldsymbol{C}_1 = H, \boldsymbol{C}_3 = T) = p(\boldsymbol{C}_1 = T, \boldsymbol{C}_3 = H) = \frac{1}{2}. \tag{3.39}$$

Now if we flip a single coin at a time, then we see that the probabilities of heads or tails are equal for any of the coins, so that we can treat each coin separately as a fair coin. But what would happen if we flipped all three coins at a time? In other words, can we use the above statistics to determine the unobservable joint statistics for the hypo-

thetical toss of all three coins together? Since the pairs of coins $\boldsymbol{C}_1$ and $\boldsymbol{C}_2$ as well as $\boldsymbol{C}_2$ and $\boldsymbol{C}_3$ are perfectly correlated, we arrive at the conclusion that the coins $\boldsymbol{C}_1$ and $\boldsymbol{C}_3$ also always give equal outcomes. So all three coins are guaranteed the same outcome. On the other hand, we know that coins $\boldsymbol{C}_1$ and $\boldsymbol{C}_3$ are perfectly *anti*-correlated. Since they cannot be both correlated *and* anti-correlated, we have arrived at a contradiction and we see that constructing a joint probability distribution for all three coins is in fact impossible. This is known as the marginal problem, which mathematicians have studied since the early part of the 20th century and has oftentimes appeared in literature [88–90].

Whenever the outcomes of two of the coins are well defined, the outcome of the third is not, striking at the heart of what we mean by counterfactual definiteness. Suppose the observables $\boldsymbol{C}_1, \boldsymbol{C}_2, \boldsymbol{C}_3$ describe measurement outcomes of physical objects. Then counterfactual definiteness assumption tells us that all outcomes must be physically determined already prior to the measurement, even if the variable determining the outcome is a hidden variable which cannot be directly measured. But nonexistence of a joint probability distribution implies that the outcomes of $\boldsymbol{C}_1, \boldsymbol{C}_2, \boldsymbol{C}_3$ cannot be determined simultaneously. Counterfactual definiteness is thus equivalent to the existence of a joint probability distribution for outcomes of all measurable random variables, whereas randomness arises merely due to the existence of the underlying hidden variables. If such a probability distribution cannot be found then realism is violated.

When a joint probability distribution does exist there are several convenient mathematical manouevers that can be executed. In particular, we may join expectation values together so that $\langle \boldsymbol{C}_1 \rangle + \langle \boldsymbol{C}_2 \rangle = \langle \boldsymbol{C}_1 + \boldsymbol{C}_2 \rangle$, as $\boldsymbol{C}_1 + \boldsymbol{C}_2$ is now a random variable in its own right (this would not be the case if there were no joint probability distribution for outcomes of $\boldsymbol{C}_1$ and $\boldsymbol{C}_2$ at the same time). Now consider the expression

$$\langle \boldsymbol{A}_1 \boldsymbol{A}_2 \rangle + \langle \boldsymbol{A}_1 \boldsymbol{B}_2 \rangle + \langle \boldsymbol{B}_1 \boldsymbol{A}_2 \rangle - \langle \boldsymbol{B}_1 \boldsymbol{B}_2 \rangle \,, \tag{3.40}$$

where $\boldsymbol{A}_1, \boldsymbol{B}_1$ are random variables describing measurement outcomes for Alice and similarly $\boldsymbol{A}_2, \boldsymbol{B}_2$ for Bob, and where each random variable has two possible outcomes, $\pm 1$. The maximum attainable value for this expression is 4, when the first three terms give $+1$ and the last term $-1$ and similarly the minimum is $-4$. However, we will see

that local realism restricts the values of this expression further.

Firstly, counterfactual definiteness allows us to treat combinations of $\boldsymbol{A}_2, \boldsymbol{B}_2$ such as $\boldsymbol{A}_2 + \boldsymbol{B}_2$ as random variables in their own right. Secondly, the locality assumption further tells us that the marginal probability distributions for $\boldsymbol{A}_1, \boldsymbol{B}_1$ cannot be changed by the choice of measurement $\boldsymbol{A}_2$ or $\boldsymbol{B}_2$ made by Bob (even though the outcomes themselves may be correlated through hidden variables that were encoded in the objects at the source). Therefore, a joint probability distribution exists for all four random variables $\boldsymbol{A}_1, \boldsymbol{B}_1, \boldsymbol{A}_2, \boldsymbol{B}_2$ and we can join the terms in Eq. (3.40) together to find

$$\langle \boldsymbol{A}_1(\boldsymbol{A}_2 + \boldsymbol{B}_2) + \boldsymbol{B}_1(\boldsymbol{A}_2 - \boldsymbol{B}_2) \rangle \,. \tag{3.41}$$

Now $\boldsymbol{A}_2$ and $\boldsymbol{B}_2$ are either equal or opposite. When they are equal, the second term vanishes and the first term is $\pm 2$. When they are opposite, the first term vanishes and the second term is $\pm 2$. Therefore, since both terms can only take values $\pm 2$, the expectation of their sum must satisfy

$$|\, \langle \boldsymbol{A}_1 \boldsymbol{A}_2 \rangle + \langle \boldsymbol{A}_1 \boldsymbol{B}_2 \rangle + \langle \boldsymbol{B}_1 \boldsymbol{A}_2 \rangle - \langle \boldsymbol{B}_1 \boldsymbol{B}_2 \rangle \,| \leq 2. \tag{3.42}$$

This is the CHSH inequality. Since local realism implies that this inequality is satisfied, we have conversely that if the inequality is violated then local realism must not hold for nature.

Notice also that in joining the expectations together to obtain Eq. (3.41) we could also have just assumed counterfactual definiteness for *all* four random variables at the same time without assuming locality. This is equivalent to assuming local counterfactual definiteness for only *local* measurement outcomes when taken together with the assumption of locality. However, if we had assumed counterfactual definiteness for only the local measurement outcomes without making the assumption of locality, then we are only allowed to transform $|\, \langle \boldsymbol{A}_1 \boldsymbol{A}_2 \rangle + \langle \boldsymbol{A}_1 \boldsymbol{B}_2 \rangle + \langle \boldsymbol{B}_1 \boldsymbol{A}_2 \rangle - \langle \boldsymbol{B}_1 \boldsymbol{B}_2 \rangle \,| = |\, \langle \boldsymbol{A}_1 (\boldsymbol{A}_2 + \boldsymbol{B}_2) \rangle + \langle \boldsymbol{B}_1 (\boldsymbol{A}_2 - \boldsymbol{B}_2) \rangle \,|$, but not any further. This expression may still attain the value $4$. Conversely, if we had made only the assumption of locality and had assumed nothing about counterfactual definiteness, then the expression is also bounded by $4$ and it was shown by Popescu and Rorhlich that this bound can actually

be attained using the so called nonlocal boxes, or Popescu-Rorhlich (PR) boxes [91].

The maximum value attainable by the CHSH inequality expression using a given two qubit state and quantum measurements may be calculated explicitly using the results in [92]. The maximum value attainable for quantum theory is given by $2\sqrt{2}$, first obtained by Cirel'son [93] and known as the *Tsirelson bound*, and the mathematically highest attainable value for any theory is 4. All probability distributions that take the form of Eq. (3.36) posses a counterfactual joint probability distribution and therefore do not violate the Ineq. (3.42). Since separable states take the form of Eq. (3.36), we therefore conclude that the violation of the CHSH inequality implies that the state is entangled.

Notice that in our discussion of the Bell inequality we at no point relied on the quantum notion of random variables being observables. We allow for general random variables and only require that the they correspond to dichotomic measurement outcomes obtained by two spacelike separated parties. The inequality therefore serves as a kind of litmus test for both theories and experiments: If violated then local realism is not satisfied by the theory in question or, in the case of experiments, nature. However, when conducting experiments particular care must be taken, as loopholes may mean that an obtained violation of the inequality might not be translatable into the violation of local realism by nature. However, we will set these issues aside in this thesis and examine the issue from a theoretical perspective, and treat violation of Bell inequalities as a consequence of the theory of quantum mechanics which might not be capable of undergoing a rigorous experimental test. For interested readers, further discsussion of Bell inequality loopholes can be found, for example, in [94–97].

It has been proposed in [98] that the principle of *information causality* - the idea that $n$ bits of classical communication and no other kind of communication can never result in more than $n$ bits of information learnt by the other party regardless of the resources shared - is behind the Tsirelson bound, as any violation of the bound implies also the violation of information causality. Information causality when $n = 0$ is commonly known as the no-signalling regime whereby lack of any classical communication or otherwise cannot result in a distant party receiving any new information. PR boxes do not violate the no-signalling assumption, but are shown to violate the more general information causality assumption.

In this section we have looked at how inequalities can be used to demonstrate the

violation of local realism in the case of two spacelike separated observers. Next we shall examine Bell inequalities in the multipartite context, where issues of genuinely and nongenuinely multipartite nonlocality will arise.

### 3.3.2 Multipartite Bell inequalities

An important difference when considering bipartite or multipartite nonlocality is in whether it is all parts of the system that violate local realism *together* or whether there exist groups of observers such that if we considered each as a single observer we would find local realism restored. To make this concept clearer, consider the case of three observers. We would like to find a local realist model for the three observers of the kind we considered for the bipartite systems in Eq. (3.36). One might write such a tripartite model as

$$p(x_1, x_2, x_3) = \int d\lambda \rho(\lambda) p_1(x_1|\lambda) p_2(x_2|\lambda) p_3(x_3|\lambda). \tag{3.43}$$

However, if for example Alice and Bob (denoted by numbers 1 and 2 above) posses a nonlocally correlated state while Charlie (system 3) possesses a local state of his own then it would follow that the probability distribution cannot be written in the form of Eq. (3.43). However, grouping Alice and Bob together and considering them as a single local observer would reduce the system to become local realistic. There are parts of the whole that can be considered to be local realistic, and are refered in the literature as bipartitions as we partitioned the system into two parts which can be considered local realistic. We therefore say that the system is not *genuinely* multipartite nonlocal, where the adverb genuinely refers to multipartite and not nonlocal.

To ensure genuinely multipartite nonlocality we should require further that no such bipartitions exist. Thus we make a requirement that for a distribution to be called genuinely multipartite nonlocal, it must not be of the form

$$p(x_1, x_2, x_3) = \sum_{k=1}^{3} q_k \int d\lambda \rho_{ij}(\lambda) p_{ij}(x_i, x_j|\lambda) p_k(x_k|\lambda), \tag{3.44}$$

where in the above sum $i, j$ take values such that $\{i, j, k\} = \{1, 2, 3\}$ and $p_{ij}$ refers to the joint distribution for parties $i$ and $j$ and $q_k \geq 0$ is a probability distribution

$\sum_k q_k = 1$. The probability model of Eq. (3.44) is less restrictive of a distribution for it to be called local realist, and consequently there are *less* distributions that do not satisfy it and which can be called genuinely multipartite nonlocal, in accordance with expectation. Svetlichny showed in [99] that if a distribution is of the said form, then it satisfies the Svetlichny inequality

$$\langle \boldsymbol{A}_1\boldsymbol{A}_2\boldsymbol{B}_3\rangle + \langle \boldsymbol{A}_1\boldsymbol{B}_2\boldsymbol{A}_3\rangle + \langle \boldsymbol{B}_1\boldsymbol{A}_2\boldsymbol{A}_3\rangle - \langle \boldsymbol{B}_1\boldsymbol{B}_2\boldsymbol{B}_3\rangle + \langle \boldsymbol{B}_1\boldsymbol{B}_2\boldsymbol{A}_3\rangle \\ + \langle \boldsymbol{B}_1\boldsymbol{A}_2\boldsymbol{B}_3\rangle + \langle \boldsymbol{A}_1\boldsymbol{B}_2\boldsymbol{B}_3\rangle - \langle \boldsymbol{A}_1\boldsymbol{A}_2\boldsymbol{A}_3\rangle \leq 4. \quad (3.45)$$

Following the arguments in [100], the fact that the model of Eq. (3.44) satisfies the Svetlichny inequality can be demonstrated by noticing that the expression in Eq. (3.45) is a sum of two terms

$$\boldsymbol{S}_3 = \boldsymbol{A}_1(\boldsymbol{A}_2\boldsymbol{B}_3 + \boldsymbol{B}_2\boldsymbol{A}_3 + \boldsymbol{B}_2\boldsymbol{B}_3 - \boldsymbol{A}_2\boldsymbol{A}_3) + \boldsymbol{B}_1(\boldsymbol{A}_2\boldsymbol{A}_3 - \boldsymbol{B}_2\boldsymbol{B}_3 \\ + \boldsymbol{B}_2\boldsymbol{A}_3 + \boldsymbol{A}_2\boldsymbol{B}_3), \quad (3.46)$$

where we should remember that in order to get the inequality the terms must be expanded and expectation value taken of each term in the sum separately. Notice therefore that whenever Alice chooses to do the measurement corresponding to random variable $\boldsymbol{A}_1$, Bob and Charlie are testing the CHSH inequality with their random variables. We shall denote the CHSH inequality polynomial as $\boldsymbol{S}_2$. When she chooses $\boldsymbol{B}_1$, they also play a CHSH game but with $\boldsymbol{A}_{2,3}$ interchanged with $\boldsymbol{B}_{2,3}$, a polynomial we shall denote as $\boldsymbol{S}'_2$. We can therefore write $\boldsymbol{S}_3$ as

$$\boldsymbol{S}_3 = \boldsymbol{A}_1\boldsymbol{S}_2 + \boldsymbol{B}_1\boldsymbol{S}'_2. \quad (3.47)$$

We could make the same claim by separating out any of the other two observers and have the remaining observers play CHSH games.

**Argument for the inequality:** Consider the bipartition $12/3$ in the model of Eq. (3.44), so that it can be written as $\int d\lambda \rho_{12} p_{12}(x_{12}|\lambda) p_3(x_3|\lambda)$. Then Bob knows which of the two CHSH games he is supposed to play with Charlie, since Alice is local to Bob and she can choose her input $\boldsymbol{A}_1$ or $\boldsymbol{B}_1$ and hence whether Bob and Charlie play $\boldsymbol{S}_2$ or $\boldsymbol{S}'_2$. However, as Bob and Charlie are not nonlocally correlated in this model, the

maximum they can attain in $\boldsymbol{S}_2$ is as before, 2. Equivalently, we can consider the partition $1/23$ or $2/13$, in which case we can express the Svetlichny inequality CHSH games determined by Charlie and arrive at the same conclusion. Therefore, the bipartition models can achieve the value of at most $2+2=4$ in the Svetlichny inequality. The model of Eq. (3.44) is a probabilistic sum of such bipartition models and since none of the terms of the sum can exceed 4, neither can the sum.

Given the above arguments, the maximum quantum mechanical value analogous to the Tsirelson bound for the CHSH inequality is $4\sqrt{2}$, since the CHSH games can achieve at most $2\sqrt{2}$ and therefore the inequality cannot exceed $4\sqrt{2}$ for quantum mechanics. The state that attains this value is the Greenberger-Horne-Zeilinger state $(|000\rangle + |111\rangle)/\sqrt{2}$, as defined in [101]. The observables required to attain this value can be found in [100].

The form of the Eq. (3.47) suggests a natural generalization of the Svetlichny inequality to the $n$-partite setting,

$$\boldsymbol{S}_n = \boldsymbol{A}_1 \boldsymbol{S}_{n-1} + \boldsymbol{B}_1 \boldsymbol{S}'_{n-1}. \tag{3.48}$$

If we take $\langle \boldsymbol{S}_n \rangle$ to mean "expand the sum and take expectation value of each of the terms separately" we can then write

$$\langle \boldsymbol{S}_n \rangle \leq 2^{n-1}. \tag{3.49}$$

We shall call this inequality the BBGL inequality after their authors, Bancal, Brunner, Gissin and Liang. The BBGL inequality can be shown to hold for systems that are not genuinely nonlocal by induction. Suppose $\langle \boldsymbol{S}_{n-1} \rangle \leq 2^{n-2}$ holds for any bipartition of $n-1$ parties. Then by the same argument as above for the three party case, but where now we join $n-1$ parties together instead of 2, the inequality $\langle \boldsymbol{S}_n \rangle \leq 2^{n-1}$ must hold for $n$ parties.

In contrast to genuinely multipartite nonlocal probability distributions, the non-genuinely multipartite nonlocal disitributions are only required not to be of the form of Eq. (3.43). This ensures they cannot be fully factored, but it may still be possible to split the subsystems among bipartitions and factor the probability distribution with respect to those. Nongenuinely multipartite nonlocality is in this sense a less powerful

form of nonlocality. We shall review several of these inequalities below. The detailed arguments for why they hold are very similar to above and will not be repeated and can also be found in the accompanying literature.

The first of these inequalities we review is the Żukovski-Brukner inequality [102]. For notational convenience, write our random variables $\boldsymbol{A}_k$ and $\boldsymbol{B}_k$ as $\boldsymbol{A}_k = \boldsymbol{q}_k^1$ and $\boldsymbol{B}_k = \boldsymbol{q}_k^2$. The Żukovski-Brukner inequality can then be written as

$$\sum_{s_1,\ldots,s_n=-1,1} \left| \sum_{k_1,\ldots,k_n=1,2} s_1^{k_1-1} s_2^{k_2-1} \ldots s_M^{k_M-1} \left\langle \boldsymbol{q}_1^{k_1} \boldsymbol{q}_2^{k_2} \cdot \ldots \cdot \boldsymbol{q}_n^{k_n} \right\rangle \right| \leq 2^n, \tag{3.50}$$

where $s_j^k = \pm 1$ are summation indices. This inequality reduces to the CHSH inequality Eq. (3.42) in the case of two parties.

Similar to Eq. (3.50), the Mermin-Ardehali-Belinskii-Klyshko (MABK) inequalities are given by [103–105]

$$\left| \sum_{s_1,\ldots,s_N=-1,1} S(s_1,\ldots s_N) \sum_{k_1,\ldots,k_N=1,2} s_1^{k_1-1} \ldots s_M^{k_N-1} \left\langle \boldsymbol{q}_1(k_1) \otimes \ldots \otimes \boldsymbol{q}_M(k_N) \right\rangle \right| \leq 2^N, \tag{3.51}$$

where $S(s_1, s_2, \ldots, s_M) = \sqrt{2} \cos\left[\frac{\pi}{4}(s_1 + \ldots + s_M - M - 1)\right]$. For the tripartite case $N = 3$ the inequality in Eq. (3.51) becomes

$$\left| \langle \boldsymbol{A}_1 \boldsymbol{B}_2 \boldsymbol{A}_3 \rangle + \langle \boldsymbol{B}_1 \boldsymbol{A}_2 \boldsymbol{B}_3 \rangle + \langle \boldsymbol{B}_1 \boldsymbol{B}_2 \boldsymbol{A}_3 \rangle - \langle \boldsymbol{A}_1 \boldsymbol{A}_2 \boldsymbol{A}_3 \rangle \right| \leq 2. \tag{3.52}$$

The violation of this inequality implies the violation of Ineq. (3.50). However, the converse is not the case. The inequalities (3.51) will be particularly useful later in this thesis.

In the context of the quantum information theory, what can we learn from Bell inequalities? Much of the quantum information theory has concerned itself with nonclassical correlations and in particular what advantages can be gained through the use of nonclassical states. Here Bell inequalities tell us that there is something profoundly different about quantum correlations, which are shown to be a fundamentally new resource in the quantum world living beyond classical correlations. One of the aims of the quantum information theory is both to exploit, as well as to give a quantita-

tive description of these resources in a way that might prove useful for applications to quantum technology. In the next chapter we provide a quantification of one of such resources, entanglement, in the presence of technical and fundamental restrictions that go beyond the usual paradigm of local operations and classical communication.

# CHAPTER 4

## Quantification of entanglement under restricted operations

Quantifying resources in the presence of restrictions underpins numerous quantum information concepts [19]. Indeed, entanglement itself first arose as a resource when protocols are viewed under the restriction of only using local operations and classical communication (LOCC) [19]. In this chapter we will further restrict operations the two parties are allowed to use, where their measurements can be selected from only a restricted repertoire of measurement operators rather than the full set. We will then show how to employ semiquantum nonlocal-games, introduced later in this chapter and earlier in the literature [30], as the quantum gauge protocols to quantify the *effective* entanglement -- a quantity which characterises how much entanglement is accessible to an observer with limitations to their measurements. Using a protocol to define quantum entanglement has a long history [10, 11], going back to the entanglement of formation and distillation. Using nonlocal games to determine the degree entanglement degrades follows the same spirit. To define effective entanglement we will first introduce a set of states in which every state can be used to play nonlocal games with at least as large payoff as is achievable with the original state, but with non-restricted measurements. We choose the least entangled state $\bar{\rho}$ from this set and define $E(\bar{\rho})$ as the effective entanglement.

We will formally define effective entanglement and prove several results charac-

terising measures of it in section 4.1, while in section 4.2 we deal with the important case where the imperfection or restriction is described by a CPM [46] and show that effective G-concurrence reduces to being proportional to the conventional G-concurrence [65, 66] with a restriction dependent scale factor. We then apply these results to quantify the entanglement of indistinguishable particles in section 4.3, concentrating on the controversial concept of single-particle entanglement [106–108], for the important examples of imperfect and super-selection rule restricted measurements.

## 4.1 Nonlocal games as gauge of effective entanglement

The definition of effective entanglement relies heavily on using the semiquantum nonlocal games as the comparison gauge. We therefore first proceed to describe the rules of semiquantum nonlocal games (also see [30] for further details). They consist of four index sets, $\mathcal{S} = \{s\}$, $\mathcal{T} = \{t\}$, $\mathcal{X} = \{x\}$ and $\mathcal{Y} = \{y\}$. The referee picks indices $s$ and $t$ at random with probabilities $p(s)$ and $q(t)$ and prepares some corresponding quantum states $\zeta^s$ and $\eta^t$, sending them to players Alice and Bob, respectively. States corresponding to different indices need not be orthogonal, which is why the game is called a semiquantum nonlocal game, to distinguish it from a more classical nonlocal game where the states corresponding to different indices are fully distinguishable. The players must separately compute the respective answers $x \in \mathcal{X}$ and $y \in \mathcal{Y}$ and send them to the referee who then computes the payoff using the payoff function $\mathfrak{p}(s,t,x,y)$. Payoff need not be positive, in which case the players must pay the referee.

Before the game begins the players may confer with one another and use any resources they like in order to coordinate the strategy. After the game has begun, however, they are not allowed to communicate. All they can do is share a joint quantum state $\rho$ and perform joint measurements, described by a POVM, on $\rho$ and the question states sent by the referee with outcomes in $\mathcal{X}$ and $\mathcal{Y}$, respectively. The average payoff Alice and Bob expect to obtain is expressed by the formula

$$\mathfrak{p}^*(\rho) = \max \sum_{s,t,x,y} p(s)q(t)\mathfrak{p}(s,t,x,y)\mu(x,y|s,t), \tag{4.1}$$

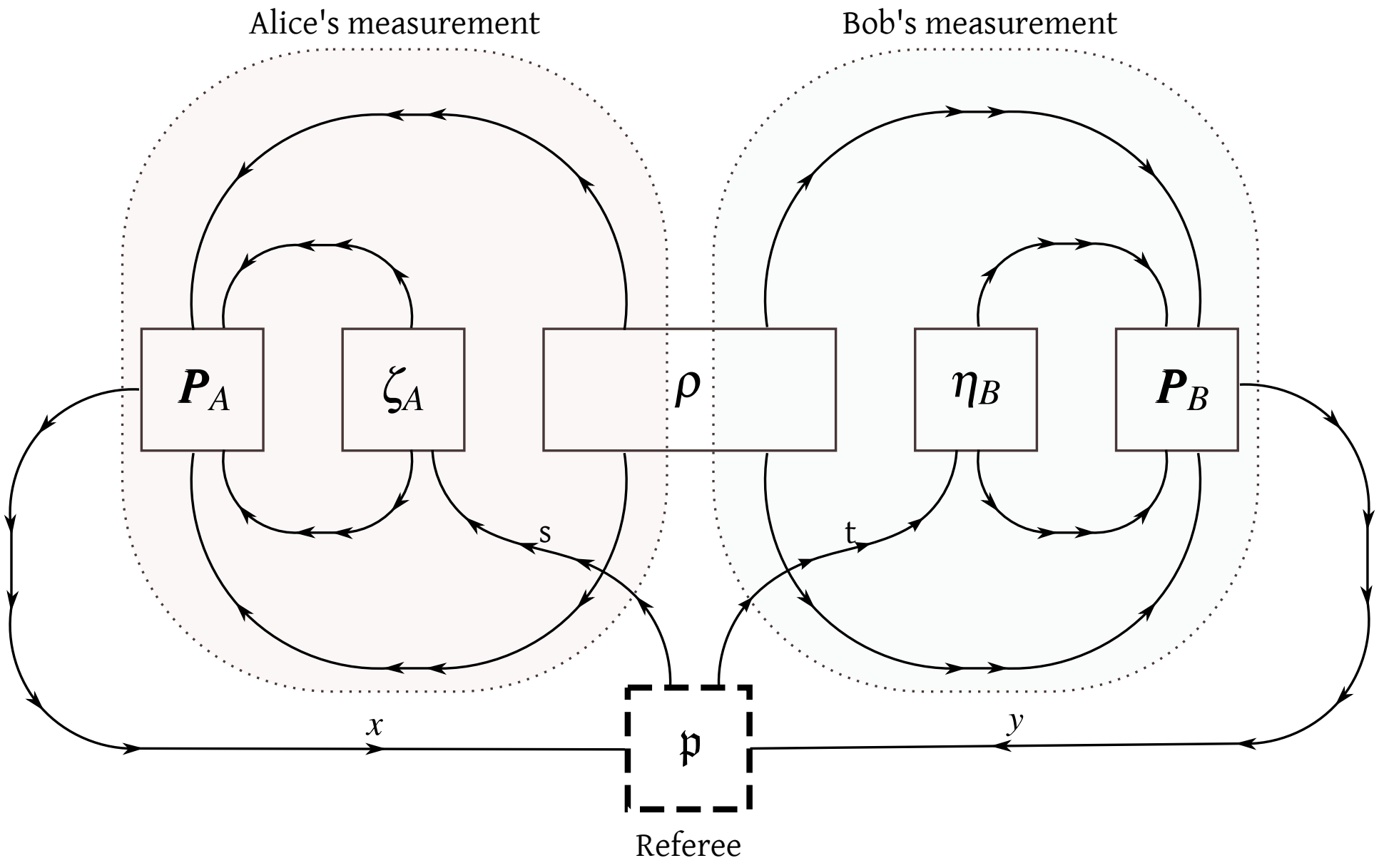


Figure 4.1: The Penrose diagram illustration of a nonlocal game. The referee sends his questions $\zeta_A$ and $\eta_B$ to the players, who conduct joint measurements on the state and the question. They then send their respective answers back to the referee, who computes the payout. As in the Penrose diagram notation, the lines represent inner products. The arrows are added to the lines to represent the flow of information.

where $\mu(x, y|s, t)$ is the joint conditional probability of obtaining outcomes $x, y$ given that the question states $\zeta^s$ and $\eta^t$ were sent and is computed using the standard quantum probability formulae (see figure 4.1). The function $\mu(x, y|s, t)$ implicitly depends on the POVMs chosen by the players and it is these POVMS that we maximize over in Eq. (4.1) to obtain $\mathfrak{p}^*(\rho)$.

We call a state $\rho_1$ *more nonlocal* than $\rho_2$, denoted as $\rho_1 \succeq \rho_2$, if and only if for every semiquantum nonlocal game $\mathfrak{p}^*(\rho_1) \geq \mathfrak{p}^*(\rho_2)$. It is then shown in [30] that $\rho_1 \succeq \rho_2$ if and only if $\rho_1$ can be transformed to $\rho_2$ using only local operations and shared randomness (LOSR). This is denoted as $\rho_1 \mapsto \rho_2$. Given an entanglement measure $E$ that is non-increasing under LOSR, we then have that

$$\rho_1 \succeq \rho_2 \Rightarrow E(\rho_1) \geq E(\rho_2). \tag{4.2}$$

Notice that this implies that whenever $\mathfrak{p}^*(\rho_1) = \mathfrak{p}^*(\rho_2)$ for all games we must have that $E(\rho_1) = E(\rho_2)$, justifying the idea that semiquantum nonlocal games act as an

entanglement gauge. LOSR transformations are a subset of LOCC operations and so any entanglement measure satisfying the standard required properties can be used (for a review of the properties of entanglement measures, see for instance [10, 11]).

Before we look at the nonlocal games in the presence of POVM restrictions, we define what we mean by an effective POVM. Suppose $\boldsymbol{P}_1, \ldots \boldsymbol{P}_2$ is a joint POVM over two Hilbert spaces, with the joint state being a product state $\rho_1 \otimes \rho_2$. Then there exists a $\rho_1$-dependent POVM acting only on $\rho_2$ with the same outcome probabilities. Writing $\boldsymbol{P}_k = \sum_l \lambda_{kl} \boldsymbol{P}^1_{kl} \otimes \boldsymbol{P}^2_{kl}$, we can obtain the effective POVM by computing $\mathrm{Tr}_1[\boldsymbol{P}_k \rho_1 \otimes \rho_2] = \sum_l \lambda_{kl} \,\mathrm{Tr}[\boldsymbol{P}^1_{kl}\rho_1]\boldsymbol{P}^2_{kl}\rho_2$, giving $\sum_l \lambda_{kl} \,\mathrm{Tr}[\boldsymbol{P}^1_{kl}\rho_1]\boldsymbol{P}^2_{kl}$ as the effective operator.

Since we want to look at the effective entanglement when allowed POVMs acting on $\rho$ alone are restricted, we therefore maximize the Eq. 4.1 over only those POVMs, whose effective POVM on $\rho$ alone belongs to a certain set of POVMs $\mathcal{R}$ describing the restrictions of our apparatus. We denote the resulting maximum average payoff as $\mathfrak{q}^*(\rho)$ (note that although notationally similar, $\mathfrak{p}^*()$ and $\mathfrak{q}^*()$ are the payoff functions, while $p$ and $q$ refer to probabilities of the referee choosing particular questions denoted by their respective indices). Denote as $\mathcal{E}$ the set of all states $\bar{\sigma}$ such that $\mathfrak{p}^*(\sigma) \geq \mathfrak{q}^*(\rho)$. In words, these are those states whose maximum average payoff function with no restrictions on POVMs is at least as great as the maximum payoff function of state $\rho$ with POVMs restricted. We then define the effective entanglement as

$$\bar{E}(\rho) = \inf\{E(\bar{\sigma}) : \bar{\sigma} \in \mathcal{E}\}. \tag{4.3}$$

The functional $\bar{E}(\rho)$ therefore gives us the least amount of entanglement that enables us to perform at least as well in any nonlocal game with no restrictions as we could with the more entangled state $\rho$ with restrictions. From the fact that $\rho \in \mathcal{E}$ it follows that $\bar{E}(\rho) \leq E(\rho)$. The infimum can be replaced by minimum whenever the set $\mathcal{E}$ is compact and the entanglement measure $E$ is continuous. This follows from a fundamental theorem of classical analysis, saying that continuous functions mapping compact sets to real numbers attain their infimum on a member of the set [47, 109]. Since $\mathcal{E} \subset \mathcal{B}(\mathcal{H})$ and $\mathcal{B}(\mathcal{H})$ is compact, $\mathcal{E}$ is also compact whenever it is closed due to the fact that closed subsets of compact sets are compact.

It is useful to have a condition that allows us to more easily find and verify when a particular functional is the effective entanglement $\bar{E}$. We now provide one such

condition.

**Theorem 4.1.** *If there exists a state* $\bar{\rho} \in \mathcal{E}$ *such that for all semiquantum nonlocal games* $\mathfrak{p}^*(\bar{\rho}) = \mathfrak{q}^*(\rho)$ *then* $\bar{E}(\rho) = E(\bar{\rho})$.

*Proof.* To see this notice that for any $\bar{\sigma} \in \mathcal{E}$ we have that $\mathfrak{p}^*(\bar{\sigma}) \geq \mathfrak{q}^*(\rho) = \mathfrak{p}^*(\bar{\rho})$ for all semiquantum nonlocal games. Therefore, $\bar{\sigma} \succeq \bar{\rho}$ and $E(\bar{\sigma}) \geq E(\bar{\rho})$ by the implication (4.2). Therefore $\bar{E}(\rho) = E(\bar{\rho})$. □

The theorem is saying that when there is a state $\bar{\rho}$ that is exactly as good for playing nonlocal games with unrestricted measurements as the state $\rho$ is with restricted measurements, this state can be considered to be the effective state that gives us the effective entanglement. Given that precisely equal performance in nonlocal games implies equal entanglement, this theorem gives a property that effective entanglement should naturally be expected to satisfy.

In the following section we will consider a particularly simple description of measurement restrictions in terms of completely positive maps, encompassing a large range of measurement errors found in modern quantum laboratories. We will provide the exact formulae for computing effective entanglement under such circumstances and show that effective G-concurrence, Eq. (3.20), is proportional to the usual G-concurrence when measurement errors occur for only one of the observers. When they occur for both, the proportionality gives us a simple to compute upper bound.

## 4.2 Imperfections and restrictions via CPMs

To define the restricted set $\mathcal{R}$ of effective POVMs we are able to measure, we require that any POVM $\{\boldsymbol{P}_k\}$ in $\mathcal{R}$ is the image of some other completely arbitrary and general POVM $\{\boldsymbol{G}_k\}$ under the action of some local CPM $\$ = \$_A \otimes \$_B$ so that

$$\boldsymbol{G}_k = \$_A^\dagger \otimes \$_B^\dagger[\boldsymbol{P}_k]. \tag{4.4}$$

Notice that this is equivalent to the CPM $\$$ acting on the state $\rho$, since

$$p_k = \mathrm{Tr}\left(\$^\dagger[\boldsymbol{P}_k]\rho\right) = \mathrm{Tr}\left(\sum_j \boldsymbol{K}_j \boldsymbol{P}_k \boldsymbol{K}_j^\dagger \rho\right) \\ = \mathrm{Tr}\left(\boldsymbol{P}_k \sum_j \boldsymbol{K}_j^\dagger \rho \boldsymbol{K}_j\right) = \mathrm{Tr}\left(\boldsymbol{P}_k \$\left[\rho\right]\right), \quad (4.5)$$

where $\boldsymbol{K}_j$ are the Kraus operators associated with the CPM $\$_A \otimes \$_B$. The Eq. (4.4) is thus the Heisenberg picture equivalent of the operation $\$_A \otimes \$_B$ acting on the state $\rho$. It is therefore unsurprising that the state $\$_A \otimes \$_B(\rho)$ is the effective state $\bar{\rho}$ that can be used to define effective entanglement.

The concept of using $\bar{\rho}$ to compute effective entanglement becomes particularly appealing when the entanglement measure $E$ used is G-concurrence in $d$ dimensions $G_d$. In this case Eqs. (3.24), (3.25) and (3.26) give us a simple linear relationship between the effective G-concurrence and the conventional G-concurrence, with the proportionality factor given by the quality $Q$, which is a function of the restriction $\$$ only. In the case when only one of the parties is affected by errors, i.e. $\$ = \mathbf{1} \otimes \$_B$, we then have that $\bar{G}_d(|\psi\rangle) = Q(\$_B)G_d(|\psi\rangle$, and an inequality for the case of two-sided errors and mixed states. Although one-sided errors may appear very restrictive, they in fact do occur in many cases, particularly quantum information protocols where only one party performs measurements, while the other party does nothing, albeit we assume here that the errors arise only during the process of measurement and that the system is otherwise well isolated from the environment. Such protocols are very common and in fact two-party LOCC state transformations can be represented as measurements done by one party and unitary operations by the other (see [9] for details). The linearity makes the effective G-concurrence, or its upper bound, particularly simple to compute and we do so analytically for several examples that we expect appear particularly often in laboratory settings.

## 4.3 Entanglement of indistinguishable particles

The underlying physical framework for entanglement is a system composed of two or more individually addressable degrees of freedom that together form a tensor product Hilbert space where the state of the complete system is described. The archetypal case is $\mathbb{C}^2 \otimes \mathbb{C}^2$ for two qubits, which is usually envisaged as arising from two localized particles with spin-$\frac{1}{2}$. Conceptually entanglement is then signalled by a lack of separability of the state of the system with respect to this tensor product structure. However, if particles are delocalized and indistinguishable this raises a significant issues for quantifying entanglement because the relevant degrees of freedom can no longer be assigned to individual particles. Instead an analysis of entanglement requires a description in terms of the second quantized field modes which the particles can occupy. The most elementary example of this problem consists of a single-particle delocalized across two distinct spatial modes $\boldsymbol{a}$ and $\boldsymbol{b}$ yielding a state [106–108]

$$\left|\Psi^{\pm}\right\rangle = \frac{1}{\sqrt{2}}(\boldsymbol{a}^{\dagger} + \boldsymbol{b}^{\dagger})\left|\text{vac}\right\rangle = \frac{1}{\sqrt{2}}\left(\left|0\right\rangle_a \left|1\right\rangle_b \pm \left|1\right\rangle_a \left|0\right\rangle_b\right),$$

where $\left|n\right\rangle_a \left|m\right\rangle_b \propto (\boldsymbol{a}^{\dagger})^n(\boldsymbol{b}^{\dagger})^m \left|\text{vac}\right\rangle$. Interpreting the Fock states $\left|0\right\rangle$ and $\left|1\right\rangle$ of either mode as Pauli $\sigma_z$ eigenstates of a qubit suggests that the second quantized form $\left|\Psi^{\pm}\right\rangle$ it is an entangled state. Yet this notion that a single-particle can truly exhibit entanglement has raised considerable controversy [110–116]. This issue is quite naturally analysed within the framework of effective entanglement since the accessibility of any correlations in states such as $\left|\Psi^{\pm}\right\rangle$ is fundamentally linked to what measurements are available. Indeed one of the central issues is whether full discrimination of the Bell basis is possible, namely if $\left|\Psi^{\pm}\right\rangle$ and

$$\left|\Phi^{\pm}\right\rangle = \frac{1}{\sqrt{2}}(1 + \boldsymbol{a}^{\dagger}\boldsymbol{b}^{\dagger})\left|\text{vac}\right\rangle = \frac{1}{\sqrt{2}}\left(\left|0\right\rangle_a \left|0\right\rangle_b \pm \left|1\right\rangle_a \left|1\right\rangle_b\right),$$

can be measured. This is essential for such states to be a resource in protocols like teleportation [117, 118]. Recently the Bell discrimination has been shown to be possible with photons when non-linear optics are used in combination with a two level atom [119]. In the following we study single-particle entanglement for the case of photons with imperfect detectors and for massive particles where a super-selection rule

physically restricts measurements.

### 4.3.1 Optical amplitude and phase damping

The case of photons provides a simple test ground for effective entanglement. Suppose that the state $|\Psi^{\pm}\rangle$ was used within an LOCC protocol where Alice implements imperfect measurements of her optical mode $\boldsymbol{a}$. With a photon counter she can measure $\boldsymbol{a}^{\dagger}\boldsymbol{a}$, or via a balanced homodyne detector with a local oscillator with a phase $\phi$ she can measure a field quadrature $X(\phi) = (\boldsymbol{a}\, e^{-i\phi} + \boldsymbol{a}^{\dagger} e^{-i\phi})/2$. Imperfections in photon counting might, for example, cause the detector not to "click" due to photon loss within the device. This is amplitude damping and can be modelled as a beam-splitter at the input port of a perfect counter which scatters an incoming photon into another unmonitored optical mode, as shown in Fig. 4.2(a). Imperfections in the measurements of field quadratures might arise due to uncertainty in the phase $\phi$. This is phase damping and can be modelled by a local oscillator subject to phase fluctuations, as shown in as shown in Fig. 4.2(b). For both types of measurements their errors, within the subspace where no more than one photon occupies the local mode, are described by a CPM of the form $\$[\rho] = \boldsymbol{E}_0 \rho \boldsymbol{E}_0^{\dagger} + \boldsymbol{E}_1 \rho \boldsymbol{E}_1^{\dagger}$. Amplitude damping has $\boldsymbol{E}_0 = |0\rangle\langle 0| + \sqrt{1-\gamma}\,|1\rangle\langle 1|$ and $\boldsymbol{E}_1 = \sqrt{\gamma}\,|0\rangle\langle 1|$, with a photon loss rate $\gamma$. Phase damping has $\boldsymbol{E}_0 = |0\rangle\langle 0| + \sqrt{1-\lambda}\,|1\rangle\langle 1|$ and $\boldsymbol{E}_1 = \sqrt{\lambda}\,|1\rangle\langle 1|$, with phase flipping elastic photon scattering occurring at a rate $(1-\sqrt{1-\lambda})/2$. In either case we can characterise the imperfect measurements via the state-channel isomorphism using $|\Psi^{\pm}\rangle$ by computing $\rho_{\$} = (\$ \otimes \mathbf{1})\,|\Psi^{\pm}\rangle\langle\Psi^{\pm}|$. The concurrence of $\rho_{\$}$ then gives the proportionality factor $Q[\$]$ between effective concurrence and the standard concurrence of any pure state as $Q[\$] = \sqrt{1-\gamma}$ and $Q[\$] = \sqrt{1-\lambda}$ for amplitude and phase damping, respectively. As expected these imperfections monotonically erode the effective entanglement accessible within a state such as $|\Psi^{\pm}\rangle$.

### 4.3.2 Massive particles subject to super-selection rules

If the modes $\boldsymbol{a}$ and $\boldsymbol{b}$ correspond to those of a massive particle then, in contrast to photons, super-selection rules impose a fundamental physical restriction on both the operations and measurements that can be performed. Specifically, massive particles

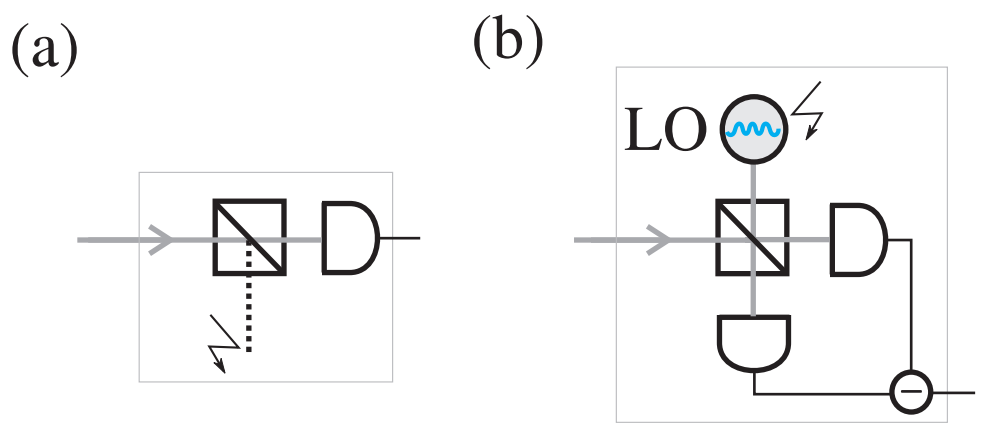


Figure 4.2: (a) A photon counter for the input mode suffering from amplitude damping noise of rate $\gamma$. The photon loss within the device is modelled by a beam-splitter with reflectivity $\gamma$. (b) A balanced homodyne device for measuring the field quadrature $X(\phi)$ of the input mode, where $\phi$ is controlled by the phase of the local oscillator (LO). Phase fluctuations of the LO cause phase damping noise in the device with a rate $\lambda$. The relation between the phase fluctuations and $\lambda$ is identical to those discussed later in Sec. 4.3.2.

are subject to Bargmann's super-selection rule for non-relativistic quantum mechanics [120, 121] which prohibits superpositions of states of different mass like those seen in the Bell states $|\Phi^{\pm}\rangle$. Strict adherence to this super-selection rule also means that the only permissible local measurements are those which commute with the number operator $\boldsymbol{a}^\dagger\boldsymbol{a}$ meaning also that measurements of local superpositions between $|0\rangle$ and $|1\rangle$ are inaccessible. With these restrictions one might reasonably question whether a single massive particle delocalized over two spatial regions in a state $|\Psi^{\pm}\rangle$ is really entangled. This issue has been hotly debated recently [122–124].

Given a state $\rho$, we say that $\sigma$ is a broken symmetry reference frame if it does not commute with the particle number operator and we intend to use it for joint measurements on $\rho\otimes\sigma$. If such a broken-symmetry reference frame is present, it can partially or fully lift the super-selection rule [19, 125–127] allowing coherences in $\rho$ to be measured. This has sparked investigations into both the quantification of entanglement in such scenarios [20, 21, 128] as well as the potential presence of single-particle quantum nonlocality [128–130]. As a result incorporating full or partial super-selection rule measurement restrictions into the quantification of entanglement is a fundamental requirement for building a meaningful entanglement measure for such systems [22, 131]. The framework of effective entanglement introduced here provides an exemplary tool in this context. Specifically, while the effective entanglement in the extreme cases of no restrictions and complete adherence to super-selection rules is currently understood, the intermediate cases permitted by a general broken-symmetry

reference frame are not.

In the case where reference frames are available which fully lift the super-selection rule restrictions then standard entanglement measures are sufficient. In the opposite case where no reference frames are present, the super-selection rule restrictions are described by a CPM, just like in the previous section. Since no coherences between different particle number sectors can be measured, the CPM must remove them from the measurement POVMs. Such CPM would take the form

$$\$\left[\rho\right] = \sum_n \Pi_n \rho \Pi_n, \tag{4.6}$$

where $\Pi_n$ is the projector onto the subspace of $n$ particles and $\rho$ is some state with fixed total number of particles. It is therefore sufficient for the above CPM to act on only one of the party's subsystems. If both Alice and Bob are affected by the same restriction, then we can get the effective state $\$[\rho] = \bar{\rho} = \bigoplus_n p_n \rho_n$ where the direct sum $\oplus$ signifies that it has block-diagonal form where $\rho_n = \Pi_n \rho \Pi_n$ and $p_n = \mathrm{Tr}\left[\Pi_n \rho\right]$. Thus, super-selection rule restricted effective entanglement measure for pure $\rho$ is given by

$$E(\bar{\rho}) = \sum_n p_n E(\rho_n), \tag{4.7}$$

where $E$ entanglement of formation. For a mixed state the above forms an upper bound. In fact this measure of entanglement for indistinguishable particles was introduced already by Wiseman and Vaccaro [22]. Here we note that the framework of effective entanglement has led naturally to the same result.

A cold-atom inspired setup provides a concrete example for exploring the general case in between these limits. Specifically, the resource state $|\Psi^{\pm}\rangle$ now describes the state of a single atom in a superposition over two tightly confined potentials, as implemented by two atomic quantum dots where the repulsive interactions between atoms is sufficiently large to prohibit double occupancy [132]. To exploit the resource state in some LOCC protocol $\Lambda$ Alice may be required to measure a super-selection rule violating superposition of the particle number states $|0\rangle$ and $|1\rangle$ of her local mode $\hat{a}$. She therefore needs access to a unitary transformation of the mode equivalent to the single-qubit rotation about the $x$-axis $\hat{R}_x(\theta) = \exp(-i\theta\sigma_x/2)$, where $\sigma_x$ is the Pauli

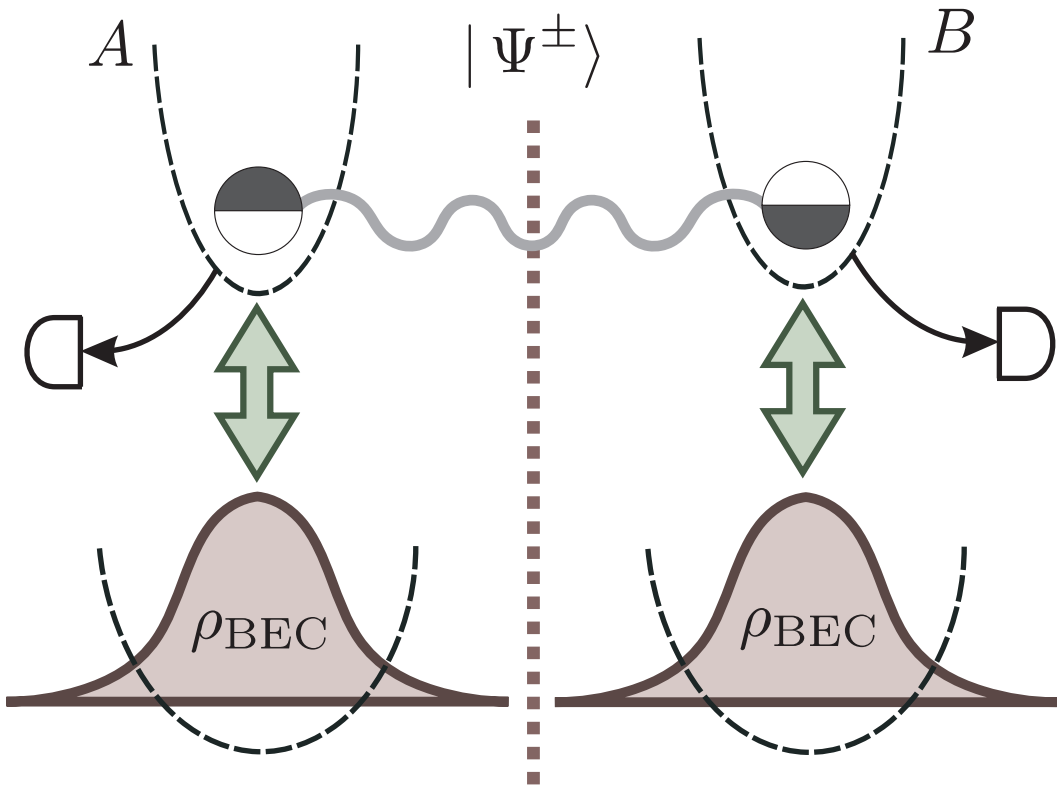


Figure 4.3: Two parties Alice $(A)$ and Bob $(B)$ share a single-particle entangled state $|\Psi^{\pm}\rangle$ whose superposition between two tight potentials is depicted as the half-filled circles. For her part of in an LOCC protocol $\Lambda$ Alice is required to measure a superposition of the particle number states $|0\rangle$ and $|1\rangle$ of her local mode $\hat{a}$. This coherent rotation is achieved by interacting her mode with an ancilla mode $\hat{c}$ which form a local BEC reference frame for a certain amount of time. She then measures the occupation of her local mode with an ideal detector. If the BEC has a large occupation and well defined phase then any rotation can be performed perfectly, breaking the super-selection rule restriction. In contrast if the local BEC has a completely uncertain phase then the rotation is completely incoherent and the super-selection rule restriction remains.

$x$ operator[1]. To implement such a rotation she exploits a local reference frame composed of an ancilla mode $\boldsymbol{c}$ in a Bose Einstein condensate (BEC) described by a mixture of coherent states

$$\rho_{\text{BEC}} = \int_0^{2\pi} d\phi p(\phi) \left||\alpha|e^{i\phi}\right\rangle \left\langle|\alpha|e^{i\phi}\right|, \tag{4.8}$$

where $|\alpha\rangle = \exp(-|\alpha|^2/2)\sum_{n=0}^{\infty}(\alpha\boldsymbol{c}^{\dagger})^n |\text{vac}\rangle /n!$ with $\alpha$ a complex number and $p(\phi)$ is the phase distribution. Owing to the global phase being unobservable the phase distribution $p(\phi)$ and its translation $p(\phi+\phi_0)$ create physically equivalent states $\rho_{\text{BEC}}$. Otherwise, depending on the structure of $p(\phi)$ the reference frame state $\rho_{\text{BEC}}$ either breaks or adheres to the number symmetry underlying the super-selection rule.

[1] Note that the mode equivalent of a single-qubit rotation about the $z$-axis $\hat{R}(\vartheta) = \exp(-i\vartheta\sigma_z/2)$ can be trivially implemented via evolution of the mode with a Hamiltonian of the form $\kappa\boldsymbol{a}^{\dagger}\boldsymbol{a}$.

The execution of the $R_x(\theta)$ rotation is then attempted by Alice jointly evolve her local mode $\boldsymbol{a}$ and the ancilla mode $\boldsymbol{c}$ via the Hamiltonian $\boldsymbol{H} = -\frac{1}{2}\Omega\left(\boldsymbol{a}^\dagger\boldsymbol{c} + \boldsymbol{c}^\dagger\boldsymbol{a}\right)$ for a given time $t$. This number-symmetric interaction drives an exchange of particles between the BEC reservoir and the resource state, after which the BEC reservoir is traced out. Such a measurement is depicted in Fig. 4.3. The effect of this evolution is best revealed by analysing one coherent state $\left||\alpha|e^{i\phi}\right\rangle$ in the mixture $\rho_{\text{BEC}}$. In the limit $|\alpha|^2 \gg 1$ this gives[2]

$$\begin{aligned} |0\rangle\left||\alpha|e^{i\phi}\right\rangle &\rightarrow \left(\cos(\omega t)|0\rangle - ie^{i\phi}\sin(\omega t)|1\rangle\right)\left||\alpha|e^{i\phi}\right\rangle, \\ |1\rangle\left||\alpha|e^{i\phi}\right\rangle &\rightarrow \left(\cos(\omega t)|1\rangle - ie^{i\phi}\sin(\omega t)|0\rangle\right)\left||\alpha|e^{i\phi}\right\rangle, \end{aligned}$$

where $\omega = \frac{1}{2}\Omega|\alpha|$. This represents a unitary evolution of the mode equivalent to the sequence of rotations $\hat{R}_z(\phi)\hat{R}_x(\omega t)$. Tracing the BEC out then yields an effective evolution of the mode as

$$\Gamma[\rho] = \int_0^{2\pi} d\phi p(\phi)\hat{R}_z(\phi)\hat{R}_x(\omega t)\rho\hat{R}_x^\dagger(\omega t)\hat{R}_z^\dagger(\phi). \tag{4.9}$$

This is equivalent to applying the unitary $\hat{R}_x(\omega t)$ to the input state as $\rho_R = \hat{R}_x(\omega t)\,\rho\,\hat{R}_x^\dagger(\omega t)$ followed by a phase-damping channel so

$$\Gamma[\rho] = (1 - |g|)\left(\mathbb{P}_0\rho_R\mathbb{P}_0 + \mathbb{P}_1\rho_R\mathbb{P}_1\right) + |g|\rho_R, \tag{4.10}$$

where $g = -i\int_0^{2\pi} d\phi p(\phi)\exp(i\phi)$, $\mathbb{P}_0 = |0\rangle\langle 0|$ and $\mathbb{P}_1 = |1\rangle\langle 1|$.

The CPM $\Gamma$ describing this measurement imperfection is not yet in the model form we restricted to in Sec. 4.2. Instead we identify the necessary map \$ as

$$\$[\rho] = \hat{R}_x^\dagger(\omega t)\Gamma[\rho]\hat{R}_x(\omega t). \tag{4.11}$$

The CPM factor in the case is $Q[\$] = |g|$, identical to a phase damping channel with a rate $\sqrt{1-\lambda} = |g|$, since the rotations $\hat{R}_x(\omega t)$ have no influence on entanglement. Thus $|g| = 0$ is a sufficient condition for vanishing effective entanglement and for pure input states it is also necessary.

[2]We write $\rightarrow$ instead of $\approx$ since exact evolution converges extremely rapidly as a function of $|\alpha|^2$ to the product form mapping shown.

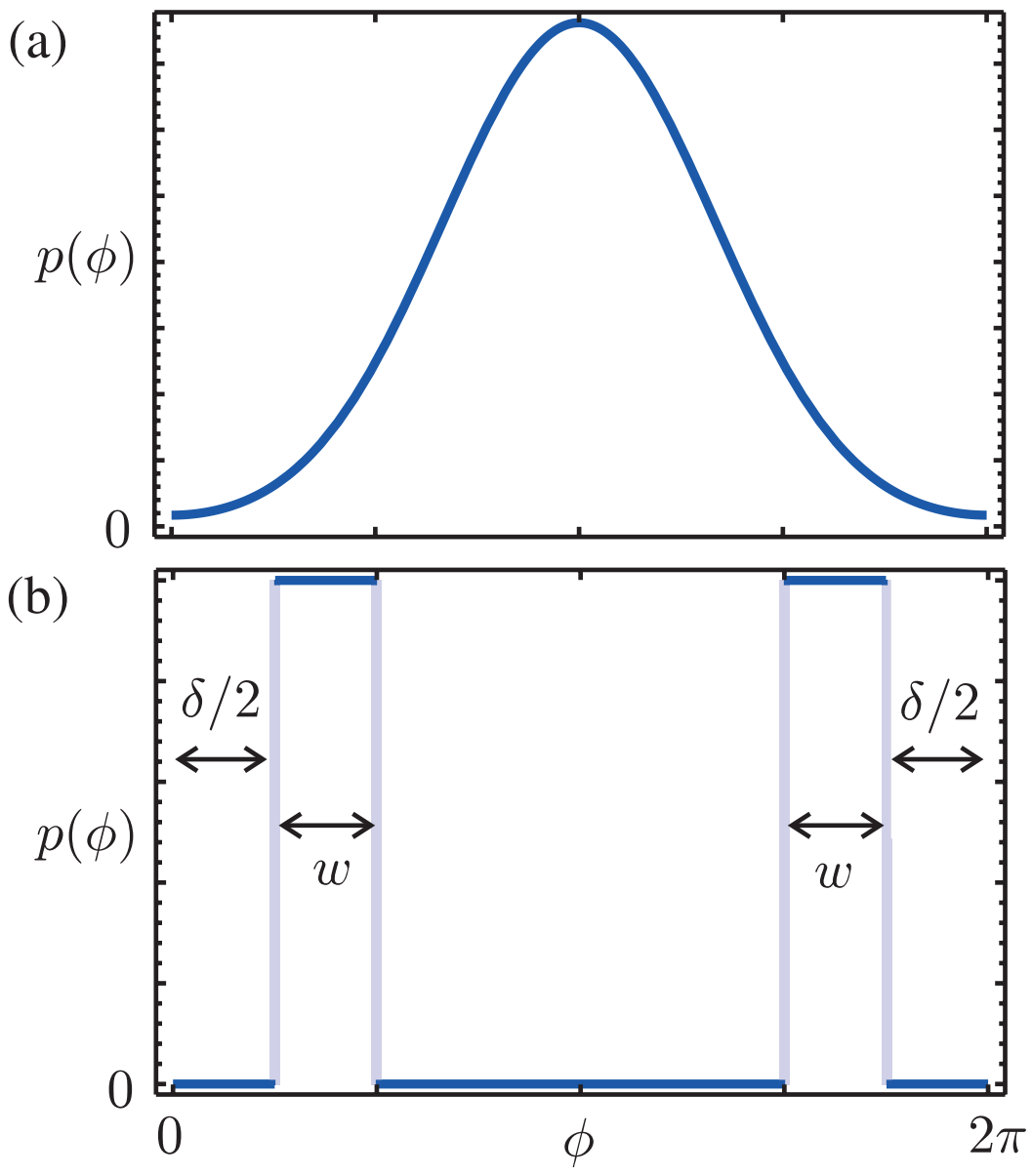


Figure 4.4: Two different examples of incoherent phase distributions $p(\phi)$ of the BEC reference frame. In (a) a wrapped normal distribution is shown which results in $|g| = \exp(-\sigma^2/2)$, where $\sigma$ is variance. In (b) sectionally constant distribution is shown which gives $|g| = (4w/\pi)\sin(w/2)\cos(w/2 + \delta/2)$, where $w, \delta$ are as shown.

If the BEC possesses a well defined phase $\phi_0$ so $p(\phi) = 2\pi\delta(\phi - \phi_0)$ then $|g| = 1$ and $\Gamma$ reduces to the unitary $\hat{R}_z(\phi_0)\hat{R}_x(\omega t)$ and Alice succeeds in rotating the state of mode $\boldsymbol{a}$ into the desired pure coherent superposition of $|0\rangle$ and $|1\rangle$. In this case the super-selection rule is fully lifted and there are no restrictions on what can be measured [19, 133]. This highlights that the BEC plays the role of a perfect local oscillator analogous to homodyne detection of field quadratures. In contrast, a completely uncertain phase $p(\phi) = (1/2\pi)$ gives $|g| = 0$ so the attempted rotation can only generate statistical mixtures of $|0\rangle$ and $|1\rangle$ in strict adherence to the super-selection rule.

For general distributions $p(\phi)$ allow $|g|$ to vary between these limiting cases yield a partial lifting of the super-selection rule restrictions. In this case it is not possible to coherently evolve into any superposition of $|0\rangle$ and $|1\rangle$ without at least some degree of mixing determined by $|g|$ Since $g$ is the average value of $\exp(i\phi)$, its absolute value is a direct measure of the amount of the reliability of the BEC as a phase reference. To

illustrate this we computed $|g|$ for a Gaussian distribution wrapped around a circle

$$p(\phi) = \frac{1}{\sigma\sqrt{2\pi}} \sum_{k=-\infty}^{\infty} \exp\left[\frac{-(\phi - \mu + 2\pi k)^2}{2\sigma^2}\right], \tag{4.12}$$

as shown in Fig. 4.4(a). We get $|g| = \exp(-\sigma^2/2)$ by integrating its uniformly convergent series over the interval $\phi \in [0, 2\pi]$ term by term. As expected, effective entanglement decreases with increasing phase uncertainty $\sigma$. We also considered the sum of two flat distributions of width $w$ on the circle with centres shifted from $\phi = 0$ and $\phi = 2\pi$ by $\delta/2$, as depicted in Fig. 4.4(b). Here we find that as long as there are no overlaps $|g| = (4w/\pi)\sin(w/2)\cos(\delta/2 + w/2)$.

The phase of the BEC is the crucial property allowing it to act as a super-selection rule breaking reference frame. Mixing a BEC of phase $\phi$ with one of phase $\phi + \pi$, by forming a state such as

$$\rho_{\text{BEC}} \propto ||\alpha|e^{i\phi}\rangle\langle|\alpha|e^{i\phi}| + ||\alpha|e^{i(\phi+\pi)}\rangle\langle|\alpha|e^{i(\phi+\pi)}|,$$

gives $|g| = 0$. However, $\rho_{\text{BEC}}$ itself still contains coherences which violate the super-selection rule. This shows that while symmetry breaking is a necessary condition for having non-vanishing effective entanglement, it is not sufficient.

We have learnt here how a nonlocal game in which entanglement is an essential resource may be used to quantify entanglement in the presence of restrictions. Now we shall turn our attention to another of the fundamental quantum phenomena, nonlocality.

# CHAPTER 5

# Multipartite nonlocality and superselection rules

As we saw in the previous chapter, particularly section 4.3, spatial field modes can exhibit entanglement in the particle number, or Fock, basis [114], just like the ordinary entangled states when appropriate reference frames are constructed. It is known from previous work that in order to attain any violation of Bell inequalities at all, the reference state must be constructed nonlocally [125] and the possibility the violation of Bell inequalities using mode entangled states is considered further in [122, 130, 134].

While the bipartite nonlocality of modes is fairly well understood, multipartite nonlocality under the influence of super-selection rules has not yet been investigated. In particularl it is not known whether the multipartite nonlocality remains genuinely multipartite and if so under what conditions. This will be the focus of this chapter. Initially we will take a selection of states known to be genuinely multipartite nonlocal when superselection rules do not apply, including W-states and Dicke states, and use multiple copies of the state to *activate* multipartite nonlocality. In such a scenario the state copies will act as a reference frame. This has the technical benefit in that it might be easier to generate multiple copies of a single state by repeating the process used to generate the first instance rather than creating a specially engineered reference frame in a separate process. In case of atomic particles, atomic beamsplitters can then be used to perform joint multi-copy measurements of the kind we propose here

[135–137]. The scheme we present allows for the violation of multipartite nonlocality in all cases we have considered, and thus demonstrates that a degree of violation of local realism is possible when multiple copies are used. This was demonstrated for the bipartite case in [129].

However, when we consider the *genuinely* multipartite nonlocality we find no violation of the related Bell inequalities. This suggests that in the cases we considered the superselection rules act to weaken the nature of multipartite nonlocality. To gain a deeper understanding of this phenomenon, we present in section 5.3 a no-go result providing a necessary condition in terms of the number of particles required for a given number of parties in order to display genuinely multipartite nonlocality. The theorem explains the lack of genuinely multipartite nonlocality in the cases we considered. On the other hand, we show that there exist states that do not satisfy the condition and can be used to maximally violate any given genuinely or nongenuinely multipartite Bell inequality. We briefly discuss the behaviour of Bell inequalities under superselection rules in section 5.1, define the notation and present our scheme in section 5.2, followed by testing of the Bell inequalities on several examples of states. Finally we will prove and discuss the implications of the no-go theorem in section 5.3.

## 5.1 Multipartite Bell-type inequalities in the presence of SSR

We introduced both bipartite and multipartite Bell inequalities in section 3.3, where we considered several expressions that can be tested to give an indication of the violation of local realism. Those expressions can be considered in general as some function $f$ of the expectation values of the combinations of random variables,

$$|f(\boldsymbol{A}_1, \boldsymbol{B}_1, \boldsymbol{A}_2, \boldsymbol{B}_2 \ldots, \boldsymbol{A}_N, \boldsymbol{B}_N)| \leq L, \tag{5.1}$$

where $L$ is the value that needs to be exceeded in order for the violation of local realism to be demonstrated and $N$ is the number of parties[1].

[1]Note that although by far the most common, the Eq. (5.1) is not the most general possible form of a Bell inequality - a more general Bell inequality might for example allow more than two possible random variables for each party. The nonlocal games introduced in the previous chapter can also be thought of

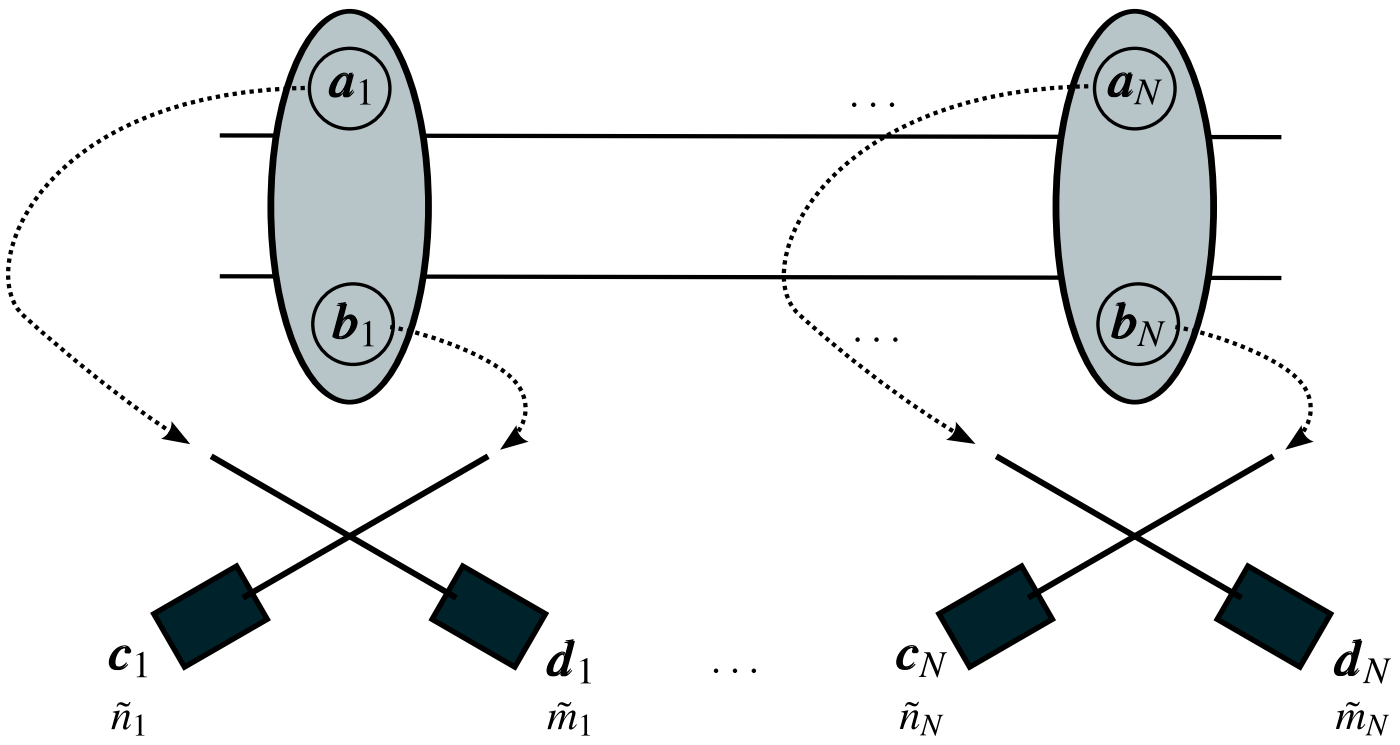


Figure 5.1: The modes from the two copies of the state, $\boldsymbol{a}_k$ and $\boldsymbol{b}_k$, are fed into a beamsplitter that outputs the modes denoted by $\boldsymbol{c}_k$ and $\boldsymbol{d}_k$, with respective particle numbers $\tilde{n}_k$ and $\tilde{m}_k$.

When the underlying quantum state is subject to particle number SSR the observables that generate the random variables are required to commute with the local number operator. This restricts the maximum possible value of the expression on the left hand side of Eq. (5.1) and raises the possibility that the SSR might degrade the observable nonlocality of the state. If despite all efforts a kind of nonlocality remains undemonstrated with SSR restricted states then the theory that such nonlocality is not present is a viable and unrefuted scientific theory. Here we investigate the predictions of quantum theory in regards to just such experiments. The inequalities of the type in Eq. (5.1) together with the stated restriction of observables form the foundation of our study.

## 5.2 Scheme introduction and results

In our scheme we shall focus on two copies of a three-party state, each with a single particle. The Fock basis at each of the modes shall be labelled as $|n_k\rangle$ for the first copy and $|m_k\rangle$ for the second copy, while the operators $\boldsymbol{a}_k$ and $\boldsymbol{b}_k$ are the corresponding annihilation operators. We allow the quantum observables $\boldsymbol{A}_k$ and $\boldsymbol{B}_k$ to act locally on all the copies simultaneously.

Each of the observers has a beamsplitter whose inputs are connected to $\boldsymbol{a}_k$ and $\boldsymbol{b}_k$

as a very general test of nonlocality, where infinitely many expressions are tested at once, essentially requiring infinitely many random variables for each party [30].

modes, as shown in figure 5.1. The output annihilation operators are denoted as $\boldsymbol{c}_k$ and $\boldsymbol{d}_k$, with the Fock basis on these modes being labelled by $|\tilde{n}_k\rangle$ and $|\tilde{m}_k\rangle$ respectively. Denoting with $\theta_k$ the mixing parameter between the two modes and $\phi_k$ their phase relationship, operators $\boldsymbol{c}_k$ and $\boldsymbol{d}_k$ can be written in terms of $\boldsymbol{a}_k$ and $\boldsymbol{b}_k$ as

$$\boldsymbol{c}_k = \cos(\theta_k)\boldsymbol{a}_k + \sin(\theta_k)e^{-i\phi_k}\boldsymbol{b}_k \tag{5.2}$$

$$\boldsymbol{d}_k = \sin(\theta_k)\boldsymbol{a}_k - \cos(\theta_k)e^{-i\phi_k}\boldsymbol{b}_k, \tag{5.3}$$

so that the state $|\tilde{n}_k, \tilde{m}_k\rangle$ can be written as

$$|\tilde{n}_k, \tilde{m}_k\rangle = \frac{\left(\boldsymbol{c}_k^\dagger\right)^{n_k}}{\sqrt{n_k!}} \times \frac{\left(\boldsymbol{d}_k^\dagger\right)^{m_k}}{\sqrt{m_k!}} |0_k, 0_k\rangle\,, \tag{5.4}$$

where $|0_k, 0_k\rangle$ is the vacuum state. The list of possible output states for two copies of single-particle input states can be seen in table 5.1. The states $|\tilde{n}_k, \tilde{m}_k\rangle$ implicitly depend on parameters $\theta_k$ and $\phi_k$. These parameters are properties of the beamsplitter and are used to switch between the different measurement settings for each observer. This could further be easily generalised to multiple copies of the state by replacing the beamsplitters with multiport devices as in [138], with their experimental implementation considered in [139].

Since the local bases have more than two possible orthogonal states, and the Bell inequalities we use rely on two measurement outcomes $\pm 1$, we need to decide which states will receive the outcome $+1$ and which $-1$. We do this by constructing the possible observables using a so called binning function, $\epsilon(\tilde{n}_k, \tilde{m}_k)$, taking values $\pm 1$. We can then write the possible observables as

$$\boldsymbol{O}_k\left(\phi_k, \theta_k\right) = \sum_{\tilde{n}_k+\tilde{m}_k=0}^{\tilde{n}_k+\tilde{m}_k=N} \epsilon(\tilde{n}_k, \tilde{m}_k)\, |\tilde{n}_k, \tilde{m}_k\rangle \langle \tilde{n}_k, \tilde{m}_k|\,. \tag{5.5}$$

We shall choose the binning function as

$$\epsilon(\tilde{n}_k, \tilde{m}_k) = (-1)^{\tilde{m}_k + \frac{1}{2}(\tilde{m}_k+\tilde{n}_k)(\tilde{m}_k+\tilde{n}_k+1)}, \tag{5.6}$$

because this choice was shown in [140] to lead to a tight and optimal test of nonlocality

| $\lvert\tilde{n}_k, \tilde{m}_k\rangle$ | Measurement in $\lvert n_k, m_k\rangle$ basis |
|---|---|
| $\lvert\tilde{0}_k, \tilde{0}_k\rangle$ | $\lvert 0_k, 0_k\rangle$ |
| $\lvert\tilde{1}_k, \tilde{0}_k\rangle$ | $\cos(\theta_k)\lvert 1_k, 0_k\rangle + \sin(\theta_k)e^{i\phi_k}\lvert 0_k, 1_k\rangle$ |
| $\lvert\tilde{0}_k, \tilde{1}_k\rangle$ | $\sin(\theta_k)\lvert 1_k, 0_k\rangle - \cos(\theta_k)e^{i\phi_k}\lvert 0_k, 1_k\rangle$ |
| $\lvert\tilde{2}_k, \tilde{0}_k\rangle$ | $\cos^2(\theta_k)\lvert 2_k, 0_k\rangle + \sqrt{2}\cos(\theta_k)\sin(\theta_k)e^{i\phi_k}\lvert 1_k, 1_k\rangle + e^{2i\phi_k}\sin^2(\theta_k)\lvert 0_k, 2_k\rangle$ |
| $\lvert\tilde{1}_k, \tilde{1}_k\rangle$ | $\sqrt{2}\cos(\theta_k)\sin(\theta_k)(\lvert 2_k, 0_k\rangle - \lvert 0_k, 2_k\rangle) - \cos(2\theta_k)e^{i\phi_k}\lvert 1_k, 1_k\rangle$ |
| $\lvert\tilde{0}_k, \tilde{2}_k\rangle$ | $\sin^2(\theta_k)\lvert 2_k, 0_k\rangle - \sqrt{2}\cos(\theta_k)\sin(\theta_k)e^{i\phi_k}\lvert 1_k, 1_k\rangle + \cos^2(\theta_k)e^{2i\phi_k}\lvert 0_k, 2_k\rangle$ |

Table 5.1: The table shows the local measurements on the single-particle state after the action of the beamsplitter, showing the dependence of states $\lvert\tilde{n}_k, \tilde{m}_k\rangle$ on $\theta_k$ and $\phi_k$.

in a bipartite case using the CHSH inequality. Because the BBGL inequality is an iterative construct using the CHSH inequality (see arguments in section 3.3.2, we conclude that the same binning function leads to a tight and optimal BBGL inequality, Eq. (3.49), in the genuinely multipartite case as well. Although we have not come across a similar proof for the nongenuinely multipartite inequalities, this choice will turn out to be sufficient in demonstrating a violation.

Since we assumed that each of the modes corresponding to $\boldsymbol{a}_k$ and $\boldsymbol{b}_k$ contain at most a single particle, we can simplify the table 5.1 by setting to zero those states that violate this assumption. This destroys the normalization of the states but preserves the outcome probabilities. Setting $\theta_k = \pi/4$ or $\theta_k = 3\pi/4$ for example can be used to set the entire $\lvert\tilde{1}_k, \tilde{1}_k\rangle$ state to zero. A similar phenomenon occurs for the states $\lvert\tilde{2}_k, \tilde{0}_k\rangle$ and $\lvert\tilde{0}_k, \tilde{2}_k\rangle$ when $\theta_k = 0, \pi/2, \pi$ etc.

Now that we have defined the notation and described our method, we shall apply it to several examples of states known to maximally violate multipartite Bell inequalities in the absence of SSR and observe to what extent, quantitatively and qualitatively, the violation is diminished.

| $N$ | Nongenuine | Bound |
|---|---|---|
| 2 | 2.41421 | 2 |
| 3 | 4.29929 | 4 |
| 4 | 8.32456 | 8 |
| 5 | 16.3915 | 16 |

Table 5.2: The table lists the values we obtained numerically for nongenuinely multipartite Bell inequalities with two copies of $N$-party W-states and the bound for $N$-partite inequalities. We include the case of $N = 2$ for comparison.

### 5.2.1 Numerical optimization for two copies of W-state

The $N$-party W-state is a symmetric single particle state, where the particle is equally likely to appear at any of the parties. It is defined by the equation

$$|W_N\rangle = \frac{1}{\sqrt{N}} \left( |100\ldots0\rangle + |0100\ldots0\rangle + \ldots + |00\ldots01\rangle \right). \tag{5.7}$$

When working in the particle-number basis the state becomes

$$|W\rangle = \frac{1}{\sqrt{N}} \sum_{k=1}^{N} \boldsymbol{a}_k^\dagger |0\rangle . \tag{5.8}$$

When $N = 3$ the W-state violates the genuinely multipartite BBGL inequality Eq. (3.49) as well as the nongenuinely multipartite MABK inequality of Eq. (3.52). When $N > 3$, the MABK inequalities of Eq. (3.51) are violated, but the BBGL inequalities are not.

We numerically maximized the inequalities over the values of $\phi_k, \theta_k$. For $N = 3$ and when enforcing SSR we find that the otherwise genuinely multipartite nonlocality is degraded to become nongenuinely multipartite (see table 5.2). The SSR thus impose a powerful restriction for this state. Although our scheme provides only for two copies, it is easily generalised to multiple copies. We found found similar results for multiple copies - genuinely multipartite nonlocality was degraded to nongenuinely multipartite.

### 5.2.2 Numerical optimization for two copies of Dicke states

Given a bitstring of length $N$ with $M$ 1s and $N-M$ 0s, a Dicke state is obtained by taking all permutations of the bits in the string and summing them in an equally weighted coherent superposition. Treating each bit in the string as a party and $0,1$ as respectively either absence or presence of an excitation, we can write the Dicke states as

$$|D_{N,M}\rangle = \frac{1}{N!}\binom{N}{M}^{1/2} \sum_{\boldsymbol{P}\in S_N} \boldsymbol{P}\Big[\bigotimes_{k=1}^{M} \boldsymbol{a}_k^\dagger\,|0\rangle\Big], \tag{5.9}$$

where $S_N$ is the group of permutations of $N$ elements and $\boldsymbol{P}$ denotes a particular permutation of $N$ elements. For $M=1$ the Dicke states reduce to W-states and are symmetric under the permutation of parties for all $M$. They were originally considered in [141] as eigenstates of the Hamiltonian describing a gas of bosonic particles confined inside a small box.

The Dicke states are numerically found to lead to the greatest violation of the BBGL inequality when half of the modes contain a particle and the other half do not for even $M$ and in general when $M=\lfloor N/2\rfloor$. Here $\lfloor x\rfloor$ is the floor function[2]. This was tested up to $N=6$. We therefore proceeded to test these states with the SSR in place. As can be seen in table 5.3, the otherwise genuinely multipartite nonlocality was again consistently degraded to the nongenuinely multipartite nonlocality when SSR were put in place.

We have thus tested two families of states, both of which demonstrated the degradation of genuinely multipartite nonlocality. We therefore asked the question whether this phenomenon is simply a consequence of the choice of the observables or the choice of using the BBGL inequality, or whether there is a more general principle that rules out the violation. We will see in the next section that there exists a general no-go theorem that rules out genuinely multipartite nonlocality for *any* choice of observables and for *any* test of genuinely multipartite nonlocality for a range of states covering the examples we considered.

---

[2]The floor function $\lfloor x\rfloor$ returns the greatest integer $n \leq x$.

| $N$ | Genuine | Nongenuine | Bound |
|---|---|---|---|
| 3 | 4 | 4.29929 | 4 |
| 4 | 4.64821 | 8.38189 | 8 |
| 5 | 8.32 | 16.5305 | 16 |

Table 5.3: Bell inequality values for Dicke states.

## 5.3 No-go theorem for SSR restricted genuinely multipartite nonlocality

Before we delve into the proof and statement of the theorem we provide a short exposition of the genuinely multipartite entanglement. A pure multipartite state of $N$ parties is called $k$-separable if and only if it can be expressed as a product of $k$ states [142, 143]

$$|\psi\rangle = |\psi_1\rangle \otimes \ldots \otimes |\psi_k\rangle. \tag{5.10}$$

Some of the states $|\psi_j\rangle$ here span multiple parties and none of the states in the product may span all of the parties for the state to be $k$-separable. Similarly to the bipartite case, a mixed multipartite state is called $k$-separable if it can be written as a convex sum of pure states, each of which is at least $k$-separable

$$\rho = \sum_j p_j |\psi_1^j\rangle\langle\psi_1^j| \otimes \ldots \otimes |\psi_k^j\rangle\langle\psi_k^j|, \tag{5.11}$$

and where $p_j$ is some probability distribution. An arbitrary state is then called *genuinely multipartite entangled* if it is not $k$ separable for any $k \geq 2$. The tensor product need not be over the same spaces in each of the terms in Eq. (5.11).

Comparing the Eq. (3.44) with Eq. (5.11), we deduce that $k$-separable states cannot be genuinely multipartite nonlocal. Conversely, genuinely multipartite entanglement is a necessary condition for genuinely multipartite nonlocality. We will use this by demonstrating that when the number of particles is smaller than the number of parties the *effective* state, defined below, is at least biseparable (i.e. $k$-separable for $k = 2$).

Similar to our treatment of effective entanglement in the previous chapter, we

define the effective state as

$$\tilde{\rho} = \sum_{n_1+\ldots+n_N=M} \mathbf{\Pi}_{n_1,\ldots,n_N}\, \rho\, \mathbf{\Pi}_{n_1,\ldots,n_N}, \tag{5.12}$$

where $\Pi_{n_1,\ldots,n_M} = \bigotimes_{j=1}^{M} |n_j\rangle\langle n_j|$ and the sum is over all partitions of integer $M$ into sums of $N$ integers. The state $\tilde{\rho}$ exhibits precisely the same outcome statistics for all permitted measurements as the state $\rho$. Since Bell inequalities are dependent only on outcome probabilities and not, for example, on the post-measurement state, we may replace the state $\rho$ with the state $\tilde{\rho}$ whenever SSR are enforced. This insight leads us to the main result of this chapter.

**Theorem 5.1.** *Whenever the SSR are enforced and the number of particles $M$ is strictly lower than the number of parties $N$, effective genuinely multipartite entanglement vanishes.*

*Proof.* We first show the theorem for a pure state $|\phi\rangle$. In general we may write such a state as

$$|\phi\rangle = \sum_{n_1+\ldots+n_N=M} \phi_{n_1,\ldots,n_N} |n_1,\ldots,n_N\rangle, \tag{5.13}$$

where we assumed that each party has only a single mode. As we will see below, there is no loss of generality associated with making this assumption.

According to Eq. (5.12), the effective state corresponding to $|\phi\rangle\langle\phi|$ is given by

$$\tilde{\rho} = \sum_{n_1+\ldots+n_M=N} \mathbf{\Pi}_{n_1,\ldots,n_M} |\phi\rangle\langle\phi|\, \mathbf{\Pi}_{n_1,\ldots,n_M}. \tag{5.14}$$

Now since $M = \sum_{j=1}^{N} n_j < N$ it must be that in each term of Eq. (5.14) at least one of $n_j$ vanishes. Therefore each of the terms can, up to a permutation, be written as $|0\rangle \otimes |\tilde{\phi}_j\rangle$ and therefore $\tilde{\rho}$ is at least biseparable.

For mixed states we have that $\rho = \sum_j \mu_j |\phi_j\rangle\langle\phi_j|$ and let us now denote with $\tilde{\rho}$ the effective state corresponding to $\rho$. Writing as $\tilde{\sigma}_j$ the effective state of $|\phi_j\rangle\langle\phi_j|$, we may write $\tilde{\rho} = \sum_j \mu_j \tilde{\sigma}_j$. But according to the argument for pure states, each of $\tilde{\sigma}_j$ is a convex combination of biseparable states and therefore the same must be true for $\tilde{\rho}$. This demonstrates the theorem for mixed states.

For the case of multiple modes per party, note that in demonstrating biseparability we used only the fact that at least one of the parties must necessarily lack a particle. This fact remains true when each party possesses multiple modes and thus the theo-

rem remains valid for an arbitrary number of modes per party. This assumption was only made in order to simplify the notation. □

As a consequence of the theorem, the state $\tilde{\rho}$ cannot violate genuinely multipartite Bell inequalities whenever $M < N$ and therefore neither can $\rho$. This implies that we need at least as many particles as we have parties in order to violate genuinely multipartite Bell inequalities, and in the case of $N = 2$ we need at least two particles to violate any Bell inequality at all. In the states we considered in the previous section we had only two particles and at least three parties. We therefore conclude that nonviolation of genuinely multipartite nonlocality was not an accident of our choice of state, observables or the exact inequality used - we would have reached the same conclusion with *any* genuinely multipartite test of nonlocality and any state, so long as the state satisfies the $M < N$ condition. Insufficient number of particles therefore always degrades the nature of nonlocality from genuinely multipartite to nongenuinely multipartite.

To see that it is necessary to make the assumption that $M < N$ in showing the theorem, we show that for any $N$ it is possible to construct a state with $M \geq N$ that maximally violates a given genuinely multipartite Bell inequality. Thus suppose the $N$-qubit state

$$|\chi\rangle = \sum_{u_1,u_2,\dots u_N} \chi_{u_1,u_2,\dots,u_N} |u_1, u_2, \dots, u_N\rangle \,, \tag{5.15}$$

where $u_1, \dots u_N$ take values in $\{0, 1\}$, violates the given genuinely multipartite Bell inequality maximally in the absence of SSR. Then we can map the state to the second quantization picture through the so called dual rail encoding, where $|0_k\rangle \mapsto \boldsymbol{a}_k^\dagger |\text{vac}_k\rangle$ and $|1_k\rangle \mapsto \boldsymbol{b}_k^\dagger |\text{vac}_k\rangle$, where $|\text{vac}_k\rangle$ is the vacuum state. This generates a local basis of equal particle-number states $|01\rangle, |10\rangle$, whose superpositions are not restricted by the SSR and therefore allowing maximum violations. We can extend this to higher dimensional states by adding an additional local mode for each additional dimension. There is, however, an important caveat. We assume here that the particles corresponding to different modes are of the same mass but are distinguishable by some other attribute such as their local momentum, location, angular momentum or some other internal state.

As an example, consider the tripartite GHZ state

$$|GHZ\rangle = \frac{1}{\sqrt{2}}(|000\rangle + |111\rangle). \tag{5.16}$$

This state maximally violates the tripartite BBGL inequality, attaining the value $4\sqrt{2}$. The dual rail encoding maps this state to

$$|GHZ\rangle \mapsto \frac{1}{\sqrt{2}}(|10,10,10\rangle + |01,01,01\rangle). \tag{5.17}$$

The transformation in Eq. (5.12) leaves the state in Eq. (5.17) invariant thus making it as powerful even in the presence of the SSR as the usual GHZ state.

In the absence of SSR, however, even a single particle can be genuinely multipartite nonlocal, as has been demonstrated experimentally in [144]. The no-go theorem helps us to understand the conditions we must fulfill in the presence of SSR in order to take full advantage of the quantum phenomena, whether our ultimate goal is to construct new quantum technology or to understand the fundamental behaviour of nature. We have studied such operational restrictions in this and the previous chapter, where we examined their impact on quantum information theoretic resources as functions of the underlying state. In the next chapter we will turn to evaluating the quantumness of operations themselves.

# CHAPTER 6

# Quantifying the nonclassicality of operations

We saw in the previous chapters that having access to an entangled or nonlocal quantum state but without the ability to operationally exploit that quantumness can prevent us from tapping into the quantum resources that are otherwise available. Equally limiting is having access to the necessary quantum operations but without the needed quantum resources present in the state. In order to attain better than classical performance, we thus require a degree of quantumness in both the states as well as the operations.

However, while the quantum resources in states are relatively well understood, quantumness as a resource present inherently in operations has so far received little attention. It has been investigated which quantum operations can produce entanglement from separable states, which led to the definition of the entangling power [145]. It has long been known that LOCC operations cannot create entanglement [10–12]. On the other hand even local operations without any classical communication are able to generate quantum discord [16, 146–148], which is possible if and only if the operation changes the local algebraic structure [149]. Several authors have also studied the evolution of quantum discord under the action of quantum operations [150–154].

Here we shall investigate the properties of quantum operations that endow them with the ability to exploit the nonclassical resources, which will lead us to the first in-

formation theoretic quantification of quantumness of an operation. Particularly important to this task is the way the classical world emerges from the quantum world through the process of einselection or decoherence (see section 2.2). While quantum world treats all superpositions of states on an equal footing, classical world is not quite so egalitarian. The sheer size of classical objects and the associated difficulty of isolating them sufficiently from the environment quickly leads to decoherence, therefore ensuring that they are constantly effectively measured by the environment. This process results in a preferred basis of states known as the *pointer states* [80] and denoted here by $|\alpha\rangle\langle\alpha| = \mathbf{\Pi}_\alpha$. The einselection operator $\mathbf{\Gamma}$ can then be written as $\mathbf{\Gamma}(\rho) = \sum_\alpha \mathbf{\Pi}_\alpha \,\rho\, \mathbf{\Pi}_\alpha$. The einselection operation sets the off-diagonal elements in the $|\alpha\rangle$ basis to zero, transforming the quantum state $\rho$ into a classical probability distribution. Classical states $\rho_c$ are precisely those states which are not affected by the einselection - i.e. the fixed points of $\mathbf{\Gamma}$, so that $\mathbf{\Gamma}(\rho_c) = \rho_c$. This paradigm explains the abundance of classical states in the universe by establishing a kind of quantum selection of states in that only those states which are unaffected by decoherence survive intact, while the rest are adapted to become more classical [38]. For a more detailed discussion of einselection see also [80].

Suppose Alice, an observer, would like to perform some operation on her quantum state tapping into its contained coherence. She must then be able to act fast enough so that the state does not decohere before she has acted and then again after she the operation has completed she must also be quick to measure the state before it decoheres. Alternatively, she may isolate her state very well from the environment, giving her more time. In this way she ensures that decoherence occurs neither before nor after she has acted. In contrast, if the decoherence is allowed to act both before and after, her operation $\$$ effectively takes the form $\$ = \mathbf{\Gamma} \circ \$ \circ \mathbf{\Gamma}$, where we use $\circ$ to denote the composition of operations. Because the Kraus operators of $\mathbf{\Gamma}$ are projectors we have that $\mathbf{\Gamma}^2 = \mathbf{\Gamma}$, which in turn implies the commutation relation $\$ \circ \mathbf{\Gamma} = \mathbf{\Gamma} \circ \$$.

While the above discussion applies only when decoherence acted both before and after the operation, we will show here that an observer equipped only with operations that satisfy the commutation relation $[\$, \mathbf{\Gamma}] = 0$ is unable to (i) generating quantum coherence between pointer states and (ii) distinguishing quantum states that posses coherence from those that do not.Such observer is thus precluded from taking advantage of quantum resources, and becomes what we term a *classical observer*. We further

show that possessing *any one* of these two abilities leads to the violation of the commutation relation. To prove these results we make use of the relative entropy (see chapter 3 for definition and properties), providing us with a quantification of quantumness of an operation based on how distinguishable are the two orderings of operations in the commutator $[\mathbf{\$}, \mathbf{\Gamma}]$. In the case of two parties spread out nonlocally, we shall consider $\mathbf{\Gamma}$ to be of the form $\mathbf{\Gamma}_1 \otimes \mathbf{\Gamma}_2$ or, when we are interested in only partially measuring nonclassicality, we allow one of the $\mathbf{\Gamma}_k$ to be replaced by the identity operation $\mathbf{1}$. This is particularly useful when we consider parallels with quantum discord, where the commutation relation satisfied by classical operation will lead us to an operational interpretation of quantum discord. Although not considered here, these concepts may also be extended to multiple parties.

## 6.1 Quantumness of an operation

We define the *quantumness of an operation* $\mathbf{\$}$ as

$$W_{\Gamma}(\mathbf{\$}) = \sup_{\rho} S\left(\mathbf{\$} \circ \mathbf{\Gamma}(\rho) \| \mathbf{\Gamma} \circ \mathbf{\$}(\rho)\right), \tag{6.1}$$

where the supremum is taken over all quantum states, but as we show later it is sufficient to maximize over pure states only. The quantity $W_{\Gamma}$ vanishes if and only if $[\mathbf{\$}, \mathbf{\Gamma}] = 0$ for all input states and measures the deviation of the commutator $[\mathbf{\$}, \mathbf{\Gamma}]$ from zero by applying both orderings to the same state and comparing the outputs[1]. While the definition of $W_{\Gamma}$ implicitly depends on $\mathbf{\Gamma}$, we will subsequently suppress the subscript in our notation when no confusion is possible. The result that an observer is classical if and only if all the available operations satisfy $W(\mathbf{\$}) = 0$ is a consequence of the following theorem, the central result of this chapter.

**Theorem 6.1.** *The quantumness of an operation $W(\mathbf{\$})$ is the sum of two independent contri-*

[1]Indeed, we can treat the maximum discrepancy of the outputs over all input states as a measure of distinguishability of operations. This relative entropy for quantum operations can be used to define more conventional properties of channels such as entropy, mutual information, conditional entropy and so on, using the same approach as we used in chapter 3.

*butions*

$$W(\$) = \sup_\rho \Big( S\big(\mathbf{\Gamma} \circ \$ \circ \mathbf{\Gamma}(\rho) \| \mathbf{\Gamma} \circ \$(\rho)\big) + S\big(\$ \circ \mathbf{\Gamma}(\rho) \| \mathbf{\Gamma} \circ \$ \circ \mathbf{\Gamma}(\rho)\big) \Big), \tag{6.2}$$

*where the supremum is over all quantum states.*

*Proof.* We start with the relative entropy featuring under maximization in $W(\$)$. Then we insert the sum of a complete set of orthonormal projectors, the Kraus operators of $\mathbf{\Gamma}$, $\sum_\alpha \mathbf{\Pi}_\alpha = \mathbf{1}$. We thus obtain

$$\begin{aligned} S\big(\$ \circ \mathbf{\Gamma}(\rho) \| \mathbf{\Gamma} \circ \$(\rho)\big) &= -S\big(\$ \circ \mathbf{\Gamma}(\rho)\big) - \mathrm{Tr}\big[\$ \circ \mathbf{\Gamma}(\rho) \log(\mathbf{\Gamma} \circ \$(\rho))\big] \\ &= -S\big(\$ \circ \mathbf{\Gamma}(\rho)\big) - \mathrm{Tr}\big[\sum_\alpha \mathbf{\Pi}_\alpha (\$ \circ \mathbf{\Gamma})(\rho) \log(\mathbf{\Gamma} \circ \$(\rho))\big]. \end{aligned} \tag{6.3}$$

where $S(\rho) = -\mathrm{Tr}\,[\rho \log(\rho)]$ is the von Neumann entropy. Next we use the fact the projective property $(\sum_\alpha \mathbf{\Pi}_\alpha)^2 = \sum_\alpha \mathbf{\Pi}_\alpha$ together with the cyclic property of the trace and the fact that $\mathbf{\Pi}_\alpha$ commutes with $\mathbf{\Gamma} \circ \$(\rho)$ and thus also with its logarithm. The above is then transformed to

$$\begin{aligned} &S\big(\$ \circ \mathbf{\Gamma}(\rho) \| \mathbf{\Gamma} \circ \$(\rho)\big) \\ &\quad = -S\big(\$ \circ \mathbf{\Gamma}(\rho)\big) - \mathrm{Tr}\big[\sum_\alpha \mathbf{\Pi}_\alpha (\$ \circ \mathbf{\Gamma}(\rho)) \mathbf{\Pi}_\alpha \log(\mathbf{\Gamma} \circ \$(\rho))\big] \\ &\quad = -S\big(\$ \circ \mathbf{\Gamma}(\rho)\big) - \mathrm{Tr}\big[\mathbf{\Gamma} \circ \$ \circ \mathbf{\Gamma}(\rho) \log(\mathbf{\Gamma} \circ \$(\rho))\big]. \end{aligned} \tag{6.4}$$

Next we add and subtract $S\big(\mathbf{\Gamma} \circ \$ \circ \mathbf{\Gamma}(\rho)\big)$ to the righthand side to get to

$$\begin{aligned} S\big(\$ \circ \mathbf{\Gamma}(\rho) \| \mathbf{\Gamma} \circ \$(\rho)\big) = S\big(\mathbf{\Gamma} \circ \$ \circ \mathbf{\Gamma}(\rho)\big) - S\big(\$ \circ \mathbf{\Gamma}(\rho)\big) \\ + S\big(\mathbf{\Gamma} \circ \$ \circ \mathbf{\Gamma}(\rho) \| \mathbf{\Gamma} \circ \$(\rho)\big). \end{aligned} \tag{6.5}$$

We now expand the entropy $S\big(\mathbf{\Gamma} \circ \$ \circ \mathbf{\Gamma}(\rho)\big)$ to give

$$\begin{aligned} S\big(\$ \circ \mathbf{\Gamma}(\rho) \| \mathbf{\Gamma} \circ \$(\rho)\big) = -\mathrm{Tr}[\mathbf{\Gamma} \circ \$ \circ \mathbf{\Gamma}(\rho) \log(\mathbf{\Gamma} \circ \$ \circ \mathbf{\Gamma}(\rho))] \\ - S\big(\$ \circ \mathbf{\Gamma}(\rho)\big) + S\big(\mathbf{\Gamma} \circ \$ \circ \mathbf{\Gamma}(\rho) \| \mathbf{\Gamma} \circ \$(\rho)\big). \end{aligned} \tag{6.6}$$

Now we expand $\mathbf{\Gamma}$ and insert the orthogonal projective operators $\mathbf{\Pi}_\alpha$ yielding

$$S\big(\$\circ\mathbf{\Gamma}(\rho)\|\mathbf{\Gamma}\circ\$(\rho)\big) = -\operatorname{Tr}[\sum_\alpha \mathbf{\Pi}_\alpha \$\circ\mathbf{\Gamma}(\rho)\mathbf{\Pi}_\alpha \log(\mathbf{\Gamma}\circ\$\circ\mathbf{\Gamma}(\rho))] \\ - S\big(\$\circ\mathbf{\Gamma}(\rho)\big) + S\big(\mathbf{\Gamma}\circ\$\circ\mathbf{\Gamma}(\rho)\|\mathbf{\Gamma}\circ\$(\rho)\big) \quad (6.7)$$

and because $\mathbf{\Pi}_\alpha$ commutes with $\mathbf{\Gamma}\circ\$\circ\mathbf{\Gamma}$, we find that

$$S\big(\$\circ\mathbf{\Gamma}(\rho)\|\mathbf{\Gamma}\circ\$(\rho)\big) = -S\big(\$\circ\mathbf{\Gamma}(\rho)\big) \\ - \operatorname{Tr}[\sum_\alpha \mathbf{\Pi}_\alpha \$\circ\mathbf{\Gamma}(\rho)\log(\mathbf{\Gamma}\circ\$\circ\mathbf{\Gamma}(\rho))] + S\big(\mathbf{\Gamma}\circ\$\circ\mathbf{\Gamma}(\rho)\|\mathbf{\Gamma}\circ\$(\rho)\big). \quad (6.8)$$

The first two terms then form another relative entropy, leading us to

$$S\big(\$\circ\mathbf{\Gamma}(\rho)\|\mathbf{\Gamma}\circ\$(\rho)\big) = S\big(\mathbf{\Gamma}\circ\$\circ\mathbf{\Gamma}(\rho)\|\mathbf{\Gamma}\circ\$(\rho)\big) \\ + S\big(\$\circ\mathbf{\Gamma}(\rho)\|\mathbf{\Gamma}\circ\$\circ\mathbf{\Gamma}(\rho)\big), \quad (6.9)$$

which after inserting the supremum over $\rho$ is the result of this theorem. □

The first term, $S\big(\mathbf{\Gamma}\circ\$\circ\mathbf{\Gamma}(\rho)\|\mathbf{\Gamma}\circ\$(\rho)\big)$, is non-zero if and only if a classical observer equipped with operation \$ is able to use \$ to distinguish between $\rho$ and its classical counterpart $\mathbf{\Gamma}(\rho)$. Therefore we call this term the *distinguishing power*. The second term, $S\big(\$\circ\mathbf{\Gamma}(\rho)\|\mathbf{\Gamma}\circ\$\circ\mathbf{\Gamma}(\rho)\big)$, is non-zero if and only if the operation \$ has turned a classical state $\mathbf{\Gamma}(\rho)$ into a non-classical state. Therefore we call this term the *generating power* (see figure 6.1). As a consequence, $W(\$) = 0$ if and only if both the generating power and the distinguishing power vanish for all states. A classical observer can therefore elevate themselves to the status of a quantum observer if and only if they posses some operation for which $W(\$) > 0$.

Crucially, the distinguishing and generating powers in Eq. (6.2) can be independently zero, which can be shown by construction. Given a quantum operation \$ where both these quantities are non-vanishing, we construct an operation $\mathbf{\Gamma}\circ\$$ for which the generating power vanishes but the distinguishing power is unchanged. Thus, $W(\mathbf{\Gamma}\circ\$)$ gives the maximum distinguishing power of \$. On the other hand, for the operation $\$\circ\mathbf{\Gamma}$, the distinguishing power vanishes while the generating power is unchanged and therefore $W(\$\circ\mathbf{\Gamma})$ is the maximum generating power of \$.

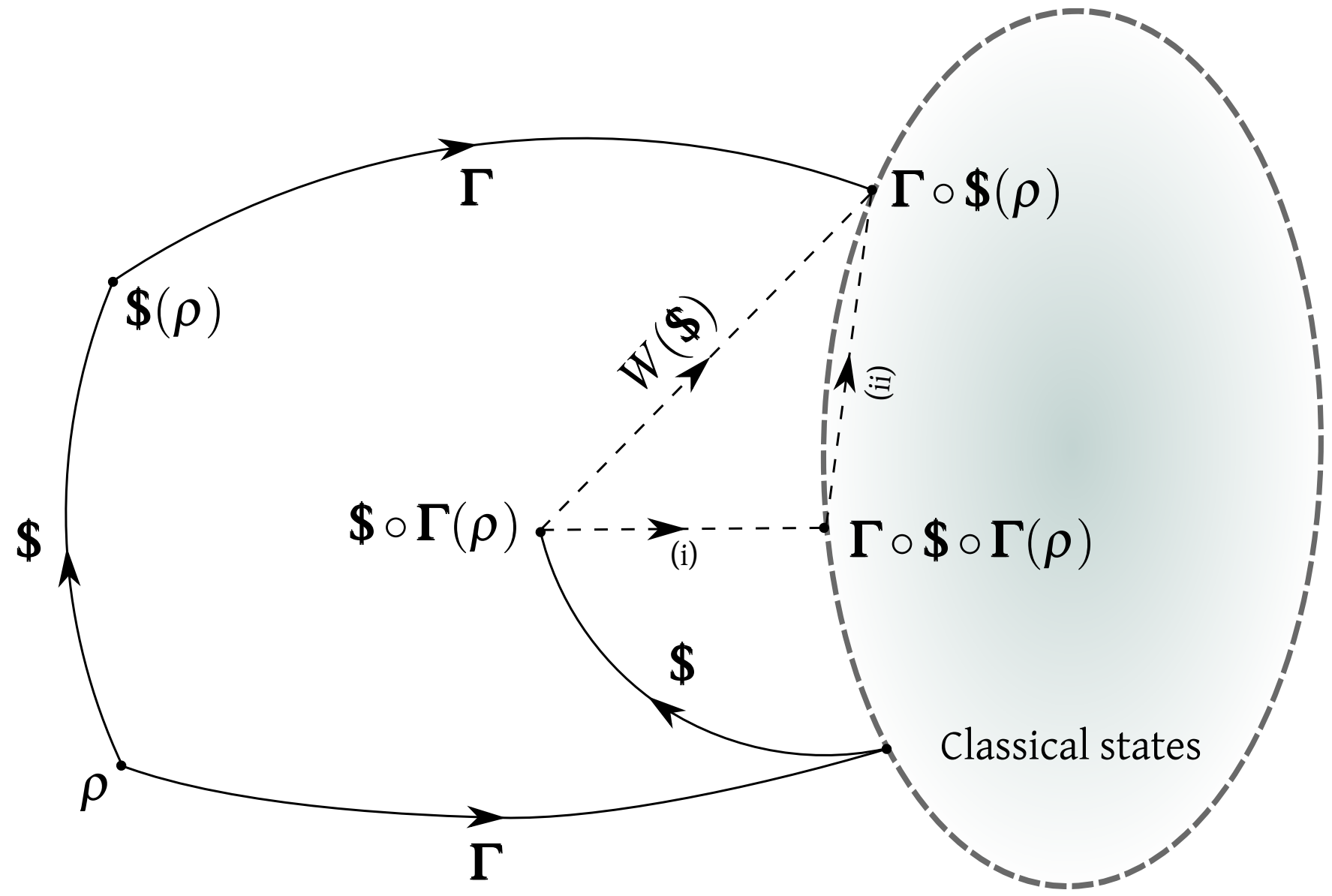

Figure 6.1: Illustration of Thm. (6.1). The dashed lines represent relative entropies corresponding to the terms from Thm. 6.1, while the solid lines represent operations. The two paths from the state $\$ \circ \mathbf{\Gamma}(\rho)$ to $\mathbf{\Gamma} \circ \$(\rho)$ are equidistant in relative entropy. Quantities (i) and (ii) represent the generating and the distinguishing power, respectively. We consider classical states to be the fixed points of the linear einselection operator $\mathbf{\Gamma}$ and as such is a simplex. This set is smaller than the set of separable [10, 11] and zero-discord states [77]. Note that our notion of classicality is stricter than that enforced by quantum discord since there is no freedom to choose the classical basis.

There are instances where both terms play essential and independent roles in a quantum protocol. As an example, consider the BB84 quantum cryptography [9]. In order to engage in the protocol, Alice must be able to prepare states in two non-orthogonal bases, which requires only the power to generate non-classical states, implying non-vanishing generating power. Bob, on the other hand, needs to be able to distinguish between classical and non-classical states in order to extract the key and detect the presence of an eavesdropper, thus requiring an operation with non-zero distinguishing power.

P1. *Extremality*. Maximum in the supremum is attained with a pure state. This similarly follows from the joint convexity of relative entropy.

P2. *Monotonicity*. Given a general operation $\$$ and a classical operation $\$_c$, then

$W(\$_c \circ \$) \leq W(\$)$ and $W(\$ \circ \$_c) \leq W(\$)$ holds, showing that the measure is non-increasing under composition.

P3. *Convexity*. The convexity follows from the joint convexity of relative entropy. Thus, given two observers with classical maps $\$_i^A$, $\$_i^B$ at their disposal, and shared source of randomness, they cannot create a nonclassical operator. In other words, if $W(\$_i^A \otimes \mathbf{1}) = 0$ and $W(\mathbf{1} \otimes \$_i^B) = 0$ then $W(\sum_i p_i \$_i^A \otimes \$_i^B) = 0$.

The proofs of the above properties are given in the following theorems. First we give the proof of property P1 that the maximum in the maximization for $W$ is always attained for a pure state.

**Theorem 6.2.** *Given* $\sup_\rho S\left(\$ \circ \mathbf{\Gamma}(\rho) \| \mathbf{\Gamma} \circ \$(\rho)\right)$*, there exists a pure state* $|\psi\rangle\langle\psi|$ *such that the supremum in Eq. (6.1) is attained when* $\rho = |\psi\rangle\langle\psi|$.

*Proof.* Imagine that we have performed maximization over only the set of pure states and found that the maximum is attained for $|\psi\rangle$. Then for some mixed state $\rho$ we can spectrally decompose it as $\rho = \sum_j \mu_j |\phi_j\rangle\langle\phi_j|$, where $|\phi_j\rangle$ are it eigenstates. Since the relative entropy is jointly convex in its arguments, Thm. (3.5), this implies that

$$\begin{aligned} S\left(\$ \circ \mathbf{\Gamma}(\rho) \| \mathbf{\Gamma} \circ \$(\rho)\right) &\leq \sum_j \mu_j S\left(\$ \circ \mathbf{\Gamma}(|\phi_j\rangle\langle\phi_j|) \| \mathbf{\Gamma} \circ \$(|\phi_j\rangle\langle\phi_j|)\right) \\ &\leq S\left(\$ \circ \mathbf{\Gamma}(|\psi\rangle\langle\psi|) \| \mathbf{\Gamma} \circ \$(|\psi\rangle\langle\psi|)\right). \end{aligned} \tag{6.10}$$

This completes the proof. □

Next we will consider the property P2, stating that the measure $W$ is non-increasing under the composition with classical maps.

**Theorem 6.3.** *If* $\$$ *is some map and* $W(\$_c) = 0$ *then* $W(\$_c \circ \$) \leq W(\$)$ *and* $W(\$ \circ \$_c) \leq W(\$)$.

*Proof.* Notice that

$$\begin{aligned} W(\$_c \circ \$) &= \sup_\rho S\left(\$_c \circ \$ \circ \mathbf{\Gamma}(\rho) \| \$_c \circ \mathbf{\Gamma} \circ \$(\rho)\right) \\ &\leq \sup_\rho S\left(\$ \circ \mathbf{\Gamma}(\rho) \| \mathbf{\Gamma} \circ \$(\rho)\right) = W(\$), \end{aligned} \tag{6.11}$$

where the last inequality is due to the monotonicity of relative entropy under completely positive operations (and thus also the strong subadditivity of the von Neumann entropy, which is equivalent to monotonicity [55]). For the reverse order

$$W(\$ \circ \$_c) = \sup_\rho S\left(\$ \circ \mathbf{\Gamma} \circ \$_c(\rho) \| \mathbf{\Gamma} \circ \$ \circ \$_c(\rho)\right) = \sup_{\$_c(\rho)} S\left(\$ \circ \mathbf{\Gamma}(\rho) \| \mathbf{\Gamma} \circ \$(\rho)\right)$$
$$\leq \sup_\rho S\left(\$ \circ \mathbf{\Gamma}(\rho) \| \mathbf{\Gamma} \circ \$(\rho)\right) = W(\$), \quad (6.12)$$

where going from second to the third line we changed the set over which we take supremum from all states to the set of states of the form $\$(\rho)$. Since this set is entirely contained in the set of all states, the inequality follows. □

Finally, the property P3 is proved in the following theorem.

**Theorem 6.4.** *Given a set of local operations $\$_w^A, \$_w^B$ such that $W(\$_w^A \otimes \mathbf{1}) = 0$ and $W(\mathbf{1} \otimes \$_w^B) = 0$ then $W(\$) = 0$ for any local operation with shared randomness of the form $\$ = \sum_w \gamma_w \$_w^A \otimes \$_w^B$.*

*Proof.* Notice that since we required that $\mathbf{\Gamma}$ be composed of local orthonormal projectors we can write it as $\mathbf{\Gamma} = \mathbf{\Gamma}^A \otimes \mathbf{\Gamma}^B$ in the bipartite case. Given that $\$_w^A$ and $\$_w^B$ commute with $\mathbf{\Gamma}^A$ and $\mathbf{\Gamma}^B$, respectively, we also have that $\$$ commutes with $\mathbf{\Gamma}$, establishing the result. □

Particularly useful is property P1, which can significantly simplify both numerical and analytical calculations of the measure. Properties P2 and P3, on the other hand, are something one would naturally expect a measure of quantumness of operations to satisfy. Property P2 tells us that combining an operation with a classical operation cannot increase its quantumness. However, we will see in the next section an example where sandwiching a classical operation in between two quantum operations can act as a catalyst. Property P3 tells us that randomly choosing a particular classical operation out of a repertoire of classical operations cannot be used to generate a nonclassical operation. We next turn our attention to examples, where a few additional interesting features are revealed.

## 6.2 Examples

We shall focus on qubits with a classical basis as $|0\rangle, |1\rangle$ and $\mathbf{\Gamma}$ implementing two-sided einselection. Firstly, we shall look at unitary operations in general, where we find an

interesting dichotomy.

**Theorem 6.5.** *When $\mathbf{\Gamma}$ acts on the entire joint Hilbert space, selecting a complete orthonormal classical basis $|k\rangle$, we have for any unitary operation $\boldsymbol{U}$ that $W(\boldsymbol{U}) = 0$ if and only if $\boldsymbol{U} = \sum_k e^{i\phi_k} |k\rangle \langle k| \boldsymbol{P}$, where $\boldsymbol{P}$ is a permutation of the classical basis and $\phi_k$ are phases. Otherwise $W(\boldsymbol{U}) = \infty$.*

*Proof.* We proceed by computing the relative entropy $S\big(\boldsymbol{U} \circ \mathbf{\Gamma}(\rho) \| \mathbf{\Gamma} \circ \boldsymbol{U}(\rho)\big)$. First we show that if $\boldsymbol{U}$ is not of the required form, then $W(\boldsymbol{U}) = \infty$. Under such assumption, there exists a non-classical state $|\phi_0\rangle$ such that $\boldsymbol{U} |\phi_0\rangle = |j\rangle$, where $|j\rangle$ is any classical basis state. The relative entropy $S(\rho\|\sigma)$ is infinite due to the term $\mathrm{Tr}[\rho \log(\sigma)]$ when the kernel of $\sigma$ has a non-zero overlap with the support of $\rho$. So suppose $|\phi_0\rangle = \sum_k \alpha_k |k\rangle$ is the expansion of $|\phi_0\rangle$ in the classical basis. Then

$$\mathbf{\Gamma}(|\phi_0\rangle \langle\phi_0|) = \sum_k |\alpha_k|^2 |k\rangle \langle k| \neq |\phi_0\rangle \langle\phi_0| , \tag{6.13}$$

since $|\phi_0\rangle$ is not classical by assumption. Thus, $\boldsymbol{U}\mathbf{\Gamma}(|\phi_0\rangle \langle\phi_0|)\boldsymbol{U}^\dagger$ will in general have support across numerous classical states besides $|j\rangle \langle j|$. However, the second argument of the relative entropy is $\mathbf{\Gamma}(\boldsymbol{U} |\phi_0\rangle \langle\phi_0| \boldsymbol{U}^\dagger) = |j\rangle \langle j|$ and thus is a state with a kernel overlapping with the support of $\boldsymbol{U}\mathbf{\Gamma}(|\phi_0\rangle \langle\phi_0|)\boldsymbol{U}^\dagger$, making the relative entropy infinite.

Secondly, we show that if $W(\boldsymbol{U}) = \infty$, then some classical state $|k\rangle$ is mapped to a non-classical state. Since the generating power is the einselected relative entropy of discord of the output state, we know that it must be bounded by $\log(d)$, where $d$ is the dimension of the joint Hilbert space. Therefore, if $W(\boldsymbol{U}) = \infty$, the distinguishing power is infinite. There exists a state $|\psi\rangle$ such that

$$S\big(\mathbf{\Gamma}(U\mathbf{\Gamma}(|\psi\rangle \langle\psi|)\boldsymbol{U}^\dagger) \| \mathbf{\Gamma}(U |\psi\rangle \langle\psi| \boldsymbol{U}^\dagger)\big) = \infty. \tag{6.14}$$

Now $|\psi\rangle$ cannot be classical, otherwise the above would vanish. So let $|\psi\rangle = \sum_k \gamma_k |k\rangle$ and $\mathbf{\Gamma}(|\psi\rangle \langle\psi|) = \sum_k |\gamma_k|^2 |k\rangle \langle k|$. Label $\boldsymbol{U} |k\rangle = |\psi_k\rangle$. Then we have the following

$$\mathbf{\Gamma}(\boldsymbol{U}\mathbf{\Gamma}(|\psi\rangle \langle\psi|)\boldsymbol{U}^\dagger) = \sum_k |\gamma_k|^2 \mathbf{\Gamma}(|\psi_k\rangle \langle\psi_k|), \tag{6.15}$$

$$\mathbf{\Gamma}(\boldsymbol{U} |\psi\rangle \langle\psi| \boldsymbol{U}^\dagger) = \sum_{k,l} \gamma_k \gamma_l^* \mathbf{\Gamma}(|\psi_k\rangle \langle\psi_l|). \tag{6.16}$$

Thus we see that if $|\psi_k\rangle$ were all classical, then the distinguishing power would vanish. Therefore we must have that at least one of the states $|\psi_k\rangle$ is not classical, showing that $\boldsymbol{U}$ is not of the form in the theorem statement. We have thus shown that $\boldsymbol{U}$ is not a

permutation matrix up to a phase if and only if $W(\boldsymbol{U}) = \infty$. Conversely, when $\boldsymbol{U}$ is a permutation matrix up to a phase, we know that $W$ vanishes. This completes the proof. □

Thus we have that $W(\boldsymbol{U}) = 0$ if and only if $\boldsymbol{U}$ is a combination of a classical permutation matrix of the classical basis states with phase shifts. We shall discuss infinite $W$ below. For standard qubit error models such as those considered in [9], we similarly find that if the errors occur in the classical basis then they have vanishing $W$. Since Pauli matrices are permutations of the classical basis up to a phase, such models include any Pauli channels on a single qubit. This includes the depolarising, bit-flip, phase-flip and the phase-damping channels. The measure $W$ also vanishes for the amplitude-damping channel, $\Xi_\gamma(\rho) = \boldsymbol{F}_1 \rho \boldsymbol{F}_1^\dagger + \boldsymbol{F}_2 \rho \boldsymbol{F}_2^\dagger$, where $\boldsymbol{F}_1 = |0\rangle\langle 0| + \sqrt{1-\gamma}\,|1\rangle\langle 1|$ and $\boldsymbol{F}_2 = \sqrt{\gamma}\,|0\rangle\langle 1|$. This can be seen by noticing that its Kraus operators correspond to permutation matrices, up to a phase, and applying the convexity property P3.

However, if we change the basis away from the classical basis, quantumness may arise. For example, consider $\boldsymbol{H} \circ \Xi \circ \boldsymbol{H}$ in which the amplitude damping channel sandwiched between Hadamard gates $\boldsymbol{H}$. In this case the measure $W$ is maximized for the state $|0\rangle$, at which point only the generating power contributes, giving the value $W(\boldsymbol{H} \circ \Xi \circ \boldsymbol{H}) = \log(a) + (a/2)\log\big((1+a)/(1-a)\big) + (b/2)\log\big((1+b)/(1-b)\big)$, where $a = \sqrt{1-\gamma}$ and $b = \sqrt{\gamma^2 - \gamma + 1}$. Notice that although the amplitude damping channel is completely classical on its own, removing it from the sequence of operations would leave only $\boldsymbol{H} \circ \boldsymbol{H} = \mathbf{1}$, making the sequence classical.

On the other hand, the Hadamard gate $\boldsymbol{H}$ on its own attains infinite quantumness. The maximum is attained for the states $|\pm\rangle = (|0\rangle \pm |1\rangle)/\sqrt{2}$, for which generating power vanishes and the distinguishing power is infinite, owing to the logarithm in the definition of relative entropy. Infinite quantumness has in itself an intuitive interpretation. In the case of the Hadamard gate, it tells us that it can be used to ascertain that the local state is classical $\rho = (|0\rangle\langle 0| + |1\rangle\langle 1|)/2$ and not $|+\rangle$ with *certainty* in a *finite* average number of measurements. This follows from the probabilistic interpretation of relative entropy (see discussion following Eq. (3.6)). The maximum $W$ continues to be attained for $|+\rangle$ even when $\boldsymbol{H}$ is followed by a depolarizing channel, $\boldsymbol{\Lambda}_\mu(\rho) = \mu\rho + (1-\mu)\mathbf{1}/d$, where $d$ is the dimension of the Hilbert space, although it is no longer infinite.

An example of a local channel generating non-classical correlations is given in Ref. [146]. The map is of the form $\$ = \mathbf{1} \otimes \$_B$, where $\$_B(\rho) = \boldsymbol{E}_1 \rho \boldsymbol{E}_1^\dagger + \boldsymbol{E}_2 \rho \boldsymbol{E}_2^\dagger$ and $\boldsymbol{E}_1 = |0\rangle\langle 0|$, $\boldsymbol{E}_2 = |+\rangle\langle 1|$ [146]. Applying the map \$ to the classical state $\sigma_c = \frac{1}{2}(|0\rangle\langle 0| \otimes |0\rangle\langle 0| + |1\rangle\langle 1| \otimes |1\rangle\langle 1|)$ leads to $\rho = \frac{1}{2}(|0\rangle\langle 0| \otimes |0\rangle\langle 0| + |1\rangle\langle 1| \otimes |+\rangle\langle +|)$, which has non-zero discord but vanishing entanglement. By computing the quantity $S(\$ \circ \mathbf{\Gamma}(\sigma_c) \| \mathbf{\Gamma} \circ \$(\sigma_c))$, we find $W(\$) = 1$, showing that this map is indeed nonclassical. Moreover the distinguishing power of this map vanishes for all states $\rho$.

It is an important question then whether we can further refine our notion of nonclassicality to distinguish between those operations for which $W$ on its own diverges. We propose here a method of regularization that can be used to compare quantumness of these operations. Thus suppose we would like to compare quantumness of two operations, $\$_1$ and $\$_2$. Then the application of the depolarizing channel $\mathbf{\Lambda}_\mu$ will make $W$ finite for $\mathbf{\Lambda}_\mu \$_1$ and $\mathbf{\Lambda}_\mu \circ \$_2$. Evaluating $\lim_{\mu \to 1} W(\mathbf{\Lambda}_\mu \circ \$_1)/W(\mathbf{\Lambda}_\mu \circ \$_2)$ then gives us a meaningful comparison. This can be seen by considering that $2^{-nW(\$)}$ gives us, in the asymptotic limit $n \to \infty$, the probability that we mistakenly believe we posses the ordering $\mathbf{\Gamma} \circ \$$ rather than $\$ \circ \mathbf{\Gamma}$. Denoting with $n_{1,2}$ the number of trials required before we attain a certain probability, we then have that the ratio $W(\$_1)/W(\$_2) = n_2/n_1$. In the limit where $W$ becomes infinite, we can therefore compute the inverse ratio of the average number of trials that we require before distinguishing $\mathbf{\Gamma} \circ \$_k$ from $\$_k \circ \mathbf{\Gamma}$ becomes certain. The depolarising channel can thus play the role of a regulator and the ratio is obtained in the limit where the regulator goes away (turns into identity). By choosing an operation to play the role of the unit of quantumness, we can use this method to determine quantumness of all other operations. The choice of the depolarizing channel as the regulator is somewhat arbitrary here and it is an important open question whether the ratio is independent of this choice.

Finally we look at an entangling operation, specifically a CNOT controlled in the $|\pm\rangle$ basis capable of generating a maximally entangled two-qubit state. As shown in Fig. 6.2 we find that when this is followed by a joint two-qubit depolarising channel $\mathbf{\Lambda}_\mu$ below $\mu = 2/3$ $W$ is maximized by the generating power alone, while above $\mu = 2/3$ it is maximized purely by the distinguishing power. Thus, as this critical point is crossed the maximum quantumness of this noisy CNOT operation switches from being exposed by its ability to generate nonclassicality to its ability to distinguish nonclassicality.

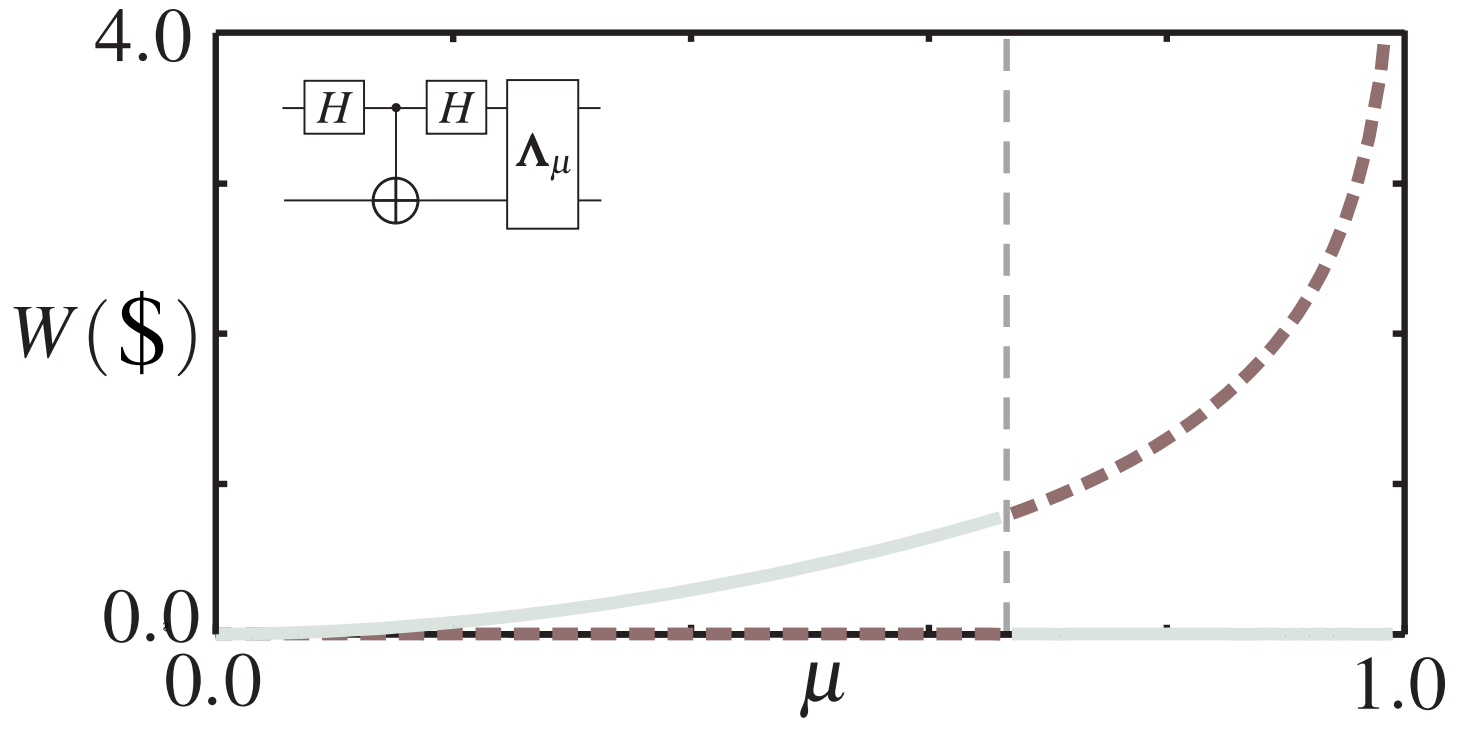

Figure 6.2: Quantumness of CNOT controlled $|\pm\rangle$ basis, followed by the depolarizing channel $\mathbf{\Lambda}_\mu$ as a function of $\mu$. We maximized $W$ and split the expression into the generating power, light blue solid line, and the distinguishing power, dashed brown line. When $\mu$ is small the action of the depolarizing channel is large and degrades distinguishability to such an extent that the generating power dominates. When the $\mu \to 1$, on the other hand, generating power is fundamentally bounded by $\log(d)$ and thus can no longer compete with the distinguishing power which experiences unbounded growth. The maximum changes from the generating to the distinguishing power at $\mu = 2/3$.

## 6.3 Interpretation of quantum discord

Suppose Alice and Bob would like to perform superdense coding using one of two types of states ordered from a source, either a quantum state $\rho$ or a cheaper, completely dephased version $\mathbf{\Gamma}(\rho)$. However, regardless of what they order, they actually receive $\$(\rho)$ or $\$ \circ \mathbf{\Gamma}(\rho)$, respectively, which accounts for transmission imperfections (see figure 6.3). The question we now ask is how much additional information can they transfer using the superdense coding protocol if they ordered the quantum state $\rho$ rather than $\mathbf{\Gamma}(\rho)$. We will show that if $\$$ is classical so $W(\$) = 0$, then the capacity difference is equal to precisely quantum discord of $\$(\rho)$, where $W$ is evaluated with $\mathbf{\Gamma} = \mathbf{1} \otimes \mathbf{\Gamma}_B$ acting on the receiver's (Bob's) side. This one-sided einselection operator is used to match the definition of the standard quantum discord [13, 14]. While this result holds in general, for simplicity we assume that $\rho = |\Phi_d\rangle\langle\Phi_d|$, the maximally entangled state $|\Phi_d\rangle = \sum_\alpha |\alpha\rangle \otimes |\alpha\rangle / \sqrt{d}$. In this case $\mathbf{\Gamma}(\rho)$ is the maximally classically correlated state (see figure 6.4).

As we discussed in chapter 3, the capacity of superdense coding using a state $\rho$ is

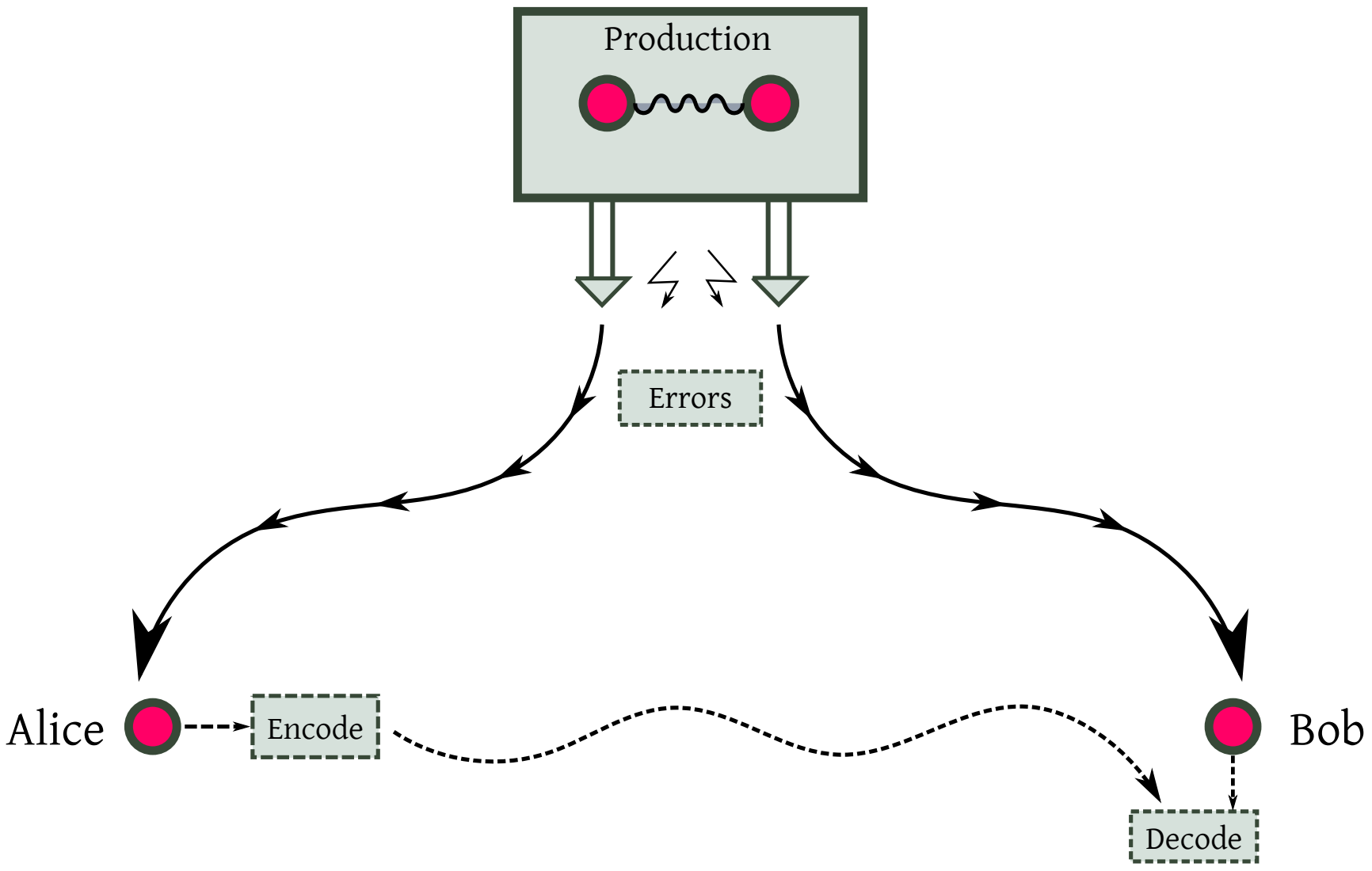


Figure 6.3: Illustration of the dense coding protocol in the presence of classical errors used to highlight the operational meaning of quantum discord. Alice and Bob order a correlated state from a factory with the intention of using it as a resource for communication. The factory offers either a quantum state $\rho$ or its much cheaper classical counterpart $\mathbf{\Gamma}(\rho)$. Regardless of which state they order, it suffers errors and instead of receiving the state they ordered they instead obtain either $\$(\rho)$ or $\$\circ\mathbf{\Gamma}(\rho)$. Whenever the errors are classical, $W(\$) = 0$, we find that the difference between the encoding capacities of the two states is equal to quantum discord $Q\big(\$(\rho)\big)$.

given by $F(\rho^{A|B}) = \log(d_A) - S(\rho^{A|B})$ (see also [60]). Employing Zurek's original definition of quantum discord, Eq. (3.29), and using basic properties of von Neumann entropy, we have that $Q_z(\rho^{A|B}) = S\big(\mathbf{\Gamma}(\rho^{A|B})\big) - S(\rho^{A|B})$. See also the related Eq. (3.35). Assuming $[\mathbf{\Gamma}, \$] = 0$, then gives

$$Q_z(\$ \left|\Phi_d\right\rangle \left\langle\Phi_d\right|^{A|B}) = F(\$ \left|\Phi_d\right\rangle \left\langle\Phi_d\right|^{A|B}) - F(\$\circ\mathbf{\Gamma} \left|\Phi_d\right\rangle \left\langle\Phi_d\right|^{A|B}). \tag{6.17}$$

Extending this to the usual definition of quantum discord $Q$, Eq. (3.30), which involves a minimization over $\mathbf{\Pi}_\alpha$, Eq. (6.17) transforms into

$$Q(\$ \left|\Phi_d\right\rangle \left\langle\Phi_d\right|^{A|B}) = F(\$ \left|\Phi_d\right\rangle \left\langle\Phi_d\right|^{A|B}) - \sup_{\Gamma} F(\$\circ\mathbf{\Gamma} \left|\Phi_d\right\rangle \left\langle\Phi_d\right|^{A|B}). \tag{6.18}$$

Thus quantum discord is the difference in the capacity of superdense coding using the maximally entangled state and the best possible classically correlated state. Our

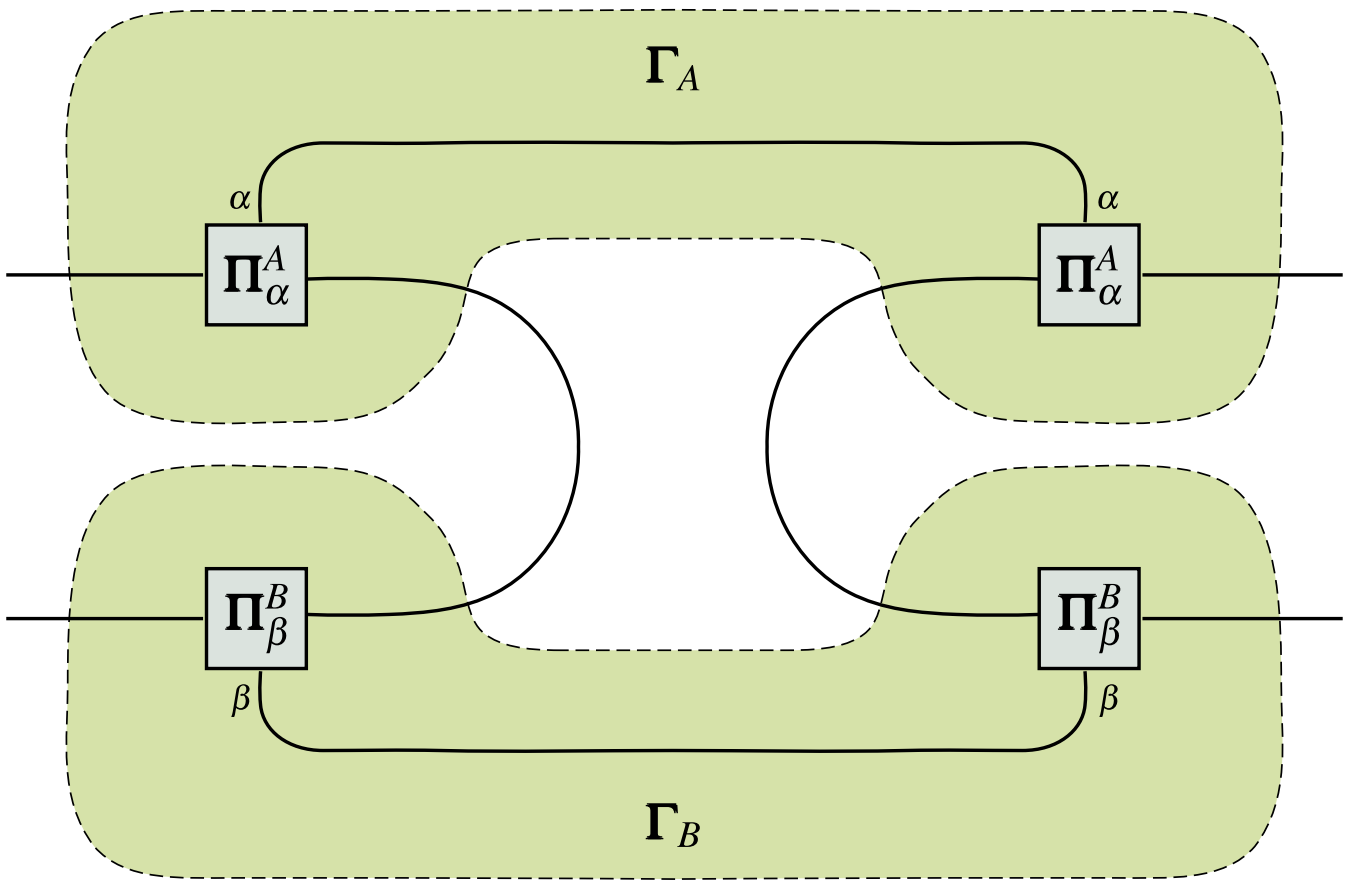


Figure 6.4: Penrose diagram of a maximally entangled state acted on by the decoherence operators $\mathbf{\Gamma}_A = \sum_\alpha \mathbf{\Pi}^A_\alpha(\cdot)\mathbf{\Pi}^A_\alpha$ and similarly for $\mathbf{\Gamma}_B$. The shaded green areas emphasise the parts of the diagram representing the operations $\mathbf{\Gamma}_A$ and $\mathbf{\Gamma}_B$.

results show that quantum advantage can be gained over the initially classical state in the presence of noise even when $\$(|\Phi_d\rangle\langle\Phi_d|)$ is unentangled. This is illustrated in Fig. (6.5) where $\$ = \Lambda_\mu$ is the depolarising channel.

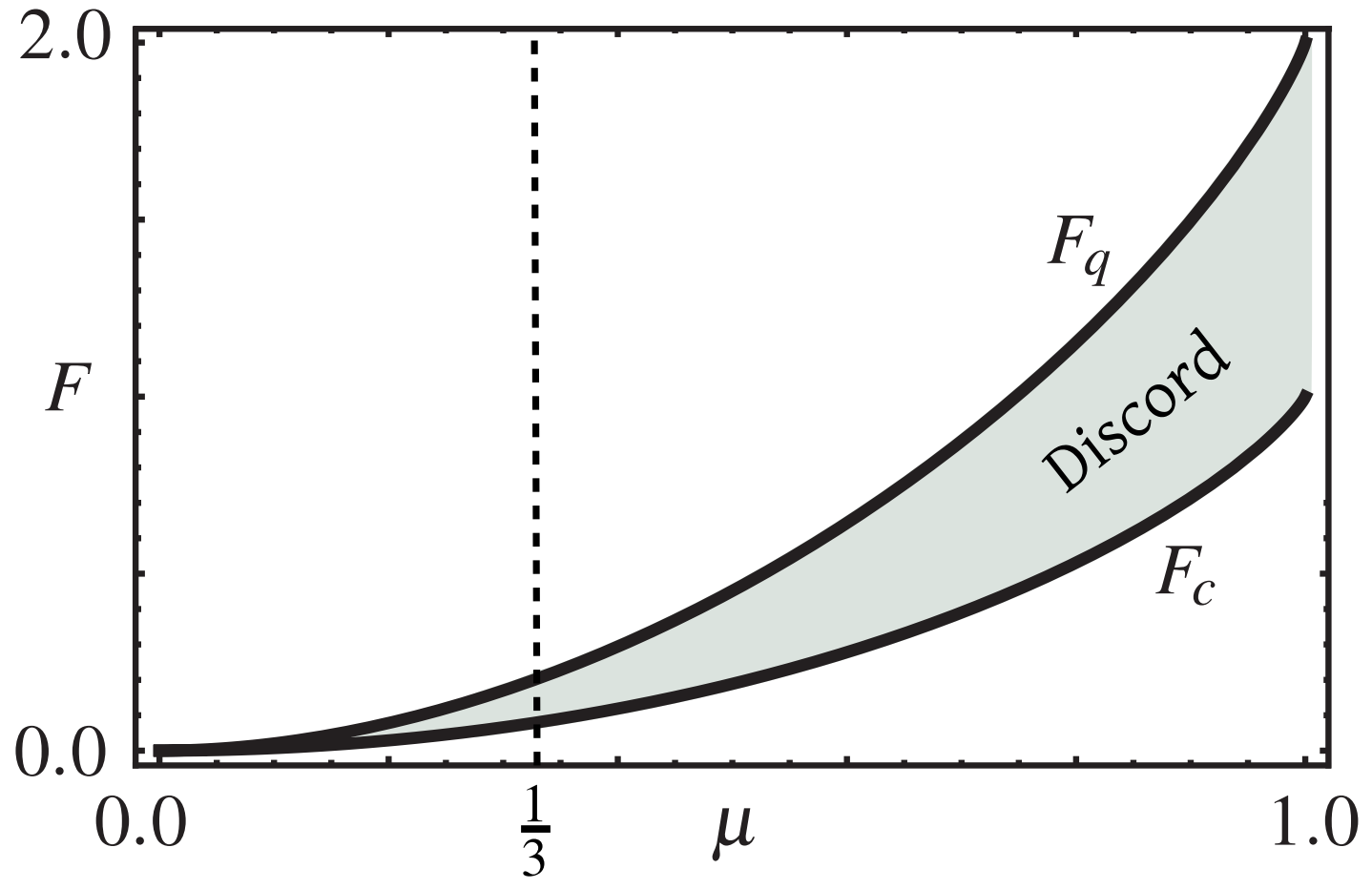


Figure 6.5: Superdense coding capacities $F_q$ using the maximally entangled state and $F_c$ using the classical maximally correlated state when the error is represented by the depolarizing channel $\$ = \mathbf{\Lambda}_\mu$. The performances $F_q$ and $F_c$ correspond to the first and the second terms of the Eq. (6.17). Quantum discord is the difference between $F_q$ and $F_c$, represented by the blue fill. The dashed line indicates the value of $\mu$ where all entanglement in the state is lost due to the depolarising channel. The difference between the two capacities remains non-zero even when all entanglement in the initially maximally entangled state is lost due to the error channel. This point is represented by the dashed line and occurs at $\mu = 1/3$.

# CHAPTER 7

## Concluding remarks

The focus of our thesis has been throughout primarily on quantifying resources and identifying unique features of quantum theory, and especially how these features persist when some of the operations the observer is normally allowed become much more expensive.

In chapter 4 we introduced an effective entanglement functional, measuring the minimum amount of entanglement needed to perform semiquantum nonlocal games with perfect measurements at least as well as with the imperfect measurements. We exploited properties of semiquantum nonlocal games that make them ideal as a gauge of the amount of entanglement in a state. We showed that whenever we can describe the restrictions through one-sided CPMs, an exact result can be obtained where the effective G-concurrence is proportional to the conventional G-concurrence, with proportionality coefficient given by the CPM dependent quality factor. For two-sided CPMs and mixed states the expression gives an upper bound. Although we have only dealt with bipartite entanglement measures in this paper, multipartite measures could similarly be treated using analogous results for multipartite entanglement evolution given in [72], again obtaining an emergent quality factor.

We should note other entanglement gauges could equally be applicable. The crucial property that enabled us to use semiquantum nonlocal games is that whenever

the maximum average payoff function $\mathfrak{p}^*(\rho) \geq \mathfrak{p}^*(\sigma)$ for all semiquantum nonlocal games, we also have $E(\rho) \geq E(\sigma)$. Any set of quantum protocols for which there exists a purely POVM dependent fidelity such that $F(\rho) \geq F(\sigma)$ for all protocols in the set implies $E(\rho) \geq E(\sigma)$ would be similarly useful for the task.

We applied this framework to describe single-particle entanglement. For the case of photons we showed how effective entanglement is attenuated by common measurement noise like amplitude and phase damping. For massive particles we considered the fundamental restrictions imposed by super-selection rules. Computing effective entanglement for strict adherence of the rule we found the entanglement of particles [22]. We further extended this work by considering measurements that can utiliize a BEC with phase uncertainty as a reservoir to partially lift the super-selection rule restriction. The multiplicative factor for concurrence was found explicitly, from which we deduce that whenever the BEC has an undefined phase effective entangled vanishes implying that the same protocol could be performed without entanglement.

We continued the theme of in chapter 5 on examining the nature of nonlocality when a particular restriction - superselection rules - is imposed on the allowed operations. We make a distinction between genuinely and nongenuinely multipartite nonlocality and show that nongenuinely multipartite nonlocality may be generated by using two copies of states in cases where a single copy would not suffice. However, these same states posses a *genuinely* multipartite nonlocality when no SSR are in force.

We therefore examined the conditions under which genuinely multipartite nonlocality is degraded to nongenuinely multipartite when SSR are put in place. We found that whenever the number of parties, $N$, is greater than the number of massive particles $M$ in an exclusively massive-particle state, no genuinely multipartite nonlocality may be observed according to the predictions of quantum mechanics. This is in sharp contrast to non-massive particles, where even a single particle can suffice.

Finally, in chapter 6 we proposed a measure of nonclassicality of quantum operations. The measure is a sum of two independent contributions, the generating power and the distinguishing power, which is non-vanishing if and only if the operation can be used by classical observers to distinguish between quantum and classical states or creates non-classical states out of a classical states. Furthermore, non-vanishing is equivalent to satisfying a commutation relation $[\mathbf{\$}, \mathbf{\Gamma}]$, where $\mathbf{\$}$ is an arbitrary quantum operation and $\mathbf{\Gamma}$ describes the action of full environmental decoherence.

Our measure satisfies several intuitive properties such as convexity and monotonicity under composition of classical maps. In addition, our results show that the einselected relative entropy of discord $Q_g(\rho) = S\big(\rho \| \mathbf{\Gamma}(\rho)\big)$, is non-increasing under the action of classical maps. This is seen by observing that for a classical operation $\$_c$, we have $Q_g\big(\$(\rho)\big) = S\left(\$_c(\rho) \| \mathbf{\Gamma} \circ \$_c(\rho)\right) = S\left(\$_c(\rho) \| \$_c \circ \mathbf{\Gamma}(\rho)\right) \leq S\left(\rho \| \mathbf{\Gamma}(\rho)\right) = Q_g(\rho)$ by monotonicity of relative entropy.

In addition to the operational interpretation of quantum discord, it is interesting to note that there is a complementarity between quantumness of operations and quantumness of states. In particular, one might define a measure of quantumness of states through $W$ as $Q_W(\rho) = \inf_{\$} W(\$)$, where the minimization is over all operations $\$$ generating the state $\rho$ from a classical state. It is clear from Thm. (6.1) that quantumness of any operation generating $\rho$ from a classical state must satisfy $W(\$) \geq Q_z(\rho)$, therefore leading us to conjecture that the lower bound is tight so that $Q_W(\rho) = Q_z(\rho)$. If true, this further implies that $\inf_{\Gamma} Q_{W_{\mathbf{\Gamma}}} = \inf_{\Gamma, \$} W_{\mathbf{\Gamma}}(\$) = Q(\rho)$, the usual quantum discord of a state, provided that $\mathbf{\Gamma}$ is non-identity on only one of the parties.

This thesis as a whole provides a step forward in forming a picture of resources in information theory. Since, as we argued in the introduction, resources exist due to scarcity of certain operations or states, understanding of resources in the presence of restrictions on top of the standard LOCC fills an important gap in our understanding.